\documentclass[aps, prd, twocolumn, superscriptaddress, showpacs]{revtex4}

\usepackage{epsf}
\usepackage{bm}
\usepackage{amssymb}
\usepackage{amsmath}
\usepackage{amsfonts}
\usepackage{dcolumn}
\usepackage{natbib}
\usepackage{color}

\newcommand{\z}{\phantom{0}}

\newcommand{\raiseentry}[1]{\smash{\raise 0.7 em \hbox{#1}}}
\newcommand{\crc}{\mbox{\large $ \circ $}}
\newcommand{\blt}{\mbox{\large $ \bullet $}}
\newcommand{\tms}{\mbox{\footnotesize $ \times $}}
\newcommand{\coconut}{\textsc{CoCoNuT}}

\newenvironment{equationarray}
{\arraycolsep 0.14 em
\begin{eqnarray}}
{\end{eqnarray}}

\newenvironment{equationarray*}
{\arraycolsep 0.14 em
\begin{eqnarray*}}
{\end{eqnarray*}}

\begin{document}

\title{The gravitational wave burst signal from core collapse of
  rotating stars}

\author{Harald Dimmelmeier}
\email{harrydee@mpa-garching.mpg.de}
\affiliation{Department of Physics, Aristotle University of
  Thessaloniki, GR-54124 Thessaloniki, Greece}

\author{Christian D.\ Ott}
\email{cott@as.arizona.edu}
\affiliation{Steward Observatory and Department of Astronomy,
  University of Arizona, Tucson AZ 85721, USA}

\author{Andreas Marek}
\email{amarek@mpa-garching.mpg.de}
\affiliation{Max Planck Institute for Astrophysics,
  Karl-Schwarzschild-Str.\ 1, D-85741 Garching, Germany}

\author{H.-Thomas Janka}
\email{thj@mpa-garching.mpg.de}
\affiliation{Max Planck Institute for Astrophysics,
  Karl-Schwarzschild-Str.\ 1, D-85741 Garching, Germany}

\date{\today}


\begin{abstract}
  We present results from detailed general relativistic simulations of
  stellar core collapse to a proto-neutron star, using two different
  microphysical nonzero-temperature nuclear equations of state as well
  as an approximate description of deleptonization during the collapse
  phase. Investigating a wide variety of rotation rates and profiles
  as well as masses of the progenitor stars and both equations of
  state, we confirm in this very general setup the recent finding that
  a generic gravitational wave burst signal is associated with core
  bounce, already known as type~I in the literature. The previously
  suggested type~II (or ``multiple-bounce'') waveform morphology does
  not occur. Despite this reduction to a single waveform type, we
  demonstrate that it is still possible to constrain the progenitor
  and postbounce rotation based on a combination of the maximum signal
  amplitude and the peak frequency of the emitted gravitational wave
  burst. Our models include to sufficient accuracy the currently known
  necessary physics for the collapse and bounce phase of core-collapse
  supernovae, yielding accurate and reliable gravitational wave signal
  templates for gravitational wave data analysis. In addition, we
  assess the possiblity of nonaxisymmetric instabilities in rotating
  nascent proto-neutron stars. We find strong evidence that in an iron
  core-collapse event the postbounce core cannot reach sufficiently
  rapid rotation to become subject to a classical bar-mode
  instability. However, many of our postbounce core models exhibit
  sufficiently rapid and differential rotation to become subject to
  the recently discovered dynamical instability at low rotation rates.
\end{abstract}


\maketitle


\section{Introduction}
\label{section:introduction}

The final event in the life of a massive star is the catastrophic
collapse of its central, electron-degenerate core composed of
iron-peak nuclei. When silicon shell burning pushes the iron core over
its effective Chandrasekhar mass, collapse is initiated by a
combination of electron capture and photo-disintegration of heavy
nuclei, both leading to a depletion of central pressure support.
Massive stars in the approximate mass range of about $ 10 $ to
$ 100 $ solar masses ($ M_\odot $) experience such a collapse phase until
their homologously contracting~\cite{goldreich_80_a, yahil_83_a} inner
core reaches densities near and above nuclear saturation density where
the nuclear equation of state (EoS) stiffens, leading to an almost
instantaneous rebound of the inner core (core bounce) into
the still supersonically infalling outer core. The hydrodynamic
supernova shock is born, travels outward in radius and mass, but
rapidly loses its kinetic energy to the dissociation of infalling
iron-group nuclei and to neutrinos that deleptonize the immediate
postshock material and stream off from these regions quasi-freely. The
shock stalls, turns into an accretion shock and must be revived to
produce the observable explosion associated with a core-collapse
supernova. Mechanisms of shock revival are still under debate (a
recent review is presented in~\cite{janka_07_a}, but see
also~\cite{burrows_07_a, burrows_07_b, bruenn_06_a}) and may involve
heating of the postshock region by neutrinos, multi-dimensional
hydrodynamic instabilities of the accretion shock, in the postshock
region, and/or in the proto-neutron star, rotation, magnetic fields,
and nuclear burning. If the shock is not revived, black-hole formation
(on a timescale of
$ \sim 1\mbox{\,--\,}2 \mathrm{\ s} $~\cite{liebendoerfer_04_a}) is
inevitable and the stellar collapse event may remain undetected by
conventional astronomy or, perhaps, appear as a gamma-ray burst if the
progenitor star has a compact enough envelope and sufficiently rapid
rotation in its central regions~\cite{woosley_06_a, dessart_08_a}.

Conventional astronomy can constrain core-collapse supernova theory
and the supernova explosion mechanism via secondary observables only,
e.g., the explosion energy, ejecta morphology, nucleosynthesis yields,
residue neutron star or black hole mass and proper motion, and pulsar
magnetic fields. Neutrinos and gravitational waves, on the other hand,
are emitted deep inside the supernova core and travel to observers on
Earth practically unscathed by intervening material. They can act as
messengers to provide first-hand and live dynamical information on the
intricate multi-dimensional dynamics of the proto-neutron star and
postshock region and may constrain directly the core-collapse
supernova mechanism. Importantly, core-collapse events that do not
produce the canonical observational astronomical signature or whose
observational display is shrouded from view can still be observed in
neutrinos and gravitational waves if occurring sufficiently close to
Earth.

Gravitational waves, in contrast to neutrinos, have not yet been
observed directly, but an international array of gravitational wave
observatories (see, e.g., \cite{aufmuth_05}) is active and taking data.
Since gravitational waves from astrophysical sources are expected
to be weak, their detection is notoriously difficult and involves
extensive signal processing and detailed analysis of the detector
output. Chances for the detection of an astrophysical event of
gravitational wave emission are significantly enhanced if accurate
theoretical knowledge of the expected gravitational wave signature
from such an event is at hand.

Theoretical predictions of the gravitational wave signature from a
core-collapse supernova are complicated, since the emission mechanisms are
very diverse. While the prospective gravitational wave burst signal from
the collapse, bounce, and the very early postbounce phase is present
only when the core rotates~\cite{mueller_82_a, moenchmeyer_91_a,
zwerger_97_a, dimmelmeier_02_a, kotake_03_a, ott_04_a, ott_07_a,
dimmelmeier_07_a}, gravitational wave signals with sizeable amplitudes
can also be expected from convective motions at postbounce times,
instabilities of the standing accretion shock, anisotropic neutrino
emission, excitation of various oscillations in the proto-neutron
star, or nonaxisymmetric rotational instabilities~\cite{rampp_98_a,
mueller_04_a, shibata_05_a, ott_05_a, ott_06_a, ott_07_a}.

In the observational search for gravitational waves from merging black
hole or neutron star binaries, powerful data analysis algorithms
such as matched filtering are applicable, as the waveform from the
inspiral phase can be modeled with high accuracy (see,
e.g., \cite{blanchet_06_a}) and gravitational wave data analysts
already have access to robust template waveforms that depend only on a
limited number of macroscopic parameters. In contrast, the complete
gravitational wave signature of a core-collapse supernova cannot be
predicted with template-level accuracy as the postbounce dynamics
involve chaotic processes (turbulence, [magneto-] hydrodynamic
instabilities) that are sensitive not only to a multitude of
precollapse parameters, but also to small-scale perturbations of any
of the hydrodynamic variables.

While the complete supernova gravitational wave signature may remain
inaccessible to template-based data analysis, a number of individual
constituent emission processes, in particular those involving
coherent global bulk dynamics and/or rotation, allow, in principle,
for accurate and robust waveform predictions that may be applied to
template-based searches. Rotating core collapse and core bounce as
well as pulsations or nonaxisymmetric rotational deformations of a
proto-neutron star constitute this group of processes. Among them,
rotating collapse and bounce is the historically most extensively
studied case (see, e.g., \cite{ott_06_b} for a historical review) and
may be the most promising for becoming robustly predictable in its
gravitational wave emission. Yet, to date, the gravitational wave
signal from rotating stellar core collapse and bounce has not
been predicted with the desired accuracy and robustness.

These deficiencies of previous simulations result from the fact that
the physically realistic modeling of core collapse requires a general
relativistic description of consistently coupled gravity and
hydrodynamics in conjunction with a microphysical treatment of the
sub- and supernuclear EoS, electron capture on heavy nuclei and free
protons, and neutrino radiation transport. Only very few
multi-dimensional general relativistic codes have recently begun to
approach these requirements~\cite{ott_07_a, dimmelmeier_07_a}. In
addition, the properties of the EoS around and above nuclear density
are not very well constrained by theory or experiments. The same
applies to the rotation rate and angular velocity profile of the
progenitor core, which are also not directly accessible by observation
and very difficult to model numerically in stellar evolution codes.
Furthermore, variations with progenitor structure and mass are to be
expected. Therefore, the influence of rotation and progenitor
structure on the collapse and bounce dynamics and thus the
gravitational wave burst signal must be investigated by extensive and
computationally expensive parameter studies.

Previous parameter studies have considered a large variety of rotation
rates and progenitor core configurations, but generally ignored
important microphysical aspects and/or the influence of general
relativity. M\"onchmeyer et al.~\cite{moenchmeyer_91_a} performed
axisymmetric Newtonian calculations with progenitor models from
stellar evolutionary studies. They employed the microphysical nuclear
EoS of Hillebrandt and Wolff~\cite{hillebrandt_85_a} and included
deleptonization via a neutrino leakage scheme and electron capture on
free protons. Capture on heavy nuclei was neglected, which resulted in
a too high electron fraction $ Y_e $ at core bounce and a consequently
overestimated inner core mass~\cite{yahil_83_a, van_riper_81_a}. In
that study a limited set of four calculations was computed and two
qualitatively and quantitatively different types of gravitational wave
burst signals were identified. Their morphology can be classified
alongside with the collapse and bounce dynamics: \emph{Type~I} signals
are emitted when the collapse of the quasi-homologously contracting
inner core is not strongly influenced by rotation, but stopped by a
\emph{pressure-dominated bounce} due to the stiffening of the EoS near
nuclear density
$ \rho_\mathrm{nuc} \approx 2 \times 10^{14} \mathrm{\ g\ cm}^{-3} $,
where the adiabatic index $ \gamma_\mathrm{eos} $ rises above
$ 4 / 3 $. This leads to a bounce with a maximum core density
$ \rho_\mathrm{max} \ge \rho_\mathrm{nuc} $. \emph{Type~II} signals
occur when centrifugal forces, which grow during contraction owing to
angular momentum conservation, are sufficiently strong to halt the
collapse, resulting in consecutive (typically multiple)
\emph{centrifugal bounces} with intermediate coherent re-expansion of
the inner core, seen as density drops by sometimes more than an order
of magnitude; thus here $ \rho_\mathrm{max} < \rho_\mathrm{nuc} $
after bounce. Type~I and~II dynamics and waveforms were also found in
the more recent Newtonian studies by Kotake et al.~\cite{kotake_03_a},
who employed a more complete leakage/capture scheme, but still
obtained too high $ Y_e $ at bounce, and by Ott et al.~\cite{ott_04_a},
who performed an extensive parameter study and for the first time also
considered variations in progenitor star structure, but neglected
deleptonization during collapse.

Zwerger and M\"uller~\cite{zwerger_97_a} carried out an extensive
two-dimensional Newtonian study of rotating collapse of idealized
polytropes in rotational equilibrium~\cite{komatsu_89_a} with a
simplified \emph{hybrid} EoS, consisting of a polytropic and a thermal
component~\cite{janka_93_a}. Electron capture during collapse was
mimicked by an instantaneous lowering of the adiabatic index
$ \gamma_\mathrm{eos} $ from its initial value of $ 4 / 3 $ to trigger
the onset of collapse. At $ \rho_\mathrm{nuc} $, the adiabatic index
was raised to $ \gtrsim 2 $ to qualitatively model the stiffening of
the nuclear EoS. Zwerger and M\"uller also obtained the previously
suggested signal types and introduced \emph{type~III} signals that
appear in a pressure-dominated bounce when the inner core has a very
small mass due to very efficient electron capture (approximated
in~\cite{zwerger_97_a} via a $ \gamma_\mathrm{eos} \lesssim 1.29 $ in
their hybrid EoS). Obergaulinger et al.~\cite{obergaulinger_06_a} also
employed the hybrid EoS, but included magnetic fields. They introduced
the additional dynamics/signal \emph{type~IV}, which occurs only in the
case of very strong precollapse core magnetization. They found that
weak to moderate core magnetization in agreement with predictions from
stellar evolution theory (see, e.g., \cite{heger_05_a}) has little
effect on the collapse and bounce dynamics and the resulting
gravitational wave signal. This finding is in agreement
with~\cite{kotake_04_a} (see also~\cite{burrows_07_b, cerda_07_a}),
where magneto-rotational collapse simulations were performed, a
smaller model set was considered, but the neutrino leakage scheme
of~\cite{kotake_03_a} was employed and it was made use of two
different microphysical EoSs to study the EoS dependence of the
collapse dynamics and gravitational wave signal.

The first extensive set of general relativistic simulations of
rotating iron core collapse to a proto-neutron star were
presented by Dimmelmeier et al.~\cite{dimmelmeier_02_a}, who employed
an analytic hybrid EoS and polytropic precollapse models in rotational
equilibrium as initial data (but see also the pioneering early work
of~\cite{nakamura_81_a}). These simulations were subsequently
confirmed in~\cite{shibata_04_a, cerda_05_a, ott_06_b,
  obergaulinger_06_b}. Dimmelmeier et al.\ studied a subset of the models
in~\cite{zwerger_97_a} in the same parameter space of rotation rate
and degree of differential rotation, and found that general
relativistic effects counteract centrifugal support and shift the
occurrence of type~II dynamics and wave signals to a higher precollapse
rotation rate at a fixed degree of differential rotation.

Recently, new general relativistic simulations of rotating core
collapse in two and three dimensions were carried out by Ott et
al.~\cite{ott_06_b, ott_07_a, ott_07_b} who included the
microphysical EoS of Shen et al.~\cite{shen_98_a}, precollapse models
from stellar evolutionary calculations as well as an approximate
deleptonization scheme~\cite{liebendoerfer_05_a}. The results of these
calculations indicate that the gravitational wave burst signal
associated with rotating core collapse is \emph{exclusively of
  type~I}. In addition, the simulations showed that rotating stellar
iron cores stay axisymmetric throughout collapse and bounce, and only
at postbounce times develop nonaxisymmetric features.

In a general relativistic two-dimensional follow-up study, Dimmelmeier
et al.~\cite{dimmelmeier_07_a, dimmelmeier_07_b} considerably extended
the number of models and comprehensively explored a wide parameter
space of precollapse rotational configurations. Even for this more general
setup they found gravitational wave signals solely of type~I form,
although for rapid precollapse rotation some of their models
experience a core bounce due to centrifugal forces only, which however
is always a \emph{single centrifugal bounce} rather than the multiple
ones observed in earlier work (see, e.g., \cite{zwerger_97_a, 
dimmelmeier_02_a, ott_04_a}). They
identified the physical conditions that lead to the emergence of this
generic gravitational wave signal type and quantified their relative
influence. These results strongly suggest that the waveform of the
gravitational wave burst signal from the collapse of rotating iron
cores in a core-collapse event is much more generic than previously
anticipated.

In this work, we extend the above study of the gravitational wave
signal from rotating core collapse and consider not only variations in
the precollapse rotational configuration, but also in progenitor
structure and nuclear EoS. In this way, we carry out the to-date
largest and most complete parameter study of rotating stellar core
collapse that includes all the (known) necessary physics to produce
reliable predictions of the gravitational wave signal associated with
rotating collapse and bounce. All our computed gravitational wave
signals are made available to the detector data analysis community in
a freely accessible waveform catalog~\cite{wave_catalog}.

We perform a large number of two-dimensional simulations with our
general relativistic core-collapse code \coconut\ and employ $ 11.2 $,
$ 15.0 $, $ 20.0 $, and $ 40.0 \, M_\odot $ (masses at zero-age main
sequence) precollapse stellar models from the stellar
evolutionary studies of Heger et al.~\cite{heger_00_a, heger_05_a}. In
addition to the EoS by Shen et al.~\cite{shen_98_a} used in our
previous studies, we also calculate models with the EoS by Lattimer
and Swesty~\cite{lattimer_91_a}. We describe in detail and explain
comprehensively the qualitative and quantitative aspects of the
collapse and bounce dynamics and the resultant gravitational wave
signal. We lay out the individual effects of general relativity,
deleptonization, precollapse stellar structure and rotational
configuration, and nuclear EoS on the gravitational wave signature
from rotating core collapse. We study the prospects for
nonaxisymmetric rotational instabilities in our postbounce cores,
which could lead to an enhancement of the gravitational wave
signature. Furthermore, we set our model gravitational radiation
waveforms in context with present and future detector technology and
assess their detectability.

This paper is organized as follows: In
Section~\ref{section:physical_model} we introduce our treatment of the
general relativistic spacetime curvature and hydrodynamics equations.
Furthermore, we introduce our variants of the two
microphysical EoS we employ, the scheme for deleptonization and
neutrino pressure contributions, our precollapse model set, and the
gravitational wave extraction technique employed.
Section~\ref{section:numerical_methods} discusses the numerical
methods used in the \coconut\ code and the computational grid setup
for the simulations presented in this paper. In
Section~\ref{section:collapse_dynamics} we present the collapse
dynamics and waveform morphology of our simulated models, while in
Section~\ref{section:density_structure_and_waveform} we investigate
the stratification of the postbounce core and its impact on the
gravitational wave signal. The detection prospects for the
gravitational wave burst from core bounce are discussed in
Section~\ref{section:detection_prospects}, while the rotational
configuration of the proto-neutron star and its susceptibility to
nonaxisymmetric rotational instabilities are examined in
Section~\ref{section:rotation_rate}. Finally, in
Section~\ref{section:summary}, we summarize and discuss our
results.

Throughout the paper we use a spacelike signature $ (-, +, +, +) $ and
units in which $ c = G = 1 $. Greek indices run from 0 to 3, Latin
indices from 1 to 3, and we adopt the standard Einstein summation
convention.


\section{Physical model and equations}
\label{section:physical_model}


\subsection{General relativistic hydrodynamics}
\label{subsection:gr_hydro}

We adopt the Arnowitt--Deser--Misner (ADM) $ 3 + 1 $ formalism of
general relativity to foliate the spacetime endowed with a four-metric
$ g_{\mu\nu} $ into spacelike hypersurfaces~\cite{york_79_a}. In this
approach the line element reads
\begin{equation}
  ds^2 = - \alpha^2 \, dt^2 +
  \gamma_{ij} (dx^i + \beta^i \, dt) (dx^j + \beta^j \, dt),
  \label{equation:line_element}
\end{equation}
where $ \alpha $ is the lapse function, $ \beta^i $ is the shift
vector, and $ \gamma_{ij} $ is the spatial three-metric induced in
each hypersurface.

The hydrodynamic evolution of a perfect fluid in general relativity
with four-velocity $ u^{\,\mu} $, rest-mass current
$ J^{\,\mu} = \rho u^{\,\mu} $, and stress-energy tensor
$ T^{\mu \nu} = \rho h u^{\,\mu} u^{\,\nu} + P g^{\,\mu \nu} $ is
determined by a system of local conservation equations,
\begin{equation}
  \nabla_{\!\mu} J^\mu = 0, \quad \nabla_{\!\mu} T^{\mu \nu} = 0,
  \label{equation:equations_of_motion}
\end{equation}
where $ \nabla_{\!\mu} $ denotes the covariant derivative with respect
to the four-metric. Here $ \rho $ is the rest-mass density,
$ h = 1 + \epsilon + P / \rho $ is the specific enthalpy, $ P $ is the
fluid pressure, and the three-velocity with respect to an Eulerian
observer moving orthogonally to the spacelike hypersurfaces is given
by $ v^{\,i} = u^{\,i} / (\alpha u^{\,0}) + \beta^i / \alpha $. We
define a set of conserved variables as
\begin{equation}
  D = \rho W,
  \quad
  S^i = \rho h W^{\,2} v^i,
  \quad
  \tau = \rho h W^{\,2} - P - D.
  \label{equation:conserved_quantities}
\end{equation}
In the above expressions $ W = \alpha u^0 $ is the Lorentz factor,
which satisfies the relation $ W = 1 / \sqrt{1 - v_i v^i} $.

The local conservation laws~(\ref{equation:equations_of_motion})
are written as a first-order, flux-conservative system of
hyperbolic equations~\cite{banyuls_97_a},
\begin{equation}
  \frac{\partial \sqrt{\gamma\,} \bm{U}}{\partial t} +
  \frac{\partial \sqrt{- g\,} \bm{F}^{\,i}}{\partial x^{\,i}} =
  \sqrt{- g\,} \bm{S},
  \label{equation:conservation_equations}
\end{equation}
with
\begin{equationarray}
  \bm{U} & = & [D, S_j, \tau, D Y_e],
  \label{equation:hydro_conservation_equation_constituents_1}
  \\
  \bm{F}^i & = & \left[ D \hat{v}^i, S_j \hat{v}^i + \delta^i_j P,
  \tau \hat{v}^i + P v^i, D Y_e \hat{v}^i \right],
  \label{equation:hydro_conservation_equation_constituents_2}
  \\
  \bm{S} & = & \biggl[ 0, \frac{T^{\mu \nu}}{2}
  \frac{\partial g_{\mu \nu}}{\partial x^j} -
  \frac{\partial P_\nu}{\partial x^j},
  T^{00} \!\left(\!\! K_{ij} \beta^i \beta^j - \beta^i
  \frac{\partial \alpha}{\partial x^i} \!\!\right)\! + \quad
  \label{equation:hydro_conservation_equation_constituents_3}
  \\
  & & \;\; T^{0i} \!\left(\!\! 2 K_{ij} \beta^j -
  \frac{\partial \alpha}{\partial x^i} \!\!\right)\! +
  T^{ij} K_{ij} - v^i \frac{\partial P_\nu}{\partial x^i}, 0 \biggr].
  \nonumber
\end{equationarray}%
Here $ \hat{v}^{\,i} = v^{\,i} - \beta^i / \alpha $, and $ g $ and
$ \gamma $ are the determinant of $ g_{\mu\nu} $ and $ \gamma_{ij} $,
respectively, with $ \sqrt{-g} = \alpha \sqrt{\gamma} $.
$ \Gamma^{\,\lambda}_{\mu \nu} $ are the four-Christoffel symbols.
Since we use a microphysical EoS that requires information on the
local electron fraction per baryon $ Y_e $, we add an advection
equation for the quantity $ D Y_e $ to the standard form of the
conservation equations~(\ref{equation:conservation_equations}).
The radiation stress due to the neutrino pressure $ P_\nu $ (as
defined in Section~\ref{subsection:deleptonization}), is included
in the form of an additive term in the source of both the momentum and
energy equations. Note also that here we use an analytically
equivalent reformulation of the energy source term in contrast to the
one presented in~\cite{dimmelmeier_02_a}.


\subsection{Metric equations in the conformal flatness approximation}
\label{subsection:metric_equations}

Using the ADM $ 3 + 1 $ formalism, the Einstein equations split into a
coupled set of first-order evolution equations for the three-metric
$ \gamma_{ij} $ and the extrinsic curvature $ K_{ij} $,
\begin{equationarray}
  \partial_t \gamma_{ij} & = & - 2 \alpha K_{ij} +
  \nabla_{\!i} \beta_j + \nabla_{\!j} \beta_i,
  \label{equation:adm_equations_1}
  \\
  \partial_t K_{ij} & = & - \nabla_{\!i} \nabla_{\!j} \alpha +
  \alpha \left( R_{ij} - 2 K_{ik} K_j^k \right) +
  \label{equation:adm_equations_2}
  \\
  & & \beta^k \nabla_{\!k} K_{ij} + K_{ik} \nabla_{\!j} \beta^k +
  K_{jk} \nabla_{\!i} \beta^k -
  \nonumber
  \\
  & & 8 \pi \alpha \left( S_{ij} - \frac{\gamma_{ij}}{2}
  \left( S_k^k - \rho_\mathrm{ADM} \right) \right),
  \nonumber
\end{equationarray}%
and constraint equations,
\begin{equationarray}
  0 & = & R - K_{ij} K^{ij} - 16 \pi \rho_\mathrm{ADM},
  \label{equation:adm_equations_3}
  \\
  0 & = & \nabla_{\!i} K^{ij} - 8 \pi S^j.
  \label{equation:adm_equations_4}
\end{equationarray}%
In the above equations, $ \nabla_{\!i} $ is the covariant derivative
with respect to the three-metric $ \gamma_{ij} $, $ R_{ij} $ is the
three-Ricci tensor and $ R $ is the scalar three-curvature. The
projection of the stress-energy tensor onto the spatial hypersurface
is $ S_{ij} = \rho h W^2 v_i v_j + \gamma_{ij} P $, the ADM energy
density is given by $ \rho_\mathrm{ADM} = \rho h W^2 - P $, and
$ S^j = \rho h W^2 v^i $ is the momentum density. In addition, we have
chosen the maximal slicing condition for which the trace of the
extrinsic curvature vanishes: $ K = 0 $.

In order to simplify the ADM metric equations and to ameliorate the
stability properties when numerically solving those equations, we
employ the conformal flatness condition (CFC) introduced
in~\cite{isenberg_78_a} and first used in a pseudo-evolutionary
context in~\cite{wilson_96_a}. In this approximation the spatial
three-metric is replaced by the conformally flat three-metric,
$ \gamma_{ij} = \phi^4 \hat{\gamma}_{ij} $, where
$ \hat{\gamma}_{ij} $ is the flat-space metric and $ \phi $ is the
conformal factor. Then the metric
equations~(\ref{equation:adm_equations_1}--\ref{equation:adm_equations_4})
reduce to a set of elliptic equations for $ \phi $, $ \alpha $, and
$ \beta^i $,
\begin{equationarray}
  \hat{\Delta} \phi & = & - 2 \pi \phi^5
  \left( E + \frac{K_{ij}K^{ij}}{16 \pi} \right),
  \label{equation:cfc_equations_1}
  \\
  \hat{\Delta} (\alpha \phi) & = & 2 \pi \alpha \phi^5
  \left( E + 2 S + \frac{7 K_{ij}K^{ij}}{16 \pi} \right),
  \label{equation:cfc_equations_2}
  \\
  \hat{\Delta} \beta^i & = & 16 \pi \alpha \phi^4 S^i +
  2 \phi^{10} K^{ij} \hat{\nabla}_j \frac{\alpha}{\phi^6} -
  \frac{1}{3} \hat{\nabla}^i \hat{\nabla}_k \beta^k, \qquad
  \label{equation:cfc_equations_3}
\end{equationarray}%
where $ \hat{\Delta} $ and $ \hat{\nabla} $ are the Laplace and
covariant derivative operators associated with the flat three-metric,
and $ S = \gamma^{ij} S_{ij} $. The CFC metric
equations~(\ref{equation:cfc_equations_1}--\ref{equation:cfc_equations_3})
do not contain explicit time derivatives, and thus the metric
components are evaluated in a fully constrained approach.

Imposing CFC in a spherically symmetric spacetime is equivalent to
solving the exact Einstein equations. For nonspherical configurations
the CFC approximation may be roughly regarded as full general
relativity without the dynamical degrees of freedom of the
gravitational field that correspond to the gravitational wave
content~\cite{york_71_a}. However, even spacetimes that do not contain
gravitational waves can be not conformally flat. A prime example are
the spacetime of a Kerr black hole~\cite{garat_00_a} or rotating
fluids in equilibrium. For rapidly rotating models of stationary
neutron stars the deviation of certain metric components from
conformal flatness has been shown to reach up to $ \sim 5\% $ in
extreme cases~\cite{cook_96_a}, while the oscillation frequencies of
such models typically deviate even less from the corresponding values
obtained in full general relativistic
simulations~\cite{dimmelmeier_06_a}. In the context of rotating
stellar core collapse the excellent quality
of the CFC approximation has been demonstrated
extensively~\cite{shibata_04_a, cerda_05_a, ott_07_a}.

Due to its fully constrained nature, the CFC approximation permits a
straightforward and numerically more robust implementation of the
metric equations in coordinate systems containing coordinate
singularities (e.g., spherical polar coordinates) compared to a Cauchy
free-evolution scheme. Furthermore, by definition it allows no
constraint violations, which is a significant benefit in cases where a
perturbation is added to the initial data. More details on the CFC
equations can be found in, e.g., \cite{dimmelmeier_02_a}.


\subsection{Equations of state}
\label{subsection:eos}

In our simulations we employ two tabulated nonzero-temperature
equations of state, the one by Shen et al.~\cite{shen_98_a, shen_98_b}
(Shen EoS), and the one by Lattimer and Swesty~\cite{lattimer_91_a}
(LS EoS). The LS EoS is based on a compressible liquid-drop
model~\cite{lattimer_85_a}. The transition from inhomogeneous to
homogeneous matter is established by a Maxwell construction, and the
nucleon-nucleon interactions are expressed by a Skyrme force. In our
version of this EoS, the incompressibility modulus of bulk nuclear
matter is taken to be $ 180 \mathrm{\ MeV} $ and the symmetry energy
parameter has a value of $ 29.3 \mathrm{\ MeV} $. In contrast, the
Shen EoS is based on a relativistic mean field model and is extended
with the Thomas--Fermi approximation to describe the homogeneous phase
of matter as well as the inhomogeneous matter composition. The
parameter for the incompressibility of nuclear matter is
$ 281 \mathrm{\ MeV} $ and the symmetry energy has a value of
$ 36.9 \mathrm{\ MeV} $.

Both EoSs employed in this study are the same as in Marek et
al.~\cite{marek_05_a} and include contributions of baryons,
electrons, positrons, and photons. Furthermore, in this study the
LS EoS has been extended to densities below
$ \rho = 5.8 \times 10^7 \mathrm{g\ cm}^{-3} $ by a smooth
transition to the Shen EoS which is tabulated down to
$ \rho = 6.4 \times 10^5 \mathrm{g\ cm}^{-3} $.

The microphysical EoS returns the fluid pressure (and additional
thermodynamic quantities) as a function of $ (\rho, T, Y_e) $, where
$ T $ is the temperature. Since the hydrodynamic
equations~(\ref{equation:conservation_equations}) operate on the
specific internal energy $ \epsilon $, we determine the corresponding
temperature $ T $ iteratively with a Newton--Raphson scheme and the
EoS table. All interpolations are carried out in tri-linear fashion
and the tables are sufficiently densely spaced to lead to an artificial
entropy increase in an adiabatic collapse by not more than
$ \sim 2\% $.


\subsection{Deleptonization and neutrino pressure}
\label{subsection:deleptonization}

Electron capture on free protons and heavy nuclei during collapse
reduces $ Y_e $ (i.e., ``deleptonizes'' the collapsing core) and
consequently decreases the size of the homologously collapsing inner
core that depends on the average value of $ Y_e $ in a roughly
quadratic way (see, e.g., \cite{martinez_06_a}). The material of the
inner core is in sonic contact and determines the dynamics and the
gravitational wave signal at core bounce and in the early postbounce
phases. Hence, deleptonization has a direct influence on the collapse
dynamics and the gravitational wave signal, and thus it is essential
to include deleptonization during collapse.

\begin{figure}[t]
  \epsfxsize = 8.6 cm
  \centerline{\epsfbox{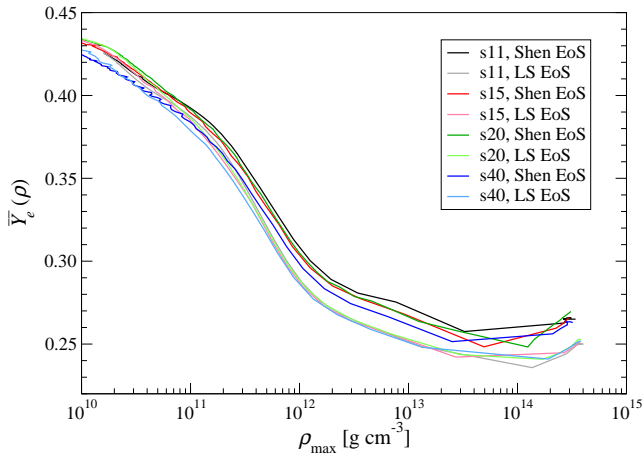}}
  \caption{Electron fraction $ \overline{Y_e\!} \, $ obtained from
    detailed spherically symmetric calculations with Boltzmann
    neutrino transport versus the maximum density $ \rho_\mathrm{max} $ 
    in the collapsing core. The EoS is encoded in dark hues for the
    Shen EoS and light hues for the LS EoS with the basis color
    specifying the progenitor mass.}
  \label{figure:y_e_profiles}
\end{figure}

Since multi-dimensional radiation-hydrodynamics calculations in
general relativity are not yet computationally feasible, in our
simulations we make use of a recently proposed approximative
scheme~\cite{liebendoerfer_05_a} where deleptonization is parametrized
based on data from detailed spherically symmetric calculations with
Boltzmann neutrino transport, for which (as in~\cite{dimmelmeier_07_a})
we take the latest available electron capture
rates~\cite{langanke_00_a}. Following the main assumption
in~\cite{liebendoerfer_05_a} that the local electron fraction for each
fluid element during the contraction phase can be modeled rather
accurately by a dependence on the density only, these simulations
yield a universal relation $ \overline{Y_e\!} \, (\rho) $.
Furthermore, we find that this relation varies only slightly with
progenitor mass, as shown in Fig.~\ref{figure:y_e_profiles}, where
models with identical progenitor but different EoS have the same
color, but different hues (e.g., dark green versus light green for the
s20 progenitor). Consequently, we utilize the $ 20.0 \, M_\odot $
progenitor to create such a profile $ \overline{Y_e\!} \, (\rho) $ for
each of the two EoSs. This profile is then used to correct the value
of $ Y_e $ obtained from the advection by an amount
\begin{equation}
  \Delta Y_e = \min \, [0, \overline{Y_e\!} \, (\rho) - Y_e]
  \label{equation:chance_of_electron_fraction}
\end{equation}
after each time integration step. This procedure assures that $ Y_e $
approaches the phenomenological input profile
$ \overline{Y_e\!}\, (\rho) $ with the constraint that $ \Delta Y_e $
must be negative. Accordingly, in order to model the entropy loss by
neutrinos escaping the collapsing core, for densities below an adopted
neutrino trapping density
$ \rho_\mathrm{tr} = 2 \times 10^{12} \mathrm{\ g\ cm}^{-3} $ the
internal specific energy $ \epsilon $ is re-adjusted at constant
$ \rho $ and $ Y_e $ such that the specific entropy per baryon $ s $
is changed by
\begin{equation}
  \Delta s = - \Delta Y_e
  \frac{\mu_p - \mu_n + \mu_e - E_\nu}{k_\mathrm{B} T},
  \label{equation:chance_of_entropy}
\end{equation}
where $ E_\nu = 10 \mathrm{\ MeV} $ is an average escape energy for
the neutrinos, $ k_\mathrm{B} $ is the Boltzmann constant and
where $ \mu_p $, $ \mu_n $, and $ \mu_e $ are the proton, neutron, and
electron chemical potentials, respectively. Note that when
equilibrium between neutrinos and matter (i.e., $ \beta $-equilibrium)
is established, this balance requires
$ \mu_\nu = \mu_p - \mu_n + \mu_e $ for the neutrino chemical
potential $ \mu_\nu $.

We stop deleptonization at the time of core bounce (i.e., as soon
as the specific entropy $ s $ per baryon exceeds $ 3 k_\mathrm{B} $ at
the outer boundary of the inner core). After core bounce, for lack of
a simple yet accurate approximation scheme for treating the further
deleptonization in the nascent proto-neutron star, we advect $ Y_e $
only passively according to the conservation
equation~(\ref{equation:conservation_equations}), although this
effectively prevents the factual cooling and contraction of the
proto-neutron star.

In all collapse phases, however, as in~\cite{liebendoerfer_05_a} we
approximate the pressure contribution of the neutrinos by that of an
ideal Fermi gas,
\begin{equation}
  P_\nu = \frac{4 \pi (k_\mathrm{B} T)^4}{3 (h c)^3}
  F_3 \! \left( \frac{\mu_\nu}{k_\mathrm{B} T} \right),
\end{equation}
with $ F_3 $ being the Fermi--Dirac function of order 3. The neutrino
pressure is included only in the regime which is optically thick to
neutrinos, which we define for densities above $ \rho_\mathrm{tr} $.


\subsection{Initial models}
\label{subsection:models}

All presupernova stellar models available to-date are end products of
Newtonian spherically-symmetric stellar evolutionary calculations from
hydrogen burning on the main sequence to the onset of core collapse by
photo-dissociation of heavy nuclei and electron captures (see,
e.g., \cite{woosley_02_a}). Here, we employ various nonrotating models
of~\cite{woosley_02_a} with zero-age main sequence masses
$ M_\mathrm{prog} = 11.2 \, M_\odot $ (core model s11.2, here for
simplicity labeled s11), $ 15.0 \, M_\odot $ (core model s15),
$ 20.0 \, M_\odot $ (core model s20), and $ 40.0 \, M_\odot $ (core
model s40). Recently, the first presupernova models that include
rotation in a one-dimensional approximate fashion have become
available~\cite{heger_00_a, heger_05_a}, and of these we select ones
with $ M_\mathrm{prog} = 15.0 \, M_\odot $ (models e15a and e15b) as
well as $ 20.0 \, M_\odot $ (core models e20a and e20b). All progenitors
have solar-metallicity (at zero-age main sequence) and we generate our
initial models by taking the data obtained from stellar evolution out
to a radius $ R_\mathrm{i} $ where the density drops to a value that
equals $ 10^{-4} $ of the initial precollapse central density
$ \rho_\mathrm{c,i} $. Selected quantities that describe the
properties of these stellar cores are summarized in
Table~\ref{table:progenitor_cores}.

\begin{table}[t]
  \caption{Properties of the iron core models used as initial data.
    $ M_\mathrm{prog} $ is the total zero-age main sequence mass of
    the progenitor star, $ M_\mathrm{core} $ and $ R_\mathrm{core} $
    are the mass and radius of the iron core, $ M_\mathrm{i} $ and
    $ R_\mathrm{i} $ are the mass and radius of the initial model on
    the computational grid, and $ \rho_\mathrm{c,i} $ is the 
    precollapse density at the center. The size of the iron core
    is determined by the condition that $ Y_e $ exceeds 0.497, while
    the initial model extends beyond the iron core to where the
    density drops to $ 10^{-4} \rho_\mathrm{c,i} $.
    $ \rho_\mathrm{c,i} $ deviates slightly from the original value of
    the models in~\cite{woosley_02_a} because of regridding to the
    more densely spaced central grid of the evolution code.}
  \label{table:progenitor_cores}
  \begin{ruledtabular}
    \begin{tabular}{lcccccc}
      Core &
      $ M_\mathrm{prog} $ &
      $ M_\mathrm{core} $ &
      $ R_\mathrm{core} $ &
      $ M_\mathrm{i} $ &
      $ R_\mathrm{i} $ &
      $ \rho_\mathrm{c,i} $ \\
      model &
      [$ M_\odot $] &
      [$ M_\odot $] &
      [$ 10^3 \mathrm{\ km} $] &
      [$ M_\odot $] &
      [$ 10^3 \mathrm{\ km} $] &
      [$ 10^9 \mathrm{\ g\ cm}^{-3} $] \\ [0.2 em]
      \hline \rule{0 em}{1.2 em}%
      s11  & $ 11.2 $ & $ 1.24 $ & $ 0.99 $ & $ 1.36 $ & $ 1.58 $ & $  17.71 $ \\
      s15  & $ 15.0 $ & $ 1.55 $ & $ 1.94 $ & $ 1.81 $ & $ 3.88 $ & $ \z6.50 $ \\
      s20  & $ 20.0 $ & $ 1.46 $ & $ 1.69 $ & $ 1.59 $ & $ 3.48 $ & $ \z8.77 $ \\
      s40  & $ 40.0 $ & $ 1.55 $ & $ 1.62 $ & $ 2.03 $ & $ 4.60 $ & $ \z3.88 $ \\ [0.5 em]
      e15a & $ 15.0 $ & $ 1.47 $ & $ 1.55 $ & $ 1.83 $ & $ 4.45 $ & $ \z5.78 $ \\
      e15b & $ 15.0 $ & $ 1.40 $ & $ 1.66 $ & $ 1.56 $ & $ 3.17 $ & $ \z8.04 $ \\
      e20a & $ 20.0 $ & $ 1.75 $ & $ 2.41 $ & $ 2.26 $ & $ 5.42 $ & $ \z4.27 $ \\
      e20b & $ 20.0 $ & $ 1.38 $ & $ 1.35 $ & $ 1.60 $ & $ 3.18 $ & $ \z7.22 $ \\
    \end{tabular}
  \end{ruledtabular}
\end{table}

We set those cores that are initially nonrotating (core models s11,
s15, s20, and s40) artificially into rotation according to the
rotation law specified in~\cite{komatsu_89_a}, where the specific
angular momentum $ j $ is given by
\begin{equation}
  j = A^2 (\Omega_\mathrm{c,i} - \Omega).
  \label{rotation_law}
\end{equation}
Here the length $ A $ parametrizes the degree of differential rotation
(stronger differentiality with decreasing $ A $) and
$ \Omega_\mathrm{c,i} $ is the precollapse value of the
angular velocity $ \Omega $ at the center. In the Newtonian limit,
this reduces to
\begin{equation}
  \Omega = \Omega_\mathrm{c,i} \frac{A^2}{A^2 + r^2 \sin^2 \theta},
  \label{newtonian_rotation_law}
\end{equation}
with $ r \sin \theta $ being the distance to the rotation axis.

In order to determine the influence of different angular momentum
distributions on the collapse dynamics, we parameterize the
precollapse rotation of our models in terms of $ A $ (A1:
$ A = 50,000 \mathrm{\ km} $, almost uniform; A2:
$ A = 1,000 \mathrm{\ km} $, moderately differential; A3:
$ A = 500 \mathrm{\ km} $, strongly differential) and
$ \Omega_\mathrm{c,i} $. The model nomenclature for the precollapse
rotation parameters is shown in Table~\ref{table:initial_models}. We
have selected the rotational configuration of the models in such a way
that for the s20 progenitor they are a representative subset of the
models investigated in~\cite{dimmelmeier_07_a, dimmelmeier_07_b}. They
reflect different properties of the collapse dynamics and the
gravitational radiation waveform discussed in that work, namely
pressure-dominated bounce with or without significant postbounce
convective overturn as well as single centrifugal bounce.

\begin{table*}[t]
  \caption{Precollapse rotation properties of the core collapse models.
    $ A $ is the differential rotation length scale,
    $ \Omega_\mathrm{c,i} $ is the precollapse angular velocity at the
    center, and $ \beta_\mathrm{i} $ is the precollapse rotation rate.
    Note that the models e15a, e15b, e20a, and e20b have a rotation
    profile from the corresponding stellar evolution calculations, while
    onto all other models an artificial rotation profile is imposed.}
  \label{table:initial_models}
  \begin{ruledtabular}
    \begin{tabular}{lccc@{\qquad\qquad}lccc@{\qquad\qquad}lccc}
      Rotating &
      $ A $ &
      $ \Omega_\mathrm{c,i} $ &
      $ \beta_\mathrm{i} $ &
      Rotating &
      $ A $ &
      $ \Omega_\mathrm{c,i} $ &
      $ \beta_\mathrm{i} $ &
      Rotating &
      $ A $ &
      $ \Omega_\mathrm{c,i} $ &
      $ \beta_\mathrm{i} $ \\
      core model &
      [$ 10^8 \mathrm{\ cm} $] &
      [$ \mathrm{rad\ s}^{-1} $] &
      [\%] &
      core model &
      [$ 10^8 \mathrm{\ cm} $] &
      [$ \mathrm{rad\ s}^{-1} $] &
      [\%]
      &
      core model &
      [$ 10^8 \mathrm{\ cm} $] &
      [$ \mathrm{rad\ s}^{-1} $] &
      [\%] \\ [0.2 em]
      \hline \rule{0 em}{1.2 em}%
      s11A1O01 & $  50.0 $ & $ \z0.45 $ & $ 0.01 $ &
      s15A1O01 & $  50.0 $ & $ \z0.45 $ & $ 0.09 $ &
      e15a     &       --- & $ \z4.18 $ & $ 0.46 $ \\
      s11A1O05 & $  50.0 $ & $ \z1.01 $ & $ 0.06 $ &
      s15A1O05 & $  50.0 $ & $ \z1.01 $ & $ 0.45 $ &
      e15b     &       --- & $ \z9.93 $ & $ 2.75 $ \\
      s11A1O07 & $  50.0 $ & $ \z1.43 $ & $ 0.12 $ &
      s15A1O07 & $  50.0 $ & $ \z1.43 $ & $ 0.91 $ &
      e20a     &       --- & $ \z3.13 $ & $ 0.28 $ \\
      s11A1O09 & $  50.0 $ & $ \z1.91 $ & $ 0.22 $ &
      s15A1O09 & $  50.0 $ & $ \z1.91 $ & $ 1.63 $ &
      e20b     &       --- & $  11.01 $ & $ 2.16 $ \\
      s11A1O13 & $  50.0 $ & $ \z2.71 $ & $ 0.43 $ &
      s15A1O13 & $  50.0 $ & $ \z2.71 $ & $ 3.26 $ \\
      s11A2O05 & $ \z1.0 $ & $ \z2.40 $ & $ 0.16 $ &
      s15A2O05 & $ \z1.0 $ & $ \z2.40 $ & $ 0.30 $ \\
      s11A2O07 & $ \z1.0 $ & $ \z3.40 $ & $ 0.31 $ &
      s15A2O07 & $ \z1.0 $ & $ \z3.40 $ & $ 0.60 $ \\
      s11A2O09 & $ \z1.0 $ & $ \z4.56 $ & $ 0.56 $ &
      s15A2O09 & $ \z1.0 $ & $ \z4.56 $ & $ 1.09 $ \\
      s11A2O13 & $ \z1.0 $ & $ \z6.45 $ & $ 1.13 $ &
      s15A2O13 & $ \z1.0 $ & $ \z6.45 $ & $ 2.18 $ \\
      s11A2O15 & $ \z1.0 $ & $ \z7.60 $ & $ 1.57 $ &
      s15A2O15 & $ \z1.0 $ & $ \z7.60 $ & $ 3.03 $ \\
      s11A3O05 & $ \z0.5 $ & $ \z4.21 $ & $ 0.20 $ &
      s15A3O05 & $ \z0.5 $ & $ \z4.21 $ & $ 0.27 $ \\
      s11A3O07 & $ \z0.5 $ & $ \z5.95 $ & $ 0.40 $ &
      s15A3O07 & $ \z0.5 $ & $ \z5.95 $ & $ 0.53 $ \\
      s11A3O09 & $ \z0.5 $ & $ \z8.99 $ & $ 0.72 $ &
      s15A3O09 & $ \z0.5 $ & $ \z8.99 $ & $ 0.96 $ \\
      s11A3O12 & $ \z0.5 $ & $  10.65 $ & $ 1.28 $ &
      s15A3O12 & $ \z0.5 $ & $  10.65 $ & $ 1.71 $ \\
      s11A3O13 & $ \z0.5 $ & $  11.30 $ & $ 1.44 $ &
      s15A3O13 & $ \z0.5 $ & $  11.30 $ & $ 1.92 $ \\
      s11A3O15 & $ \z0.5 $ & $  13.31 $ & $ 2.00 $ &
      s15A3O15 & $ \z0.5 $ & $  13.31 $ & $ 2.67 $ \\ [0.5 em]
      s20A1O01 & $  50.0 $ & $ \z0.45 $ & $ 0.05 $ &
      s40A1O01 & $  50.0 $ & $ \z0.45 $ & $ 0.13 $ \\
      s20A1O05 & $  50.0 $ & $ \z1.01 $ & $ 0.25 $ &
      s40A1O05 & $  50.0 $ & $ \z1.01 $ & $ 0.64 $ \\
      s20A1O07 & $  50.0 $ & $ \z1.43 $ & $ 0.50 $ &
      s40A1O07 & $  50.0 $ & $ \z1.43 $ & $ 1.28 $ \\
      s20A1O09 & $  50.0 $ & $ \z1.91 $ & $ 0.90 $ &
      s40A1O09 & $  50.0 $ & $ \z1.91 $ & $ 2.31 $ \\
      s20A1O13 & $  50.0 $ & $ \z2.71 $ & $ 1.80 $ &
      s40A1O13 & $  50.0 $ & $ \z2.71 $ & $ 4.62 $ \\
      s20A2O05 & $ \z1.0 $ & $ \z2.40 $ & $ 0.25 $ &
      s40A2O05 & $ \z1.0 $ & $ \z2.40 $ & $ 0.36 $ \\
      s20A2O07 & $ \z1.0 $ & $ \z3.40 $ & $ 0.50 $ &
      s40A2O07 & $ \z1.0 $ & $ \z3.40 $ & $ 0.72 $ \\
      s20A2O09 & $ \z1.0 $ & $ \z4.56 $ & $ 0.90 $ &
      s40A2O09 & $ \z1.0 $ & $ \z4.56 $ & $ 1.30 $ \\
      s20A2O13 & $ \z1.0 $ & $ \z6.45 $ & $ 1.80 $ &
      s40A2O13 & $ \z1.0 $ & $ \z6.45 $ & $ 2.60 $ \\
      s20A2O15 & $ \z1.0 $ & $ \z7.60 $ & $ 2.50 $ &
      s40A2O15 & $ \z1.0 $ & $ \z7.60 $ & $ 3.62 $ \\
      s20A3O05 & $ \z0.5 $ & $ \z4.21 $ & $ 0.25 $ &
      s40A3O05 & $ \z0.5 $ & $ \z4.21 $ & $ 0.29 $ \\
      s20A3O07 & $ \z0.5 $ & $ \z5.95 $ & $ 0.50 $ &
      s40A3O07 & $ \z0.5 $ & $ \z5.95 $ & $ 0.57 $ \\
      s20A3O09 & $ \z0.5 $ & $ \z8.99 $ & $ 0.90 $ &
      s40A3O09 & $ \z0.5 $ & $ \z8.99 $ & $ 1.03 $ \\
      s20A3O12 & $ \z0.5 $ & $  10.65 $ & $ 1.60 $ &
      s40A3O12 & $ \z0.5 $ & $  10.65 $ & $ 1.84 $ \\
      s20A3O13 & $ \z0.5 $ & $  11.30 $ & $ 1.80 $ &
      s40A3O13 & $ \z0.5 $ & $  11.30 $ & $ 2.07 $ \\
      s20A3O15 & $ \z0.5 $ & $  13.31 $ & $ 2.50 $ &
      s40A3O15 & $ \z0.5 $ & $  13.31 $ & $ 2.87 $ \\
    \end{tabular}
  \end{ruledtabular}
\end{table*}

Note that models with the same rotation specification (but different
progenitor mass or EoS) have an identical angular velocity profile,
while the precollapse rotation rate
$ \beta_\mathrm{i} = T_\mathrm{i} / |W|_\mathrm{i} $, which is the
precollapse ratio of rotational energy to gravitational energy,
varies. We have decided to compare models with identical initial
angular velocity $ \Omega_\mathrm{c,i} $ and not precollapse rotation
rate $ \beta_\mathrm{i} $, as the latter quantity is rather sensitive
to the chosen core radius $ R_\mathrm{core} $ in the case of (almost)
uniform rotation.

The models that are based on progenitor calculations including
rotation (core models e15a, e15b, e20a, and e20b) are mapped onto our
computational grids under the assumption of constant rotation on
cylindrical shells of constant distance to the rotation axis.
We also point out that due to the one-dimensional nature, none of the
considered models are in rotational equilibrium. Still, in slowly
rotating initial models this effect is small and thus negligible.
For more rapidly rotating models, which exhibits the strongest
deviation from rotational equilibrium, the collapse proceeds more
slowly due to stabilizing centrifugal forces, and hence the star
has more time for the adjustment to the appropriate angular
stratifications for its rate of rotation.

In this study, we focus on the collapse of massive presupernova iron
cores with at most moderate differential rotation and precollapse
rotation rates that except for the most slowly rotating models lead to
proto-neutron stars that are probably spinning too fast to yield spin
periods of cold neutron stars in agreement with observationally inferred
injection periods of young pulsars into the $ P / \dot{P} $
diagram~\cite{heger_05_a, ott_06_c}. However, they may be highly
relevant in the collapsar-type gamma-ray burst
scenario~\cite{ott_06_c, woosley_06_b, dessart_08_a}.


\subsection{Gravitational wave extraction}
\label{subsection:wave_extraction}

We employ the Newtonian quadrupole formula in the first-moment of
momentum density formulation as discussed, e.g.,
in~\cite{dimmelmeier_02_a, shibata_03_a, dimmelmeier_05_a} to extract
the gravitational waves generated by nonspherical accelerated fluid
motions. It yields the quadrupole wave amplitude
$ A_{20}^\mathrm{E2} $ as the lowest order term in a multipole
expansion of the radiation field into pure-spin tensor
harmonics~\cite{thorne_80_a}. The wave amplitude is related to the
dimensionless gravitational wave strain $ h $ in the equatorial plane
by
\begin{equation}
  h = \frac{1}{8} \sqrt{\frac{15}{\pi}} \frac{A^\mathrm{E2}_{20}}{r} =
  8.8524 \times 10^{-21} \frac{A^\mathrm{E2}_{20}}{10^3 \mathrm{\ cm}}
  \frac{10 \mathrm{\ kpc}}{r},
\end{equation}
where $ r $ is the distance to the emitting source.

We point out that although the quadrupole formula is not gauge
invariant and is only valid in the Newtonian slow-motion limit, for
gravitational waves emitted by pulsations of rotating NSs (i.e., in
astrophysical situations comparable to collapsing stellar cores at
bounce in terms of compactness and rotation rates) it yields results
that agree very well in phase and to $ \sim 10 \mbox{\,--\,} 20\% $ in
amplitude with more sophisticated methods~\cite{shibata_03_a,
nagar_07_a}.

In order to assess the prospects for detection by current and planned
interferometer detectors and to specifically address the issue of
detection range and expected event rates, we calculate the
dimensionless \emph{characteristic} gravitational wave strain
$ h_\mathrm{c} $ of the signal according to~\cite{thorne_87_a}. We
first perform a Fourier transform of the gravitational wave strain
$ h $,
\begin{equation}
  \hat{h} =
  \int_{-\infty}^\infty \!\! e^{2 \pi i f t} h \, dt.
  \label{eq:waveform_fourier_tranform}
\end{equation}
To obtain the (detector dependent) integrated characteristic signal
frequency
\begin{equation}
  f_\mathrm{c} =
  \left( \int_0^\infty \!
  \frac{\langle \hat{h}^2 \rangle}{S_h}
  f \, df \right)
  \left( \int_0^\infty \!
  \frac{\langle \hat{h}^2 \rangle}{S_h}
  df \right)^{-1}
  \label{eq:characteristic_frequency}
\end{equation}
and the integrated characteristic signal strain
\begin{equation}
  h_\mathrm{c} =
  \left( 3 \int_0^\infty \!
  \frac{S_{h\mathrm{\,c}}}{S_h} \langle \hat{h}^2 \rangle
  f \, df \right)^{1/2}\!\!\!\!\!\!\!\!,
  \label{eq:characteristic_amplitude}
\end{equation}
the power spectral density $ S_h $ of the detector is needed (with
$ S_{h\mathrm{\,c}} = S_h (f_\mathrm{c}) $). We approximate the
average $ \langle \hat{h}^2 \rangle $ over randomly distributed
angles by $ \hat{h}^2 $, assuming optimal orientation of the
interferometer detector. From Eqs.~(\ref{eq:characteristic_frequency},
\ref{eq:characteristic_amplitude}) the signal-to-noise ratio can be
computed as
$ \mathrm{SNR} = h_\mathrm{c} / [h_\mathrm{rms} (f_\mathrm{c})] $,
where $ h_\mathrm{rms} = \sqrt{f S_h} $ is the value of the rms strain
noise (i.e., the theoretical sensitivity window) for the detector.


\section{Numerical methods}
\label{section:numerical_methods}

We perform all simulations using the \coconut\ code described in
detail in~\cite{dimmelmeier_02_a, dimmelmeier_05_a}. The equations of
general relativistic hydrodynamics are solved in semi-discrete
fashion. The spatial discretization is performed by means of a
high-resolution shock-capturing (HRSC) scheme employing a second-order
accurate finite-volume discretization. We make use of the
Harten--Lax--van Leer--Einfeldt (HLLE) flux formula for the local
Riemann problems and piecewise-parabolic reconstruction (PPM) of the
primitive variables $ (\rho, v^i, \epsilon) $ at cell interfaces. For
a review of such methods in general relativistic hydrodynamics,
see~\cite{font_03_a}. The time integration and coupling with curvature
are carried out with the method of lines~\cite{hyman_76_a} in
combination with a second-order accurate explicit Runge--Kutta
scheme. Once the state vector $ \bm{U} $ is updated in time, the
primitive variables are recovered from the conserved ones given in
Eq.~(\ref{equation:conserved_quantities}) through an iterative
Newton--Raphson method. Note that the component associated to $ Y_e $
in the system~(\ref{equation:conservation_equations}) of hydrodynamic
evolution equations is treated as a passive advection equation, which
does not contribute to the characteristic structure in the form of
eigenvalues and eigenvectors required by some flux solvers.

To numerically solve the metric equations we utilize an iterative
nonlinear solver based on spectral methods. The spectral grid of the
metric solver is split into 6 radial domains with 33 radial and 17
angular collocation points each. The combination of HRSC methods for
the hydrodynamics and spectral methods for the metric equations (the
\emph{Mariage des Maillages\/} or ``grid wedding'' approach) in a
multidimensional numerical code has been presented in detail
in~\cite{dimmelmeier_05_a}. Even when using spectral methods the
calculation of the spacetime metric from the
system~(\ref{equation:cfc_equations_1}--\ref{equation:cfc_equations_3})
of elliptic equations is computationally expensive. Hence, in our
simulations the metric is updated only once every 100/10/50
hydrodynamic time steps before/during/after core bounce, and
extrapolated in between. The numerical adequacy of this procedure has
been tested and discussed in detail in~\cite{dimmelmeier_02_a}.

In this study we focus on the gravitational wave signal associated
with core bounce. As demonstrated by~\cite{ott_07_a, ott_06_b},
effects that may break rotational symmetry are most likely unimportant
in this context. Hence, we assume axisymmetry and in addition impose
symmetry with respect to the equatorial plane.

The \coconut\ code utilizes Eulerian spherical coordinates
$ \{r, \theta\} $, and for the computational grid we choose 250
logarithmically-spaced, centrally-condensed radial zones with a
central resolution of $ 250 \mathrm{\ m} $ and 45 equidistant angular
zones covering $ 90^\circ $. A small part of the grid is covered by an
artificial low-density atmosphere extending beyond the core's outer
boundary defined where $ \rho \le 10^{-4} \rho_\mathrm{c,i} $.

We also note that we have performed extensive resolution tests with
different grid resolutions to ascertain that the grid setup specified
above is appropriate for our simulations.


\section{Collapse dynamics and waveform morphology}
\label{section:collapse_dynamics}


\subsection{Generic waveform type}
\label{subsection:generic_waveform}

\begin{figure}[t]
  \epsfxsize = 8.6 cm
  \centerline{\epsfbox{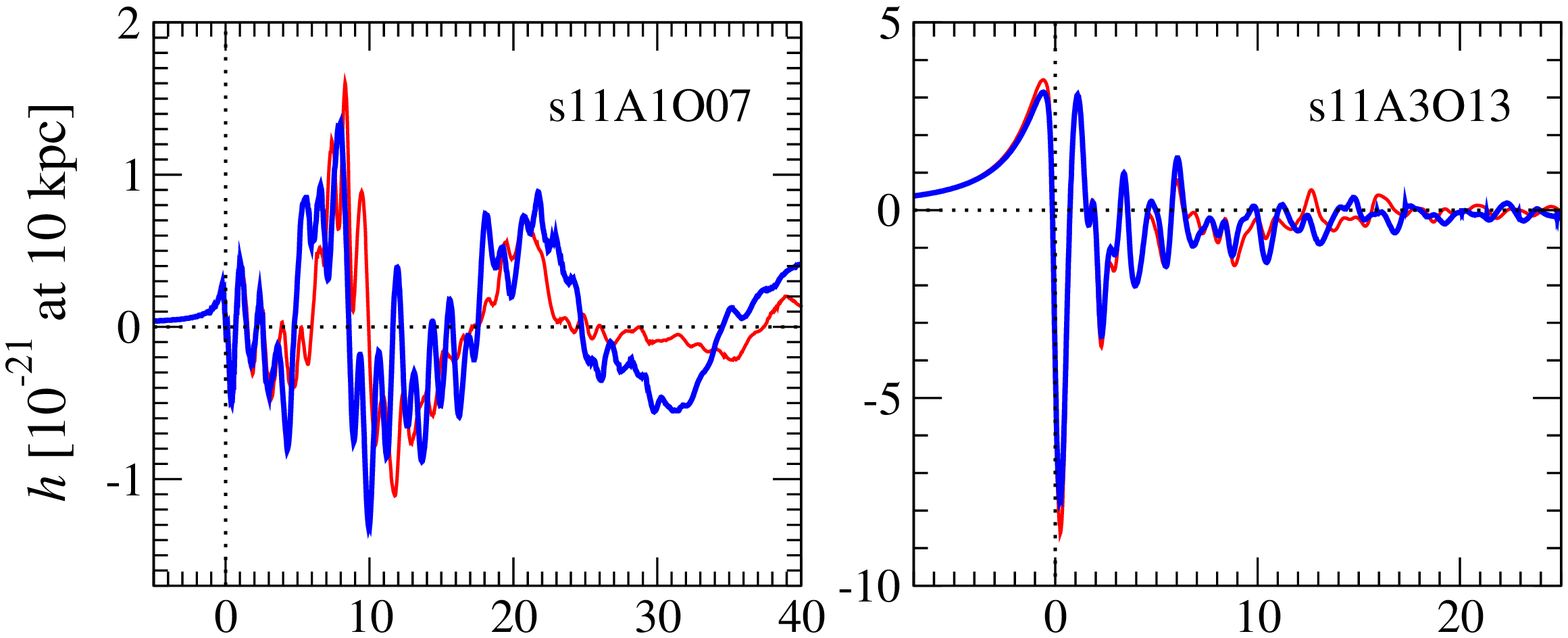}}
  ~ \\
  \epsfxsize = 8.6 cm
  \centerline{\epsfbox{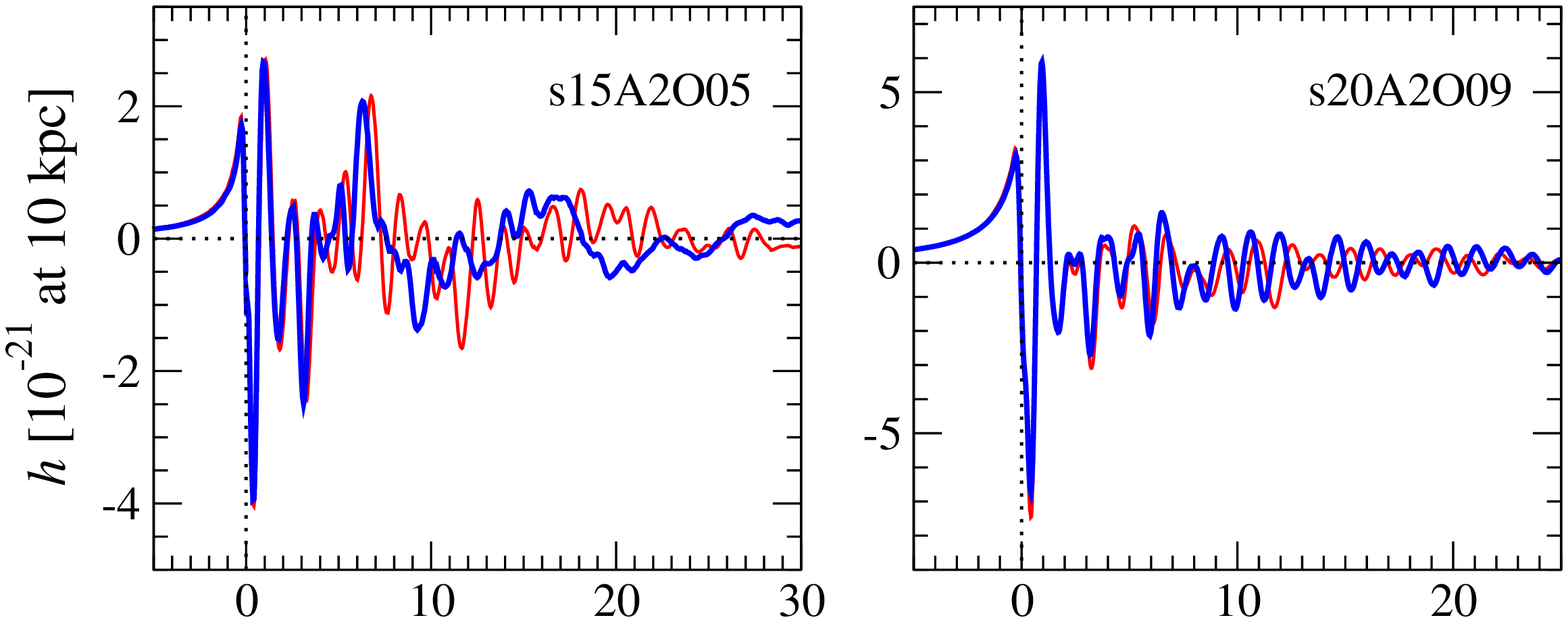}}
  ~ \\
  \epsfxsize = 8.6 cm
  \centerline{\epsfbox{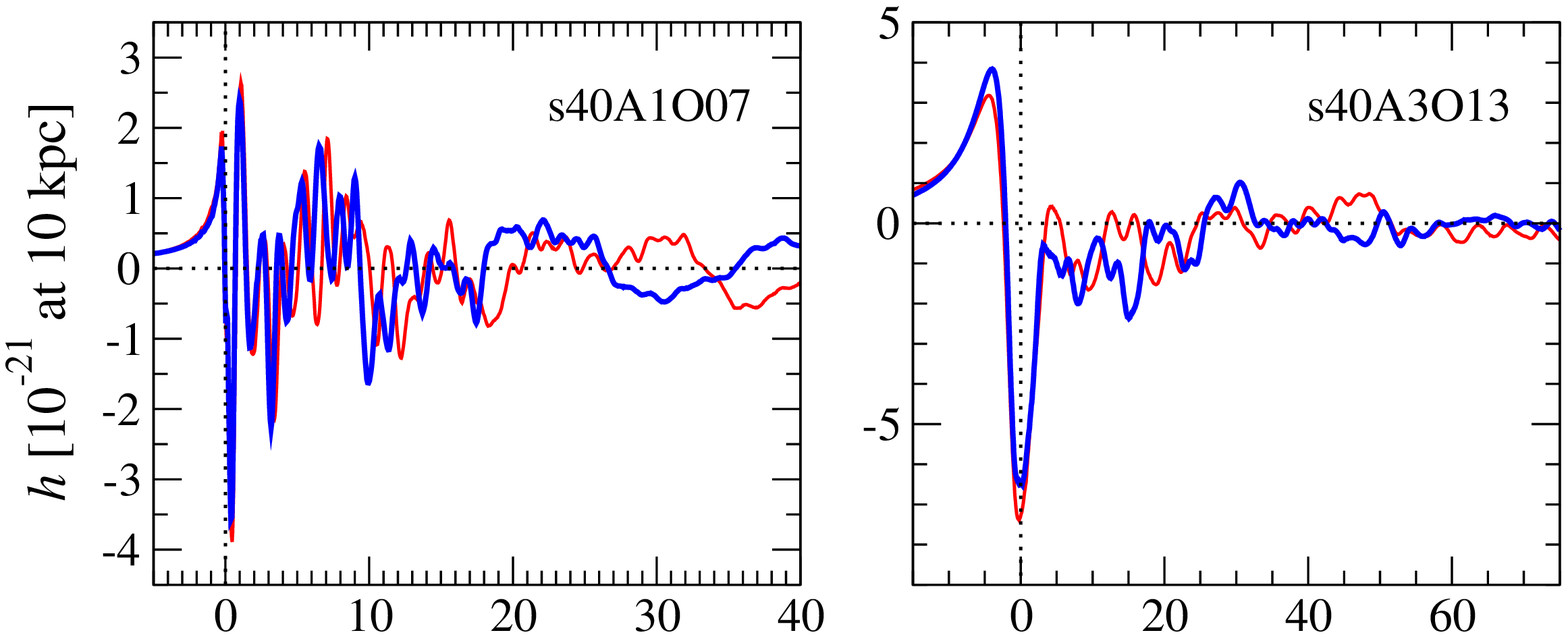}}
  ~ \\
  \epsfxsize = 8.6 cm
  \centerline{\epsfbox{figure_01_b.eps}}
  \caption{Time evolution of the gravitational wave amplitude $ h $
    for representative models with different precollapse rotation profiles
    using the Shen EoS (red lines) or LS EoS (blue lines). Models with
    slow and almost uniform precollapse rotation (e.g., s11A1O07)
    develop considerable prompt postbounce convection visible as a
    dominating lower-frequency contribution in the waveform, while the
    waveform for both models with moderate rotation (e.g., s11A3O13,
    s15A2O05, s20A2O09, s40A1O07, or e15a) and rapidly rotating models
    which undergo a centrifugal bounce (e.g., s40A3O13 or e20b)
    exhibit an essentially regular ring-down. Time is normalized to
    the time of bounce $ t_\mathrm{b} $.}
  \label{figure:generic_collapse_type_waveform}
\end{figure}

We begin our discussion with an analysis of the gravitational
radiation waveform emitted during core bounce as an indicator for the
influence of the EoS, the progenitor structure, and the precollapse
rotational configuration on the collapse and bounce dynamics. In
Fig.~\ref{figure:generic_collapse_type_waveform}, we present example
waveforms for representative collapsing cores selected from the
investigated parameter space of models (i.e., less or more massive
progenitors with slow or rapid precollapse rotation, varying
degree of differential rotation, and using either the Shen EoS or LS
EoS). The waveforms of all models are of type~I, hence exhibit a
positive prebounce rise and a large negative peak, followed by a
ring-down. In the light of a considerably extended parameter space in
terms of EoS and progenitor mass of the rotating core collapse models
investigated in this work, we thus confirm the observation presented
in~\cite{ott_07_a, ott_07_b, dimmelmeier_07_a, dimmelmeier_07_b} that
in general relativistic gravity all models with microphysics exhibit
gravitational wave burst signals of type~I.

\begin{figure}[t]
  \epsfxsize = 8.6 cm
  \centerline{\epsfbox{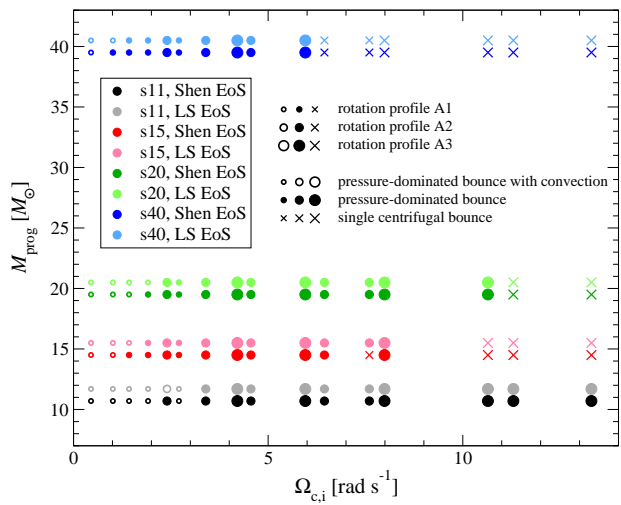}}
  \caption{Collapse dynamics of all investigated models in the
    parameter space of precollapse rotational configuration (specified
    by the precollapse angular velocity $ \Omega_\mathrm{c,i}
    $ at the center and the precollapse differential rotation length
    scale $ A $), progenitor mass $ M_\mathrm{prog} $, and EoS. Models
    marked by unfilled/filled circles undergo a pressure-dominated
    bounce with/without significant early postbounce convection, while
    models marked with crosses show a single centrifugal bounce. The
    EoS is encoded as in Fig.~\ref{figure:y_e_profiles}, while
    small/medium/large symbols represent the precollapse rotation
    parameter A1/A2/A3. For better visibility, the symbols for the
    same $ M_\mathrm{prog} $ but different EoS are spread a bit in the
    vertical direction. Note also that in this and the following plots
    that encode the precollapse rotational configuration in the form of
    the parameter $ \Omega_\mathrm{i} $, we refrain from including
    models e15a, e15b, e20a, and e20b, as these have a precollapse
    rotation profile that is not given by the analytic rotation
    law~(\ref{rotation_law}).}
  \label{figure:collapse_dynamics}
\end{figure}

As already inferred in~\cite{dimmelmeier_07_a, dimmelmeier_07_b}, this
signal type can be classified into three subtypes, which, however, do
have in common the same qualitative features of a type~I waveform
described above:
\renewcommand{\labelenumi}{(\arabic{enumi})}
\begin{enumerate}
\item For a slowly rotating core, prompt convective overturn at early
  postbounce times after the pressure-dominated bounce adds a
  significant low-frequency contribution to the regular ring-down
  signal (see, e.g., model s11A1O07).
\item In the case of moderately rapid rotation which still leads to a
  pressure-dominated bounce, this convection is effectively suppressed
  due to the growing influence of angular momentum
  gradients~\cite{fryer_00_a, cerda_07_a} and does not strongly stand
  out in the postbounce ring-down signal (see, e.g., models s11A3O13,
  s15A2O05, s20A2O09, s40A1O07, or e15a).
\item If rotation is sufficiently rapid, the core bounces at
  subnuclear or only slightly supernuclear densities due to the
  increased effects of centrifugal forces, reflected by a significant
  widening of the bounce peak of the waveform and an overall lower
  frequency of the signal (see, e.g., models s40A3O13 or e20b).
\end{enumerate}
Fig.~\ref{figure:generic_collapse_type_waveform} also demonstrates
that for comparable precollapse rotational configuration (as specified
by the parameters $ A $ and $ \Omega_\mathrm{i} $) the impact of the
EoS on the collapse dynamics and, hence, the gravitational wave signal
is small. In Table~\ref{figure:collapse_dynamics} we mark each model
with its type of collapse dynamics, and in
Fig.~\ref{figure:collapse_dynamics} we encode that type in the
parameter space spanned by rotational configuration, progenitor
mass/model, and EoS.

For our set of collapse models, only in four cases (models s11A1O13,
s15A1O07, s20A1O09, and s40A1O05) the LS EoS yields a signal with
dominant convective contribution while the Shen EoS does not, and only
a single model (s15A2O15) changes its collapse type from a centrifugal
bounce to a pressure-dominated bounce when replacing the Shen EoS with
the LS EoS. However, Fig.~\ref{figure:collapse_dynamics} shows that
the transition between the three different collapse and waveform
subtypes occurs at different precollapse rotational
configurations for the various progenitor masses. This is a
consequence of differences in the mass $ M_\mathrm{ic,b} $ of the
inner core at bounce as discussed in
Section~\ref{subsection:eos_and_progenitor_influence}.

The growth of the strong prompt \emph{early} postbounce convection in
slowly rotating models depends sensitively on the seed perturbations
resulting from the numerical scheme/grid. In nature, prompt convection
will be influenced by random small-scale to large-scale variations in
the final stages of silicon burning that are frozen in during
collapse. We point out that the duration of the prompt postbounce
convection is most likely overestimated in our approach, since in full
postbounce radiation-hydrodynamics calculations, energy deposition by
neutrinos in the immediate postshock region rapidly smoothes out the
negative entropy gradient left behind by the shock (see, e.g.,
\cite{mueller_04_a, buras_06_a}) and quickly damps this early
convective instability.

\begin{table*}[t]
  \caption{Summary of relevant quantities from the rotating collapse
    of all iron core models. $ \rho_\mathrm{max,b} $ is the maximum
    density in the core at the time of bounce, $ |h|_\mathrm{max} $ is
    the peak value of the gravitational wave amplitude for the burst
    signal, while $ \beta_\mathrm{b} $ and $ \beta_\mathrm{pb} $ are
    the rotation rates at
    the time of bounce and late after core
    bounce, respectively. Models marked by unfilled/filled circles
    undergo a pressure-dominated bounce with/without significant early
    postbounce convection, while models marked with crosses show a
    single centrifugal bounce. The values left/right of the vertical
    separator ($ | $) are for the Shen/LS EoS.}
  \label{table:collapse_models}
  \begin{ruledtabular}
    \begin{tabular}{l@{}ccccc@{\qquad}l@{}ccccc}
      Collapse &
      &
      $ \rho_\mathrm{max,b} $ &
      $ |h|_\mathrm{max} $ &
      $ \beta_\mathrm{b} $ &
      $ \beta_\mathrm{pb} $ &
      Collapse &
      &
      $ \rho_\mathrm{max,b} $ &
      $ |h|_\mathrm{max} $ &
      $ \beta_\mathrm{b} $ &
      $ \beta_\mathrm{pb} $ \\
      \raiseentry{model} &
      &
      $ \displaystyle \left[ \frac{10^{14}}{\mathrm{g\ cm}^{-3}} \right] $ &
      $ \displaystyle \left[ {\lower0.5ex\hbox{$ 10^{-21} $ at} \atop \hbox{$ 10 \mathrm{\ kpc}$ }} \right] $ &
      \raiseentry{[\%]} &
      \raiseentry{[\%]} &
      \raiseentry{model} &
      &
      $ \displaystyle \left[ \frac{10^{14}}{\mathrm{g\ cm}^{-3}} \right] $ &
      $ \displaystyle \left[ {\lower0.5ex\hbox{$ 10^{-21} $ at} \atop \hbox{$ 10 \mathrm{\ kpc}$ }} \right] $ &
      \raiseentry{[\%]} &
      \raiseentry{[\%]} \\ [1.0 em]
      \hline \rule{0 em}{1.2 em}%
      s11A1O01 & $ \crc|\crc $ & $ 3.24|4.43 $ & $   0.05|  0.05 $ & $ \z0.1|\z0.1 $ & $ \z0.1|\z0.1 $ &
      s15A1O01 & $ \crc|\crc $ & $ 3.28|4.43 $ & $ \z0.20|\z0.20 $ & $ \z0.2|\z0.2 $ & $ \z0.3|\z0.3 $ \\
      s11A1O05 & $ \crc|\crc $ & $ 3.23|4.41 $ & $   0.26|  0.25 $ & $ \z0.3|\z0.3 $ & $ \z0.4|\z0.5 $ &
      s15A1O05 & $ \crc|\crc $ & $ 3.17|4.20 $ & $ \z0.98|\z0.97 $ & $ \z1.0|\z1.0 $ & $ \z1.3|\z1.2 $ \\
      s11A1O07 & $ \crc|\crc $ & $ 3.22|4.35 $ & $   0.51|  0.49 $ & $ \z0.6|\z0.6 $ & $ \z0.8|\z0.8 $ &
      s15A1O07 & $ \blt|\crc $ & $ 3.12|4.13 $ & $ \z1.84|\z1.84 $ & $ \z2.0|\z1.9 $ & $ \z2.4|\z2.6 $ \\
      s11A1O09 & $ \crc|\crc $ & $ 3.17|4.21 $ & $   0.95|  0.90 $ & $ \z1.1|\z1.1 $ & $ \z1.3|\z1.4 $ &
      s15A1O09 & $ \blt|\blt $ & $ 2.97|3.88 $ & $ \z3.11|\z3.08 $ & $ \z3.4|\z3.4 $ & $ \z3.7|\z4.0 $ \\
      s11A1O13 & $ \blt|\crc $ & $ 3.11|4.13 $ & $   1.77|  1.76 $ & $ \z2.0|\z2.0 $ & $ \z2.4|\z2.6 $ &
      s15A1O13 & $ \blt|\blt $ & $ 2.86|3.56 $ & $ \z5.35|\z5.01 $ & $ \z6.2|\z6.1 $ & $ \z6.0|\z6.6 $ \\
      s11A2O05 & $ \crc|\crc $ & $ 3.16|4.18 $ & $   1.30|  1.35 $ & $ \z1.4|\z1.5 $ & $ \z1.6|\z1.7 $ &
      s15A2O05 & $ \blt|\blt $ & $ 2.95|3.76 $ & $ \z4.04|\z3.94 $ & $ \z4.1|\z4.1 $ & $ \z3.8|\z4.3 $ \\
      s11A2O07 & $ \blt|\blt $ & $ 3.02|3.92 $ & $   2.47|  2.52 $ & $ \z2.8|\z2.8 $ & $ \z2.9|\z3.0 $ &
      s15A2O07 & $ \blt|\blt $ & $ 2.81|3.44 $ & $ \z6.84|\z6.33 $ & $ \z7.5|\z7.4 $ & $ \z6.7|\z6.8 $ \\
      s11A2O09 & $ \blt|\blt $ & $ 2.94|3.69 $ & $   4.08|  4.00 $ & $ \z4.8|\z4.8 $ & $ \z4.6|\z4.7 $ &
      s15A2O09 & $ \blt|\blt $ & $ 2.58|3.05 $ & $ \z8.61|\z7.83 $ & $  11.8| 11.6 $ & $  10.3| 10.4 $ \\
      s11A2O13 & $ \blt|\blt $ & $ 2.76|3.35 $ & $   6.68|  6.09 $ & $ \z8.5|\z8.4 $ & $ \z7.9|\z8.0 $ &
      s15A2O13 & $ \blt|\blt $ & $ 2.14|2.33 $ & $ \z7.07|\z6.21 $ & $  18.2| 17.5 $ & $  15.6| 15.5 $ \\
      s11A2O15 & $ \blt|\blt $ & $ 2.66|3.15 $ & $   7.72|  7.01 $ & $  10.9| 10.8 $ & $ \z9.8|\z9.9 $ &
      s15A2O15 & $ \tms|\blt $ & $ 1.80|1.90 $ & $ \z4.01|\z3.73 $ & $  20.1| 19.7 $ & $  17.9| 18.1 $ \\
      s11A3O05 & $ \blt|\blt $ & $ 3.02|3.88 $ & $   2.96|  3.05 $ & $ \z3.2|\z3.2 $ & $ \z2.8|\z3.0 $ &
      s15A3O05 & $ \blt|\blt $ & $ 2.82|3.46 $ & $ \z7.65|\z7.27 $ & $ \z7.2|\z7.4 $ & $ \z5.7|\z5.8 $ \\
      s11A3O07 & $ \blt|\blt $ & $ 2.89|3.60 $ & $   5.33|  5.30 $ & $ \z5.9|\z6.0 $ & $ \z5.1|\z5.2 $ &
      s15A3O07 & $ \blt|\blt $ & $ 2.55|2.94 $ & $  10.06|\z9.55 $ & $  12.8| 12.7 $ & $  10.0|\z9.9 $ \\
      s11A3O09 & $ \blt|\blt $ & $ 2.71|3.25 $ & $   8.42|  7.66 $ & $ \z9.7|\z9.7 $ & $ \z8.0|\z8.2 $ &
      s15A3O09 & $ \blt|\blt $ & $ 2.17|2.31 $ & $ \z9.74|\z8.48 $ & $  18.7| 18.1 $ & $  14.8| 14.6 $ \\
      s11A3O12 & $ \blt|\blt $ & $ 2.46|2.75 $ & $   8.92|  7.84 $ & $  14.9| 14.7 $ & $  12.3| 12.3 $ &
      s15A3O12 & $ \tms|\tms $ & $ 1.15|1.26 $ & $ \z5.68|\z4.82 $ & $  21.1| 21.0 $ & $  18.3| 18.9 $ \\
      s11A3O13 & $ \blt|\blt $ & $ 2.36|2.64 $ & $   8.62|  7.73 $ & $  16.1| 15.8 $ & $  13.2| 13.2 $ &
      s15A3O13 & $ \tms|\tms $ & $ 0.72|0.84 $ & $ \z5.33|\z4.52 $ & $  21.3| 21.3 $ & $  18.9| 19.6 $ \\
      s11A3O15 & $ \blt|\blt $ & $ 2.10|2.23 $ & $   7.21|  6.32 $ & $  19.4| 18.6 $ & $  16.2| 15.8 $ &
      s15A3O15 & $ \tms|\tms $ & $ 0.25|0.30 $ & $ \z4.84|\z4.53 $ & $  22.2| 22.3 $ & $  20.3| 21.1 $ \\ [0.5 em]
      s20A1O01 & $ \crc|\crc $ & $ 3.28|4.41 $ & $   0.13|  0.13 $ & $ \z0.1|\z0.1 $ & $ \z0.2|\z0.2 $ &
      s40A1O01 & $ \crc|\crc $ & $ 3.29|4.42 $ & $ \z0.50|\z0.42 $ & $ \z0.4|\z0.4 $ & $ \z0.5|\z0.6 $ \\
      s20A1O05 & $ \crc|\crc $ & $ 3.21|4.23 $ & $   0.63|  0.64 $ & $ \z0.7|\z0.7 $ & $ \z0.9|\z1.0 $ &
      s40A1O05 & $ \blt|\crc $ & $ 3.13|4.14 $ & $ \z2.12|\z1.92 $ & $ \z1.9|\z1.8 $ & $ \z2.1|\z2.3 $ \\
      s20A1O07 & $ \crc|\crc $ & $ 3.17|4.23 $ & $   1.19|  1.28 $ & $ \z1.3|\z1.3 $ & $ \z1.6|\z1.9 $ &
      s40A1O07 & $ \blt|\blt $ & $ 2.96|3.89 $ & $ \z3.89|\z3.57 $ & $ \z3.5|\z3.5 $ & $ \z3.7|\z4.5 $ \\
      s20A1O09 & $ \blt|\crc $ & $ 3.10|4.11 $ & $   2.20|  2.12 $ & $ \z2.3|\z2.3 $ & $ \z2.6|\z3.0 $ &
      s40A1O09 & $ \blt|\blt $ & $ 2.85|3.64 $ & $ \z5.97|\z5.37 $ & $ \z5.9|\z5.8 $ & $ \z5.7|\z6.5 $ \\
      s20A1O13 & $ \blt|\blt $ & $ 2.95|3.77 $ & $   3.81|  3.68 $ & $ \z4.3|\z4.3 $ & $ \z4.5|\z5.0 $ &
      s40A1O13 & $ \blt|\blt $ & $ 2.63|3.22 $ & $ \z8.30|\z7.07 $ & $  10.2|\z9.9 $ & $ \z9.4|\z9.5 $ \\
      s20A2O05 & $ \blt|\blt $ & $ 3.03|3.94 $ & $   2.89|  2.89 $ & $ \z3.0|\z3.0 $ & $ \z2.9|\z3.1 $ &
      s40A2O05 & $ \blt|\blt $ & $ 2.81|3.57 $ & $ \z7.43|\z6.79 $ & $ \z6.8|\z6.7 $ & $ \z5.7|\z5.8 $ \\
      s20A2O07 & $ \blt|\blt $ & $ 2.90|3.63 $ & $   5.04|  4.87 $ & $ \z5.5|\z5.5 $ & $ \z5.1|\z5.3 $ &
      s40A2O07 & $ \blt|\blt $ & $ 2.57|3.06 $ & $ \z9.95|\z8.74 $ & $  11.8| 11.6 $ & $ \z9.9|\z9.9 $ \\
      s20A2O09 & $ \blt|\blt $ & $ 2.75|3.31 $ & $   7.46|  6.73 $ & $ \z9.0|\z8.8 $ & $ \z8.0|\z8.3 $ &
      s40A2O09 & $ \blt|\blt $ & $ 2.22|2.44 $ & $ \z9.22|\z7.80 $ & $  17.3| 16.7 $ & $  14.3| 14.3 $ \\
      s20A2O13 & $ \blt|\blt $ & $ 2.42|2.75 $ & $   7.83|  7.07 $ & $  14.8| 14.4 $ & $  12.8| 12.9 $ &
      s40A2O13 & $ \tms|\tms $ & $ 0.91|1.28 $ & $ \z4.04|\z3.30 $ & $  20.4| 20.6 $ & $  19.2| 19.0 $ \\
      s20A2O15 & $ \blt|\blt $ & $ 2.20|2.37 $ & $   7.00|  6.10 $ & $  17.8| 17.2 $ & $  15.4| 15.2 $ &
      s40A2O15 & $ \tms|\tms $ & $ 0.27|0.40 $ & $ \z3.51|\z3.51 $ & $  21.1| 21.4 $ & $  20.3| 21.0 $ \\
      s20A3O05 & $ \blt|\blt $ & $ 2.92|3.62 $ & $   5.59|  5.53 $ & $ \z5.6|\z5.7 $ & $ \z4.6|\z4.7 $ &
      s40A3O05 & $ \blt|\blt $ & $ 2.65|3.21 $ & $  10.19| 10.07 $ & $  10.5| 10.6 $ & $ \z7.6|\z7.7 $ \\
      s20A3O07 & $ \blt|\blt $ & $ 2.70|3.20 $ & $   9.50|  8.72 $ & $  10.1| 10.2 $ & $ \z8.0|\z8.1 $ &
      s40A3O07 & $ \blt|\blt $ & $ 2.21|2.47 $ & $  10.29| 10.09 $ & $  17.2| 16.8 $ & $  12.7| 12.5 $ \\
      s20A3O09 & $ \blt|\blt $ & $ 2.38|2.63 $ & $   9.67|  8.65 $ & $  15.7| 15.4 $ & $  12.5| 12.4 $ &
      s40A3O09 & $ \tms|\tms $ & $ 1.69|1.72 $ & $ \z7.45|\z7.72 $ & $  21.6| 21.4 $ & $  16.9| 16.8 $ \\
      s20A3O12 & $ \blt|\blt $ & $ 1.93|2.00 $ & $   6.52|  5.98 $ & $  21.0| 20.3 $ & $  17.6| 17.1 $ &
      s40A3O12 & $ \tms|\tms $ & $ 0.33|0.40 $ & $ \z7.36|\z6.34 $ & $  22.5| 22.8 $ & $  19.2| 20.1 $ \\
      s20A3O13 & $ \tms|\tms $ & $ 1.77|1.79 $ & $   5.35|  4.98 $ & $  21.3| 20.8 $ & $  18.1| 18.0 $ &
      s40A3O13 & $ \tms|\tms $ & $ 0.23|0.28 $ & $ \z7.40|\z6.51 $ & $  22.9| 23.4 $ & $  19.8| 20.7 $ \\
      s20A3O15 & $ \tms|\tms $ & $ 0.65|0.75 $ & $   4.62|  3.78 $ & $  21.6| 21.5 $ & $  19.7| 20.2 $ &
      s40A3O15 & $ \tms|\tms $ & $ 0.09|0.11 $ & $ \z6.90|\z6.90 $ & $  24.4| 25.1 $ & $  21.5| 22.4 $ \\ [0.5 em]
      e15a     & $ \blt|\blt $ & $ 2.66|3.25 $ & $   9.85|  8.30 $ & $ \z9.7|\z9.5 $ & $ \z7.6|\z7.8 $ \\
      e15b     & $ \tms|\tms $ & $ 1.61|1.69 $ & $   3.62|  3.57 $ & $  20.2| 20.1 $ & $  18.0| 19.0 $ \\
      e20a     & $ \blt|\blt $ & $ 2.69|3.35 $ & $   9.41|  8.09 $ & $ \z8.7|\z8.5 $ & $ \z6.4|\z6.4 $ \\
      e20b     & $ \tms|\tms $ & $ 1.41|1.50 $ & $   6.40|  5.54 $ & $  21.0| 20.4 $ & $  18.3| 17.4 $ \\
    \end{tabular}
  \end{ruledtabular}
\end{table*}


\subsection{Influence of general relativity and deleptonization}
\label{subsection:gr_and_deleptonization_influence}

The general type of collapse and bounce dynamics of the core, i.e.,
pressure-dominated or centrifugal bounce, can be influenced (provided
that the description of gravity and neutrino effects are identical) by
the progenitor core stratification and thermodynamic structure, the
amount and precollapse distribution of angular momentum, and the
properties of the EoS in the density regime just below the stiffening
threshold~\cite{dimmelmeier_07_a, dimmelmeier_07_b}. These
conditions influence the mass $ M_\mathrm{ic,b} $ of the homologously
contracting inner core at bounce, which in turn determined the region
that is dynamically relevant at bounce and sets the initial size of
the proto-neutron star.

\begin{figure}[t]
  \epsfxsize = 8.6 cm
  \centerline{\epsfbox{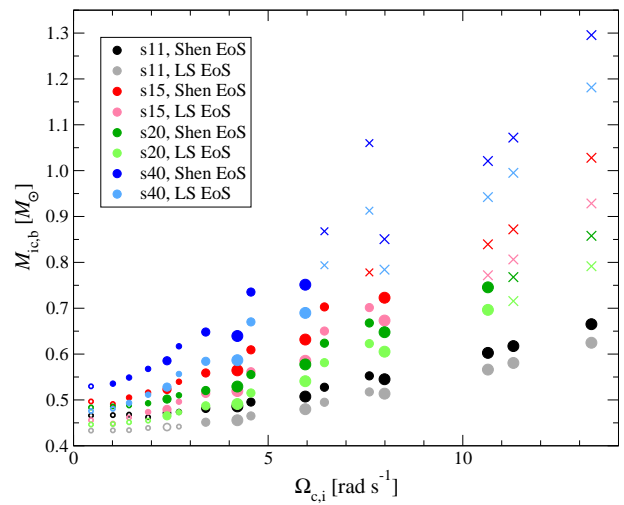}}
  \caption{Mass $ M_\mathrm{ic,b} $ of the inner core at the time of
    bounce for all models versus the precollapse initial central
    angular velocity $ \Omega_\mathrm{c,i} $. The progenitor model, the
    EoS, the initial rotation parameter $ A $, and the collapse
    dynamics are encoded as in Fig.~\ref{figure:collapse_dynamics}.}
  \label{figure:m_rest_inner_core}
\end{figure}

\begin{figure}[t]
  \epsfxsize = 8.6 cm
  \centerline{\epsfbox{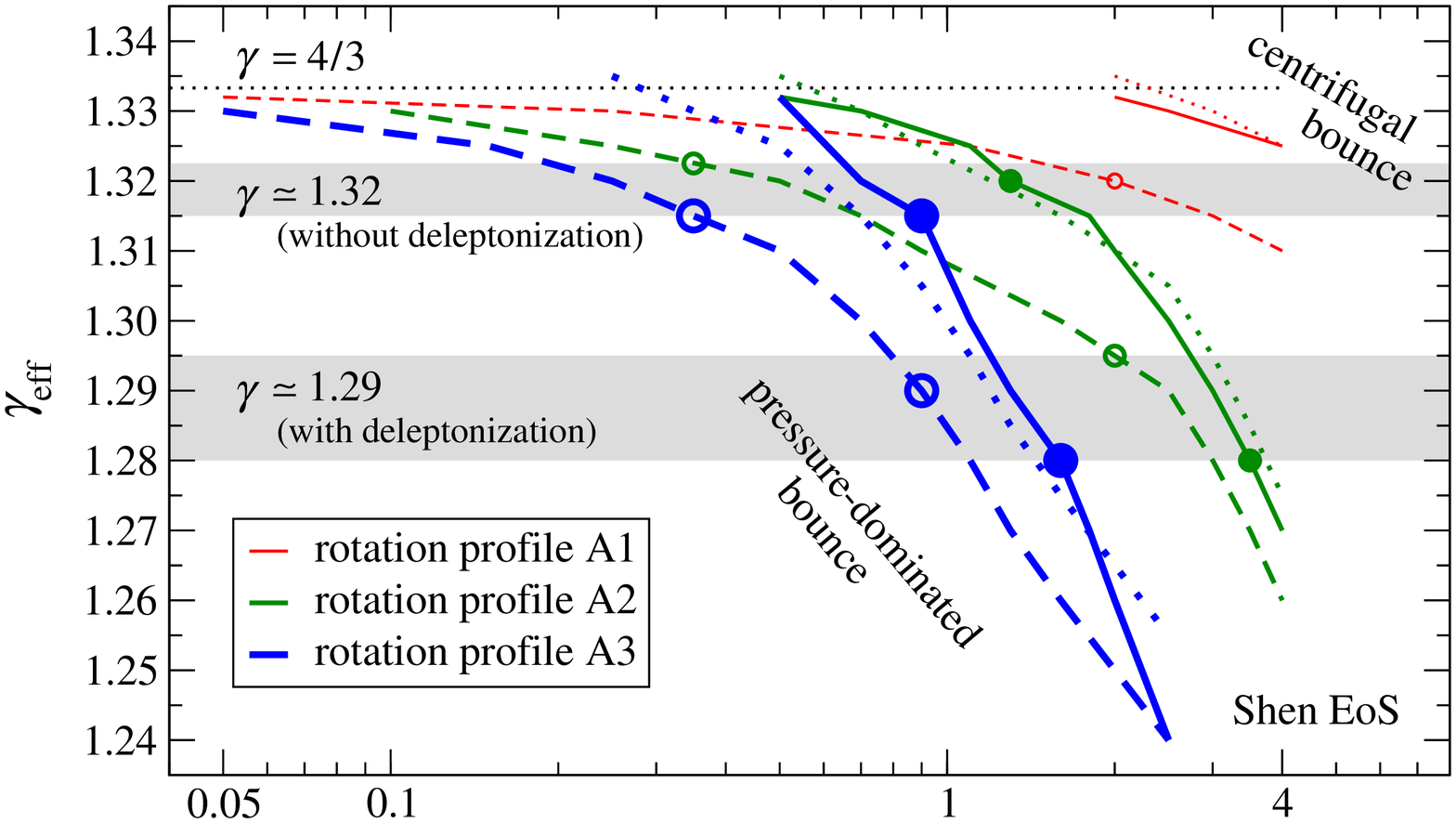}}
  ~ \\
  \epsfxsize = 8.6 cm
  \centerline{\epsfbox{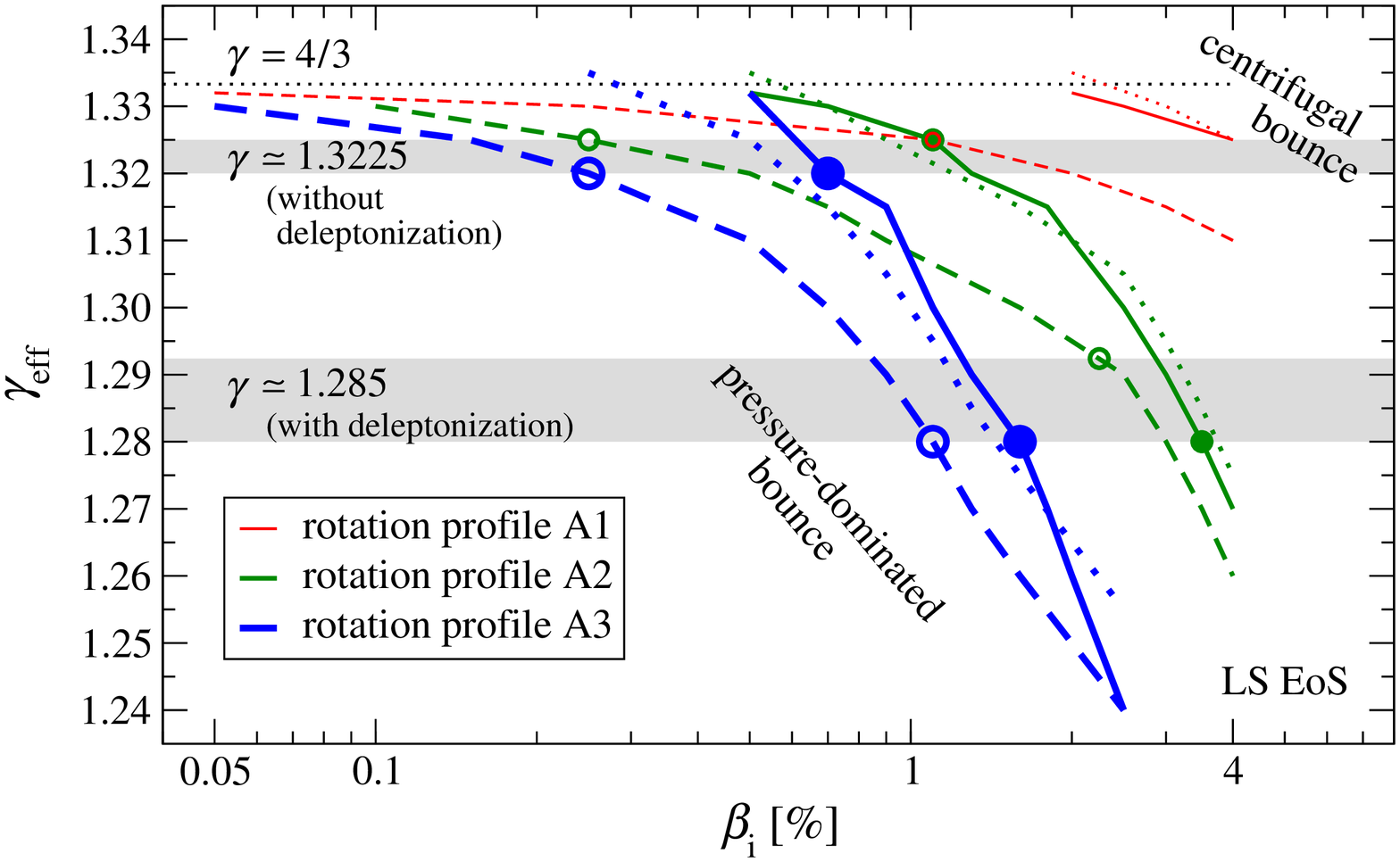}}
  \caption{Boundary between pressure-dominated and centrifugal
    bounce in the $ \gamma_\mathrm{eff} $--$ \beta_\mathrm{i} $
    plane for s20 progenitor models using the hybrid EoS in Newtonian
    gravity (dashed lines) and general relativity (solid lines). The
    curved dotted lines show the Newtonian results shifted by
    $ - \Delta \gamma_\mathrm{gr} = 0.015 $. The transition points
    for models using the microphysical EoS without and with
    deleptonization, again for Newtonian gravity (circles) and general
    relativity (bullets), lie in the shaded areas around
    $ \gamma_\mathrm{eos,Shen} \simeq 1.32 $ and
    $ \gamma_\mathrm{eff,Shen} \simeq 1.29 $, respectively,
    for the Shen EoS (top panel) and around
    $ \gamma_\mathrm{eos,LS} \simeq 1.3225 $ and
    $ \gamma_\mathrm{eff,LS} \simeq 1.285 $, respectively,
    for the LS EoS (bottom panel).}
  \label{figure:comparison_with_polytrope_gr_and_newtonian}
\end{figure}

In Fig.~\ref{figure:m_rest_inner_core} we show the resulting variation
of $ M_\mathrm{ic,b} $ with $ \Omega_\mathrm{c,i} $, progenitor model,
precollapse differential rotation parameter $ A $, EoS, and collapse type
(encoded via symbols as in Fig.~\ref{figure:collapse_dynamics}). The
details of the variation of $ M_\mathrm{ic,b} $ with progenitor, EoS
and rotational configuration will be discussed in
Section~\ref{subsection:eos_and_progenitor_influence}. In the
following, without loss of generality, we focus on a single progenitor
and discuss the influence of general relativity and deleptonization on
the collapse dynamics and the gravitational wave burst signal along
the lines of the discussion in~\cite{dimmelmeier_07_a,
  dimmelmeier_07_b}.

In order to assess the individual influence of relativistic effects
and deleptonization, and to explain the absence of type~II and~III
burst signals in microphysical general relativistic models,
in~\cite{dimmelmeier_07_a, dimmelmeier_07_b} we compared collapse
models of the s20 progenitor using the Shen EoS and a description for
deleptonization with models using a simple hybrid
polytropic/$ \gamma $-law EoS. We selected the adiabatic index
$ \gamma_\mathrm{eos} $ of these simple models in such a way that the
transition between pressure-dominated bounce and centrifugal bounce
occurs at the same precollapse rotation rate $ \beta_\mathrm{i} $ as
for the microphysical models. With this method we were able to
demonstrate that the influence of deleptonization can be approximated
by a correction $ \Delta \gamma_\nu \simeq 0.03 $ that must be applied
to the estimate of the average EoS adiabatic index
$ \gamma_\mathrm{eos,Shen} \simeq 1.32 $ in the density interval
between $ 10^{12} $ and $ 10^{14} \mathrm{\ g\ cm}^{-3} $. This leads
to a \emph{generic} value for the effective adiabatic index
$ \gamma_\mathrm{eff,Shen} = \gamma_\mathrm{eos,Shen} - \Delta
\gamma_\nu \simeq 1.32 - 0.03 = 1.29 $, practically independent of the
precollapse rotational configuration, both in Newtonian gravity and
general relativity (where relativistic effects are accounted for by a
correction $ \Delta \gamma_\mathrm{gr} \simeq - 0.015 $). A graphic
representation of this argument is shown in the top panel of
Fig.~\ref{figure:comparison_with_polytrope_gr_and_newtonian}, that is
identical to Fig.~2 in~\cite{dimmelmeier_07_a} and Fig.~4
in~\cite{dimmelmeier_07_b}, and which we include here for
completeness.

The estimate $ \gamma_\mathrm{eff} \simeq 1.29 $ for microphysical
models also allows us to explain the suppression of multiple
centrifugal bounces with an associated type~II waveform in a
straightforward way, since this type of collapse occurs only in the
respective hybrid EoS models with an effective adiabatic index that is
much closer to $ 4 / 3 $, i.e., $ \gamma_\mathrm{eff} \ge 1.31 $.
Rapid collapse dynamics that is characterized by a type~III burst
signal is also not realized in microphysical models of massive star
collapse, as it requires a mass of the inner core at bounce
$ M_\mathrm{ic,b} \lesssim 0.3 \, M_\odot $ \cite{zwerger_97_a} which
is considerably smaller than those found in microphysical models with
any of our progenitors, for which we find
$ M_\mathrm{ic} \gtrsim 0.4 \, M_\odot $ (see
Fig.~\ref{figure:m_rest_inner_core}, and also the discussion in
Section~\ref{subsection:eos_and_progenitor_influence}). However,
in~\cite{dessart_06_a, ott_06_b} is was suggested that rapid collapse
dynamics and a type~III burst signal may be associated with very
efficient electron capture in the accretion-induced collapse of
massive, rapidly rotating white dwarfs.

Finally, we point out that calculations with $ \gamma_\mathrm{eff} $
used in the hybrid EoS have the tendency to underestimate the mass
$ M_\mathrm{ic,b} $ at bounce compared to the full microphysical
treatment. This is a consequence of the fact that in these
calculations $ \gamma_\mathrm{eff} $ is kept constant throughout the
collapse, leading to a reduction of the inner core mass
$ M_\mathrm{ic} $ already at much earlier collapse stages than in
microphysical models. The underestimated $ M_\mathrm{ic,b} $, in turn,
leads to gravitational wave burst signals from bounce in those simple
models that are quantitatively or even qualitatively incorrect (as in
the case of type~III signals which do not occur in microphysical
models). Hence, while useful for understanding the collapse dynamics,
the $ \gamma_\mathrm{eff} $ approach cannot replace the full
microphysical treatment with a nonzero-temperature microphysical EoS
and deleptonization as performed in the present work.


\subsection{Influence of the equation of state and progenitor model}
\label{subsection:eos_and_progenitor_influence}

\begin{figure}[t]
  \epsfxsize = 8.6 cm
  \centerline{\epsfbox{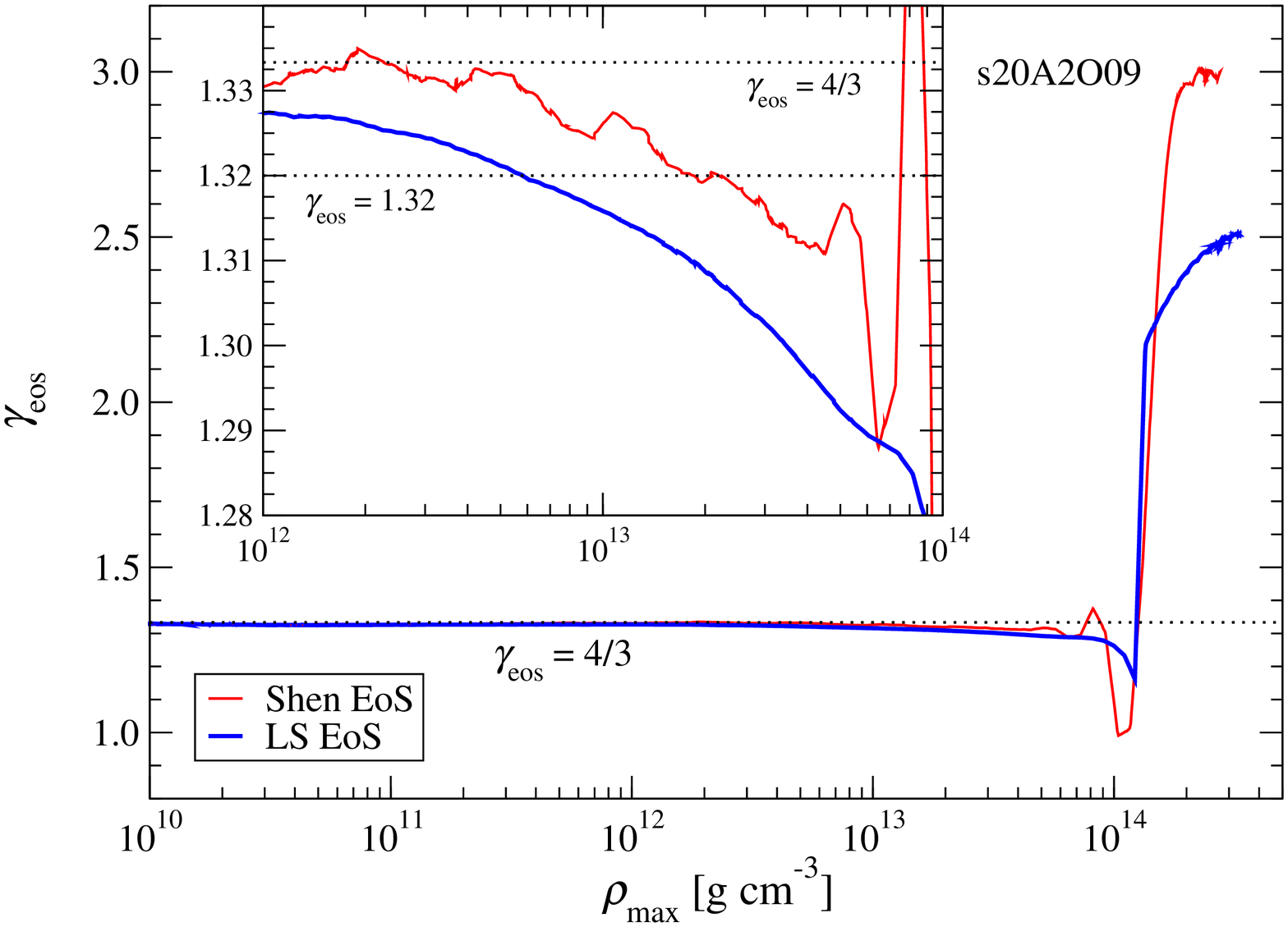}}
  \caption{Adiabatic index $ \gamma_\mathrm{eos} $ of the
    Shen EoS (red line) and LS EoS (blue line) versus the maximum
    density $ \rho_\mathrm{max} $ in the collapsing core for model
    s20A2O09. Although $ \rho_\mathrm{max} $, which for this model is
    located in the center of the core, does not follow a trajectory of
    constant specific entropy, $ s $ is still approximately conserved in the
    prebounce phase. Inset: Magnified view of $ \gamma_\mathrm{eos} $
    in the dynamically most relevant density range between $ 10^{12} $
    and $ 10^{14} \mathrm{\ g\ cm}^{-3} $. The average value of
    $ \gamma_\mathrm{eos} $ in this density regime is roughly $ 1.32 $
    for both EoSs.}
  \label{figure:gamma_eos_versus_rho_max}
\end{figure}

At densities below $ \rho_\mathrm{nuc} $ the total fluid pressure is
dominated by the contribution from the degenerate electrons, hence the
two microphysical EoSs should lead to rather similar dynamics in the
infall phase of collapse. This is also reflected in the very similar
behavior of their adiabatic indices $ \gamma_\mathrm{eos} $ as shown
in Fig.~\ref{figure:gamma_eos_versus_rho_max}.

\begin{figure}[t]
  \epsfxsize = 8.6 cm
  \centerline{\epsfbox{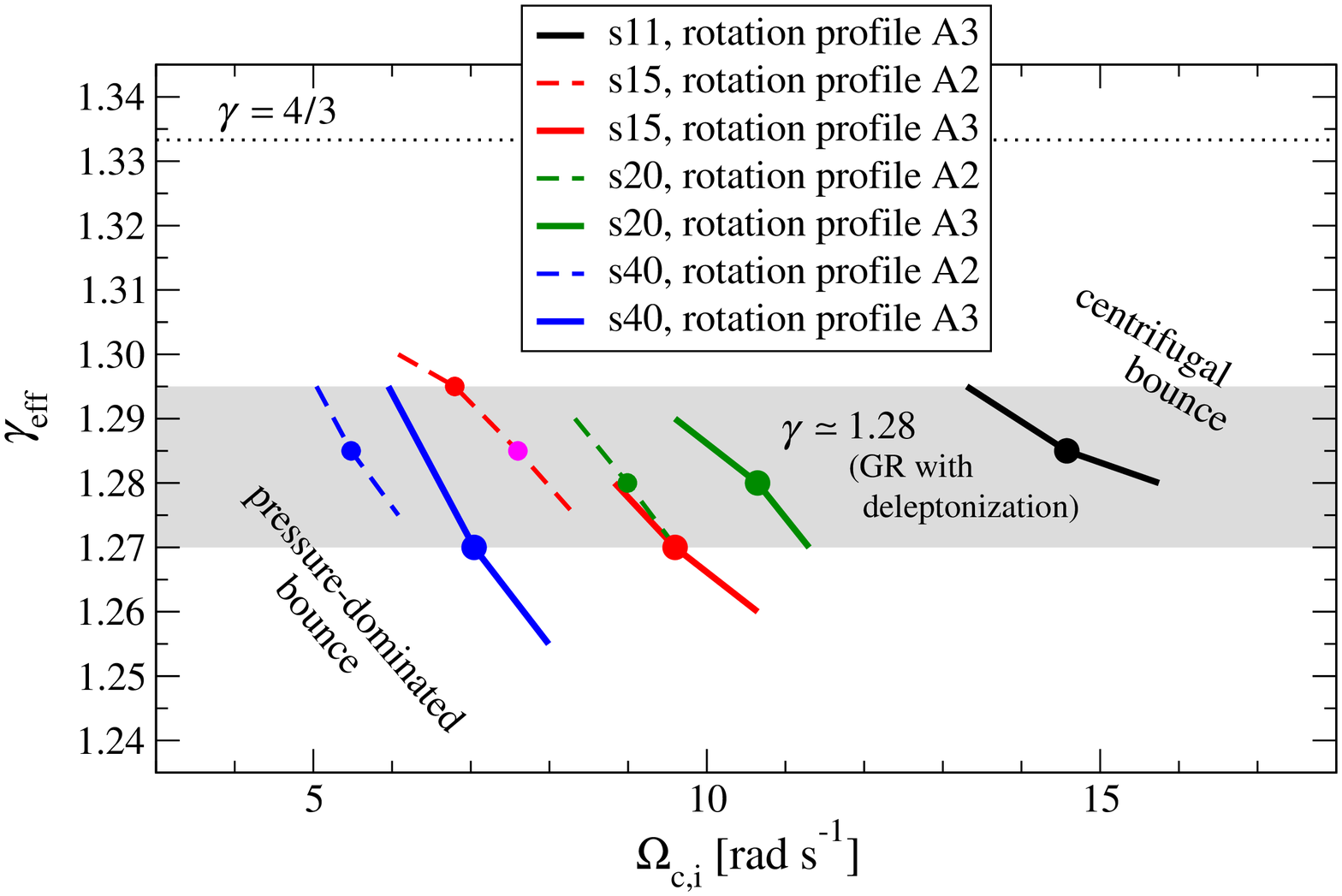}}
  \caption{Boundary between pressure-dominated and centrifugal
    bounce in the $ \gamma_\mathrm{eff} $--$ \Omega_\mathrm{c,i} $
    plane for models of all progenitors using the hybrid EoS in
    general relativity. The transition points for models using the
    microphysical EoS with deleptonization (bullets) lie in the shaded
    area around $ \gamma_\mathrm{eff} \simeq 1.28 $. Except for the
    rotation profile A2 of the s15 progenitor, the locations of these
    points are identical for the two EoSs. Note that for the A1
    profiles of any progenitor (and the A2 profile for the s11
    progenitor) we do not observe a centrifugal bounce for any
    value of $ \Omega_\mathrm{c,i} $. In contrast to
    Fig.~\ref{figure:comparison_with_polytrope_gr_and_newtonian} we
    use here the $ \Omega_\mathrm{c,i} $ instead of
    $ \beta_\mathrm{i} $ as parameter for the precollapse rotational
    configuration (see the discussion in
    Section~\ref{subsection:models}).}
  \label{figure:gamma_eff_versus_omega_0_shen_and_ls}
\end{figure}

In the bottom panel of
Fig.~\ref{figure:comparison_with_polytrope_gr_and_newtonian} we
demonstrate that the same influence of general relativistic effects and
deleptonization as discussed in the previous
Section~\ref{subsection:gr_and_deleptonization_influence} applies for
the s20 progenitor when the LS EoS is used instead of the Shen EoS. We
obtain values of $ \gamma_\mathrm{eos,LS} \simeq 1.3225 $ for the
adiabatic index of the EoS (without deleptonization) and
$ \gamma_\mathrm{eff,LS} \simeq 1.285 $ for the effective adiabatic
index (including deleptonization), which is in very close agreement
with the values deduced from the simulations using the Shen EoS. As
shown in Fig.~\ref{figure:gamma_eff_versus_omega_0_shen_and_ls}, now
only for general relativistic gravity, there is some spread of
$ \gamma_\mathrm{eff} $ with progenitor mass/model, but on average, we
find $ \gamma_\mathrm{eff} \simeq 1.28 $ for the s11, s15, and
s40 progenitor models.

Again following the line of arguments presented
in~\cite{dimmelmeier_07_a, dimmelmeier_07_b}, the combination of a low
effective adiabatic index $ \gamma_\mathrm{eff} < 1.31 $ and a high
inner core mass $ M_\mathrm{ic} \gtrsim 0.4 \, M_\odot $ at bounce
results in a type~I gravitational wave burst signal for all our
models, independent of the EoS or progenitor model. Note that creating
Figs.~\ref{figure:comparison_with_polytrope_gr_and_newtonian}
and~\ref{figure:gamma_eff_versus_omega_0_shen_and_ls} we have
performed additional simulations of microphysical models that are more
narrowly spaced in $ \Omega_\mathrm{i,c} $ (and correspondingly in
$ \beta_\mathrm{i} $) than the ones listed in
Table~\ref{table:initial_models}. As a result, those figures reveal a
small dependence of the transition between pressure-dominated bounce
and centrifugal bounce (i.e., the location of the bullets and circles
in the direction of the abscissa) on the EoS, which is generally not
apparent from Table~\ref{table:initial_models}.

Although the sensitivity of the deleptonization and collapse dynamics
on the progenitor mass and EoS is only small,
Fig.~\ref{figure:m_rest_inner_core} still reveals a dependence of the
inner core mass $ M_\mathrm{ic,b} $ at bounce both on the EoS and (in
particular) on rotation. Furthermore, $ M_\mathrm{ic,b} $ varies
non-monotonically with the progenitor mass $ M_\mathrm{prog} $.  In
the absence of rotation, $ M_\mathrm{ic,b} $ is solely determined by
the mean trapped lepton fraction $ Y_\mathrm{lep} = Y_e + Y_\nu $ and
specific entropy $ s $ in the inner core~\cite{goldreich_80_a,
van_riper_81_a, yahil_83_a, burrows_83_a} with a roughly quadratic
dependence on both quantities. Since we employ the same $ Y_e (\rho) $
parametrization (based on model s20) for all models with the same EoS,
the variations in $ M_\mathrm{ic,b}$ are caused by differences in the
specific entropy in the precollapse iron core. Generally, the specific
entropy in the iron core increases with progenitor mass, but, in
particular in the mass range of $ \sim 18\mbox{\,--\,}25 \, M_\odot $, the
relationship of progenitor mass and specific core entropy can be
non-monotonic (see, e.g., \cite{woosley_02_a}).  However, note that
the systematics for $ M_\mathrm{ic,b} $ with progenitor mass seen in
Fig.~\ref{figure:m_rest_inner_core} are possibly overemphasized by our
$ Y_e (\rho) $ parametrization and may be less pronounced in full
radiation transport simulations which remain to be carried out in the
future.

For a rotating collapse, the variations of $ M_\mathrm{ic,b} $ with
progenitor mass are amplified, while the overall systematics are
preserved. Obviously, a more massive and hence more extended inner
core is more susceptible to the influence of centrifugal forces (which
scale proportional to the radius $ r $) than a less massive and thus
more compact inner core. This behavior is confirmed by
Fig.~\ref{figure:m_rest_inner_core}, which depicts the dependence of
the mass $ M_\mathrm{ic,b} $ of the inner core at bounce on the
precollapse central angular velocity $ \Omega_\mathrm{c,i} $, the EoS,
and the differential rotation parameter $ A $. Models with comparably
large precollapse iron core specific entropy (and also large iron core
mass) and, thus, larger $ M_\mathrm{ic,b} $ already in the nonrotating
case, show a more pronounced increase of $ M_\mathrm{ic,b} $ with
rotation than models with lower precollapse specific entropy (and also
smaller iron core mass). The scaling of $ M_\mathrm{ic,b} $ with
$ \Omega_\mathrm{c,i} $, at fixed differential rotation parameter
$ A $, is linear for small to intermediate $ \Omega_\mathrm{c,i} $ and
turns approximately quadratic for the most rapidly rotating
configurations. On the other hand, when increasing the degree of
differential rotation $ A $ at fixed $ \Omega_\mathrm{c,i} $,
$ M_\mathrm{ic,b} $ decreases since then the angular velocity in the
outer parts of the inner core and consequently centrifugal support
drops.

We also observe that the impact of the EoS on the mass of the inner
core manifests itself only via an almost constant positive relative
increase in $ M_\mathrm{ic,b} $ of $ \sim 10\% $ when changing from
the LS EoS to the Shen EoS, practically independent of rotation and
progenitor mass (see Fig.~\ref{figure:m_rest_inner_core}). Again, the
mean electron (respectively lepton) fraction and specific entropy in
the inner core are the key to understanding these systematics. The
representative s20 progenitor model used to parametrize $ Y_e (\rho) $
in this study yields minima for $ Y_e $ in the center of the core at
bounce of $ \sim 0.249 $ and $ \sim 0.241 $ for the Shen EoS and the
LS EoS, respectively. This relative difference of $ \sim 3.3\% $
translates into a difference in $ M_\mathrm{ic,b} $ of $ \sim 7\% $,
assuming that the mass of the inner core scales quadratically with
$ Y_e $, which slightly underestimates the actual change. We attribute
the remaining difference to variations in the specific entropy $ s $
of the inner core at bounce due to the slightly more efficient
electron capture in the models with the LS EoS.

We point out that the progenitor models e15a, e15b, e20a, and e20b,
which already come with a rotation profile from the stellar evolution
calculation, are very well represented in terms of collapse dynamics,
waveform, and postbounce rotation state by members of our model set
with an artificially added precollapse rotation profile, specifically
the models s15A2O09, s15A2O15, s20A2O09, and s20A2O15. For this reason
we refrain from separately discussing those special models in the
entire paper.


\subsection{Influence of differential rotation}
\label{subsection:differential_rotation}

Increasing the degree of differential rotation by lowering the value
of the differential rotation parameter $ A $ at fixed precollapse
central angular velocity $ \Omega_\mathrm{c,i} $ results in less
centrifugal support in outer core regions and, as already pointed out
in Section~\ref{subsection:eos_and_progenitor_influence}, in a smaller
mass $ M_\mathrm{ic,b} $ of the inner core at bounce. Consequently, a
higher value of $ \Omega_\mathrm{c,i} $ is necessary for a stronger
differentially rotating precollapse core to become significantly
affected by centrifugal forces during the collapse. This is confirmed
by Fig.~\ref{figure:gamma_eff_versus_omega_0_shen_and_ls} which
displays the systematics of the transition between pressure-dominated
and centrifugal bounce for our set of progenitors and the A2 and A3
rotation profiles. Compared to the transition values of
$ \Omega_\mathrm{c,i} $ for the A2 profile, the A3 profile requires a
roughly $ 20\mbox{\,--\,}40\% $ higher $ \Omega_\mathrm{c,i} $
(varying slightly with progenitor model) for a transition from
pressure-dominated to centrifugal bounce.

\begin{figure}[t]
  \epsfxsize = 8.6 cm
  \centerline{\epsfbox{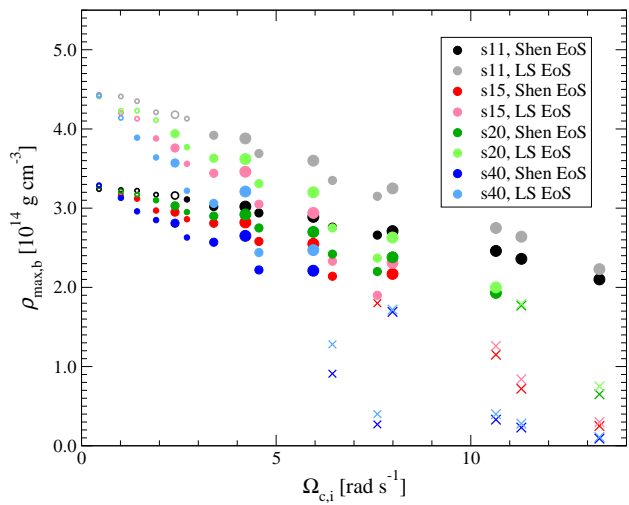}}
  \caption{Maximum density $ \rho_\mathrm{max,b} $ in the star at the
    time of bounce for all models versus the precollapse
    central angular velocity $ \Omega_\mathrm{c,i} $. For moderately
    fast or rapidly rotating models, which do not exceed nuclear
    density at bounce, $ \rho_\mathrm{max,b} $ is almost identical for
    the Shen EoS (dark hues) and the LS EoS (light hues), while for
    slowly rotating models the difference reaches $ \Delta
    \rho_\mathrm{max,b} \simeq 10^{14} \mathrm {\ g\ cm}^{-3} $ in the
    nonrotating limit. The progenitor mass, the EoS, the precollapse
    rotation parameter $ A $, and the collapse dynamics are encoded as
    in Fig.~\ref{figure:collapse_dynamics}.}
  \label{figure:rho_max_b_versus_omega_cap_c_ini}
\end{figure}

In previous extensive parameter studies of rotating stellar core
collapse (see, e.g., \cite{dimmelmeier_02_a, ott_04_a, kotake_03_a})
the effect of differential rotation was studied in model sequences in
the parameter space spanned by the precollapse differential rotation
parameter $ A $ and the precollapse rotation rate $ \beta_\mathrm{i} $.
At a constant $ \beta_\mathrm{i} $, more differentially rotating
models require a larger $ \Omega_\mathrm{c,i} $ than less
differentially rotating ones and experience core bounce at lower
densities. Hence, at fixed $ \beta_\mathrm{i} $,
more differentially rotating models are generally more affected by
centrifugal effects. Our s20 model series is constructed as a sequence
of fixed $ \beta_\mathrm{i} $ for each of the rotation profiles A1, A2,
and A3 (see Table~\ref{table:initial_models}), and therefore permits
a direct comparison with preceeding work. Our results confirm
\emph{qualitatively} the previously identified systematics (see
Table~\ref{table:collapse_models}). However, in contrast to more
simplistic simulations, the combination of general relativity and
deleptonization in our models weakens the overall
impact of centrifugal effects on the collapse dynamics (see
Section~\ref{subsection:gr_and_deleptonization_influence}), and
consequently leads to much smaller \emph{quantitative} changes in the
characteristic collapse variables (such as $ \rho_\mathrm{max,b} $,
$ |h|_\mathrm{max} $, $ M_\mathrm{ic,b} $, or $ \beta_\mathrm{b} $)
when varying the degree of differential rotation.


\section{Structure of the postbounce core and impact on the wave
  signal}
\label{section:density_structure_and_waveform}


\subsection{Equation of state at supernuclear densities and maximum
  density in the core}
\label{subsection:eos_at_supernuclear_densities}

\begin{figure}[t]
  \epsfxsize = 8.6 cm
  \centerline{\epsfbox{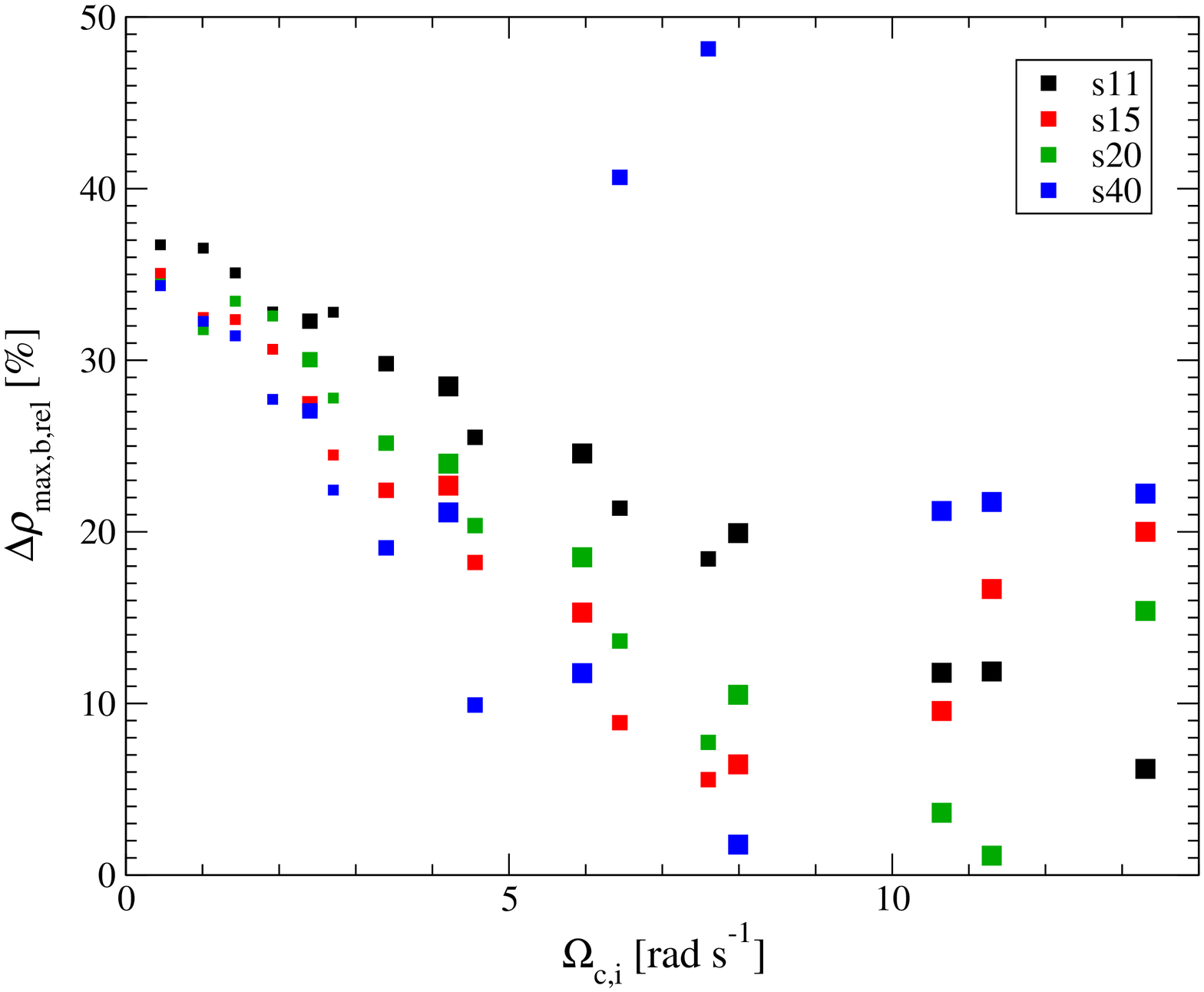}}
  \caption{Relative change $ \Delta \rho_\mathrm{max,b,rel} $ of the
    maximum density at bounce when changing from the Shen EoS to the
    LS EoS for all models versus the precollapse central
    angular velocity $ \Omega_\mathrm{c,i} $. The progenitor mass and
    the precollapse differential rotation parameter $ A $ are encoded
    as in Fig.~\ref{figure:collapse_dynamics}, while the collapse
    dynamics are not specified.}
  \label{figure:delta_rho_max}
\end{figure}

From Table~\ref{table:collapse_models} it is apparent that the change
from the Shen EoS to the LS EoS in an otherwise identical model
results systematically in an increase of the peak maximum density
$ \rho_\mathrm{max,b} $ at bounce, i.e.,
$ \rho_\mathrm{max,b,LS} > \rho_\mathrm{max,b,Shen} $. This result is
in agreement with the previous Newtonian study of Kotake et
al.~\cite{kotake_04_a} who compared simulations of a single model 
carried out with the Shen EoS and the LS EoS.

For centrifugally bouncing models, which only marginally exceed or
even remain below $ \rho_\mathrm{nuc} $ at bounce, the \emph{absolute}
increase in the maximum core density at bounce is small, exhibiting a
maximum $ \Delta \rho_\mathrm{max,b} = \rho_\mathrm{max,b,LS} -
\rho_\mathrm{max,b,Shen} = 0.13 \times 10^{14} \mathrm{\ g\ cm}^{-3} $
for model s40A3O09 (leaving aside the exceptional models s40A2O13 and
s40A2O15 which we will discuss separately later). This is another
manifestation of the similarity of the two microphysical EoSs at
subnuclear densities (see also
Section~\ref{subsection:eos_and_progenitor_influence}).

For slowly or at most moderately fast rotating models that undergo
pressure-dominated bounce and whose center exceeds supernuclear
density at (and also after) bounce, $ \Delta \rho_\mathrm{max,b} $ can
amount up to $ 1.19 \times 10^{14} \mathrm{\ g\ cm}^{-3} $ for model
s11A1O01, the most slowly rotating model of the s11 model series. This
strong impact of the EoS can be readily explained by the fact that at
supernuclear densities the LS EoS is considerably softer than the Shen
EoS. Fig.~\ref{figure:gamma_eos_versus_rho_max} shows a difference in
the adiabatic index $ \gamma_\mathrm{eos} $ between the two
microphysical EoSs (for a representative model) of about
$ \Delta \gamma_\mathrm{eos} \simeq - 0.5 $ at those densities, where
nuclear forces have an essential impact on the EoS properties.

The large effect of the EoS seen in $ \rho_\mathrm{max,b} $ in models
where this quantity exceeds $ \rho_\mathrm{nuc} $ does not contradict
our observation that the EoS has little impact on the collapse
dynamics, since once the core plunges into the supernuclear density
regime, where stronger differences in the two microphysical EoSs
emerge, the mass $ M_\mathrm{ic,b} $ of the inner core at bounce is
already fixed and the bounce dynamics (pressure-dominated or
centrifugal) is already determined.

\begin{figure}[t]
  \epsfxsize = 8.6 cm
  \centerline{\epsfbox{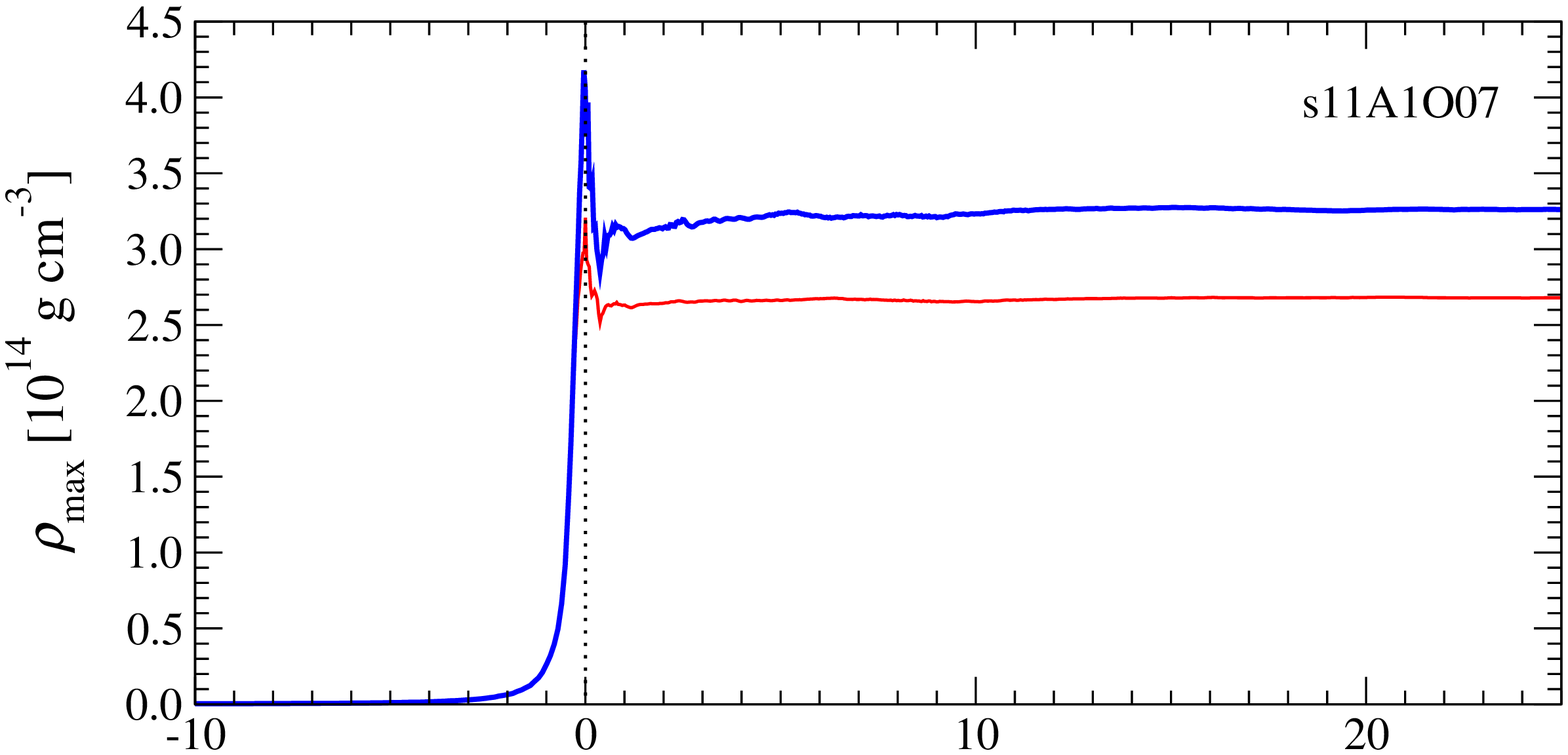}}
  ~ \\
  \epsfxsize = 8.6 cm
  \centerline{\epsfbox{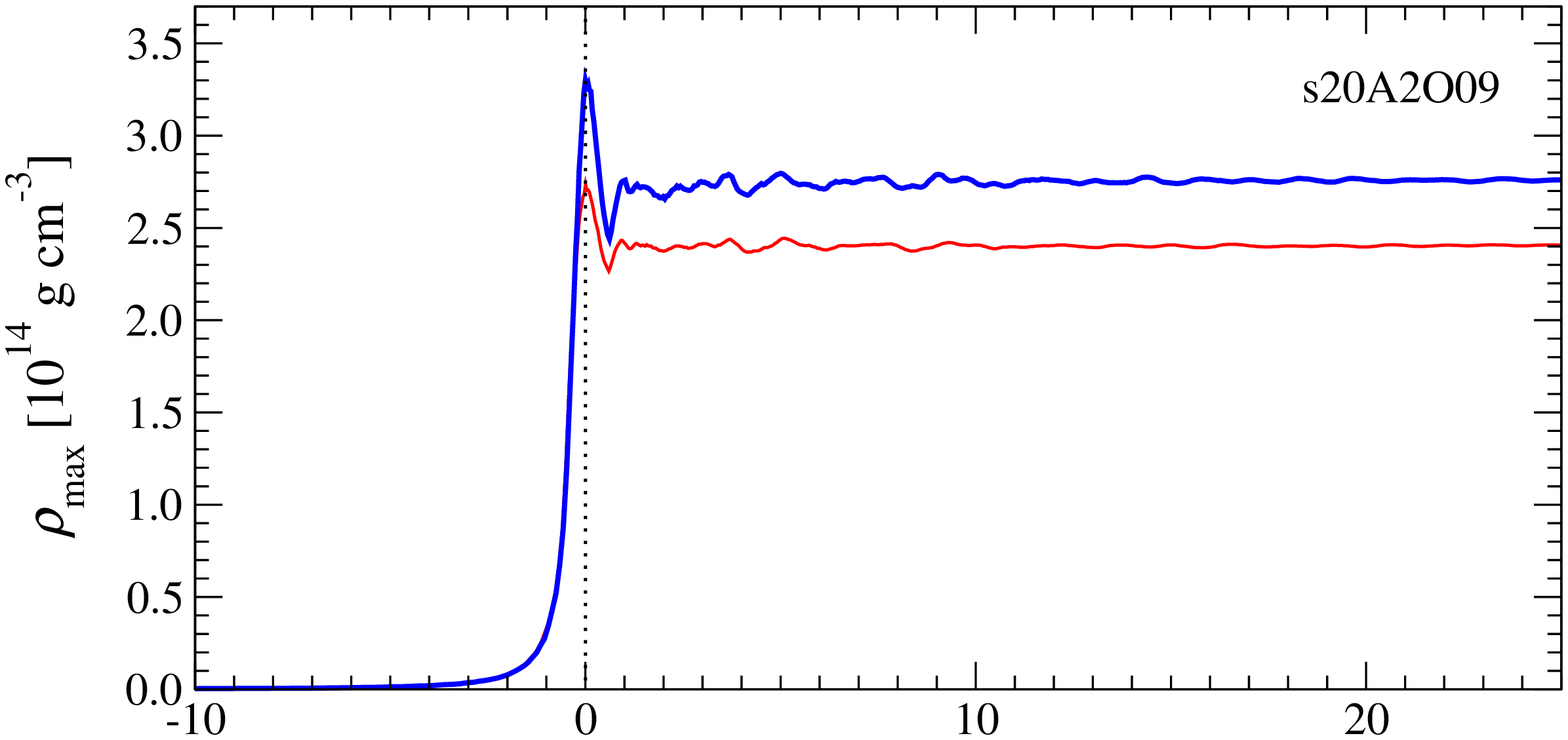}}
  ~ \\
  \epsfxsize = 8.6 cm
  \centerline{\epsfbox{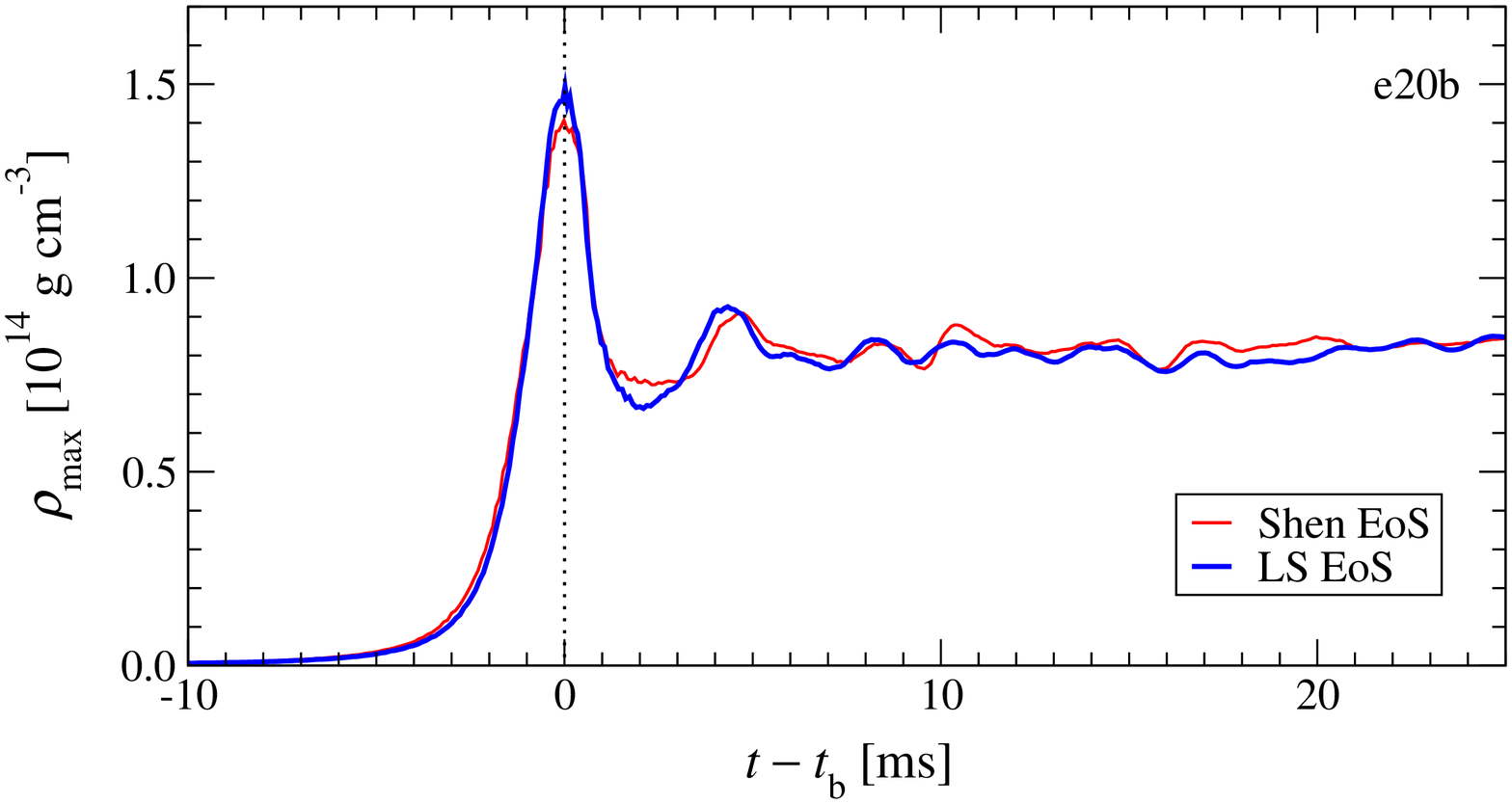}}
  \caption{Time evolution of the maximum density $ \rho_\mathrm{max} $
    for representative models with different precollapse rotation profiles
    using the Shen EoS (red lines) or LS EoS (blue lines). While
    models with at most moderate precollapse rotation rates (e.g., s11A1O07
    or s20A2O09) undergo a pressure-dominated bounce at supernuclear
    densities, rapidly rotating models (e.g., e20b) experience a
    centrifugal bounce.}
  \label{figure:generic_collapse_type_density}
\end{figure}

The impact of the EoS on $ \rho_\mathrm{max,b} $ is also visualized in
Fig.~\ref{figure:rho_max_b_versus_omega_cap_c_ini}. As expected, for
moderately or rapidly rotating models, whose central parts do not
reach high supernuclear densities at bounce, the difference in
$ \rho_\mathrm{max,b} $ gradually decreases.
Fig.~\ref{figure:rho_max_b_versus_omega_cap_c_ini} also reveals that
the two models s40A2O13 and s40A2O15 (marked by two dark and light
blue crosses at intermediate values of $ \Omega_\mathrm{c,i} $,
respectively) are the ones that undergo a clear centrifugal bounce
for both EoSs with the lowest value of $ \Omega_\mathrm{c,i} $ of all
models with the A2 rotation profile.

The convergence of $ \rho_\mathrm{max,b} $ for the two microphysical
EoS with increasing rotation can also be observed in the
\emph{relative} difference $ \Delta \rho_\mathrm{max,b,rel} =
\rho_\mathrm{max,b,LS} / \rho_\mathrm{max,b,Shen} - 1 $ shown in
Fig.~\ref{figure:delta_rho_max}, which starting from $ \Delta
\rho_\mathrm{max,b,rel} \simeq 35\mbox{\,--\,}40\% $ in the
nonrotating limit first declines linearly with $ \Omega_\mathrm{c,i} $
until it levels off at roughly constant values. However, the largest
values are obtained with
$ \Delta \rho_\mathrm{max,b,rel} \simeq 40\% $ and $ 48\% $ for the
rapidly rotating and centrifugally bouncing models s40A2O13 and
s40A2O15, emphasizing their exceptional nature. This particular
behavior results from a combination of two effects, exhibited by only
these two models in our entire model set. First, when switching from
the Shen EoS to the LS EoS the inner core mass $ M_\mathrm{ic,b} $ at
bounce significantly decreases (see
Fig.~\ref{figure:m_rest_inner_core}). Hence in the LS EoS variant the
two models experience weaker rotational support (in particular with
the differential rotation parameter A2; see also
Section~\ref{subsection:eos_and_progenitor_influence}). Second, the
two models bounce in a density regime (see
Table~\ref{table:collapse_models} and
Fig.~\ref{figure:gamma_eos_versus_rho_max}) where the LS EoS exhibits
a smaller $ \gamma_\mathrm{eos} $ than the Shen EoS, resulting in less
pressure support when the LS EoS is used. The combination of weaker
rotational support and pressure support when using the LS EoS can then
readily explain the excess in $ \rho_\mathrm{max,b,LS} $ compared to
$ \rho_\mathrm{max,b,Shen} $ in the two exceptional s40 models.

A higher value of the maximum density $ \rho_\mathrm{max} $ in the
collapsed core for the LS EoS is not limited to the time of bounce,
but typically also remains in the nascent proto-neutron star at later
postbounce times, as shown in
Fig.~\ref{figure:generic_collapse_type_density} for models
representing the three collapse type and waveform subclasses (see
Section~\ref{subsection:generic_waveform}). Only very rapidly and thus
centrifugally bouncing models such as model e20b in
Fig.~\ref{figure:generic_collapse_type_density} have a time evolution
of $ \rho_\mathrm{max} $ that is practically independent of the chosen
EoS.

We point out that in our discussion we always make use of the maximum
density $ \rho_\mathrm{max} $ instead of the central density
$ \rho_\mathrm{c} $, since, after bounce, some of the most rapidly
rotating and thus centrifugally bouncing models develop a slightly
toroidal density structure with an off-center density maximum that is
at most $ 20\% $ larger than $ \rho_\mathrm{c} $. This is much less
extreme than for models with the simplified hybrid EoS treatment,
where the maximum density was found to be several orders of magnitude
larger than the central density in extreme cases~\cite{zwerger_97_a,
  dimmelmeier_02_a}.


\subsection{Structure of the postcollapse core and peak waveform
  amplitude}
\label{subsection:waveform_peak_amplitude}

\begin{figure}[t]
  \epsfxsize = 8.6 cm
  \centerline{\epsfbox{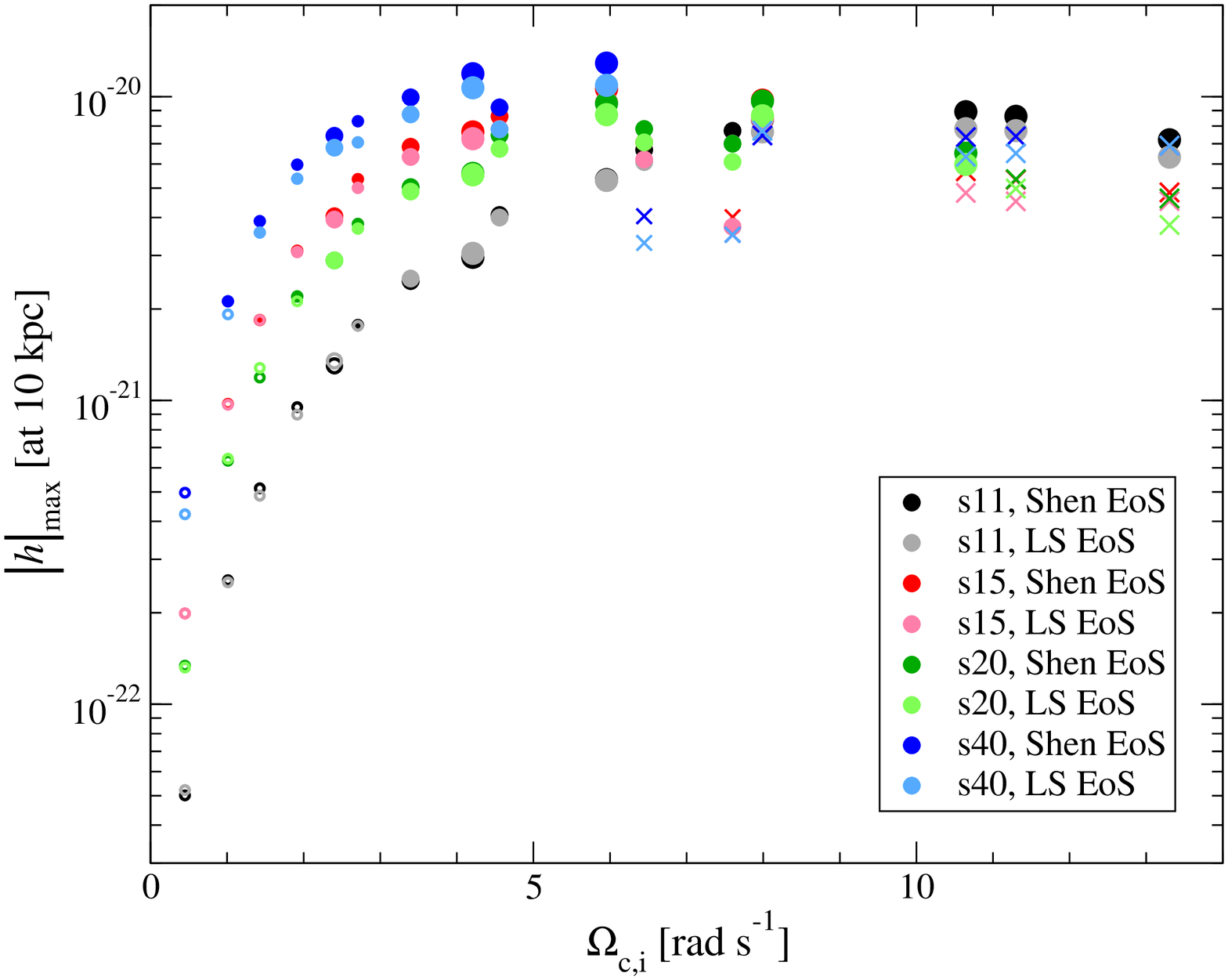}}
  \caption{Peak value $ |h|_\mathrm{max} $ of the gravitational wave
    amplitude at $ 10 \mathrm{\ kpc} $ distance for the burst signal
    (neglecting possibly larger contributions from postbounce
    convection at later times) for all models versus initial
    precollapse central angular velocity $ \Omega_\mathrm{c,i} $. The
    progenitor mass, the EoS, the precollapse differential rotation
    parameter $ A $, and the collapse dynamics are encoded as in
    Fig.~\ref{figure:collapse_dynamics}.}
  \label{figure:h_max_versus_omega_cap_c_ini}
\end{figure}

Since the LS EoS leads to higher central densities at bounce, one can
on the one hand expect higher gravitational wave peak amplitudes in
the burst signal from core bounce, as a denser and more compact core
should yield in an increase of the contribution to the quadrupole
moment from the central parts of the core. Furthermore, the associated
shorter dynamical times also lead to a more rapid time variation
in the quadrupole formula. On the other hand, the higher compactness
of the inner core of a model run with the LS EoS results in lower
densities at intermediate and large radii than in the less compact
core of the corresponding model with the Shen EoS. In turn, this may
lead to an effectively smaller \emph{total} quadrupole moment and thus
to a decrease in the signal amplitude compared to the counterpart
model with the Shen EoS. We now test which of these two competing
effects dominates in our models.

In Fig.~\ref{figure:h_max_versus_omega_cap_c_ini} we show the peak
value $ |h|_\mathrm{max} $ of the gravitational wave amplitude for the
burst signal from core bounce (see also
Table~\ref{table:collapse_models}), where we neglect any possibly larger
contributions at later times for models with strong prompt postbounce
convection. For slowly or at most moderately rapidly rotating cores,
$ |h|_\mathrm{max} $ rises steeply with increasing
$ \Omega_\mathrm{c,i} $ and covers a range of more than two orders of
magnitude for our selection of initial models. For rapid rotation,
when centrifugal forces become dynamically important and can be the
dominant factor at bounce, the peak amplitude $ |h|_\mathrm{max} $
saturates and even decreases again at very high
$ \Omega_\mathrm{c,i} $. This behavior is a consequence of centrifugal
support, which prevents such rapidly spinning cores from reaching high
densities and more extreme compactness as well as being subject to
short variations of the quadrupole moment (see also the discussion in
Section~\ref{section:rotation_rate} and in~\cite{ott_04_a}).

\begin{figure}[t]
  \epsfxsize = 8.6 cm
  \centerline{\epsfbox{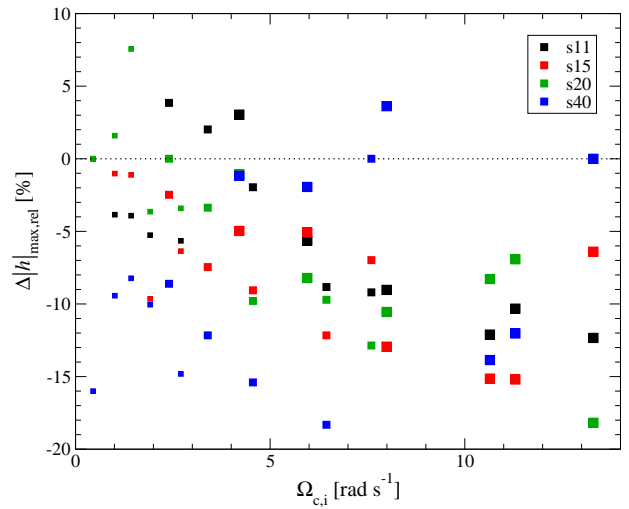}}
  \caption{Relative change $ \Delta |h|_\mathrm{max,rel} $ of the peak
    value $ |h|_\mathrm{max} $ of the gravitational wave amplitude for
    the burst signal when changing from the Shen EoS to the LS EoS for
    all models versus the precollapse central angular velocity
    $ \Omega_\mathrm{c,i} $. The progenitor mass and the precollapse
    differential rotation parameter $ A $ are encoded as in
    Fig.~\ref{figure:collapse_dynamics}, while the collapse dynamics
    are not specified, as for some models it depends on the EoS.}
  \label{figure:delta_h_max}
\end{figure}

For each precollapse rotational configuration (i.e., at constant
$ \Omega_\mathrm{c,i} $ and differential rotation parameter $ A $ in
Fig.~\ref{figure:h_max_versus_omega_cap_c_ini}), the value of
$ |h|_\mathrm{max} $ depends indirectly on the mass
$ M_\mathrm{prog} $ of the progenitor via the mass $ M_\mathrm{ic,b} $
of the inner core at bounce. As already discussed in
Section~\ref{subsection:eos_and_progenitor_influence},
$ M_\mathrm{ic,b} $ does not depend in a monotonic way on
$ M_\mathrm{prog} $, but for our standard model set increases in the
order of the progenitor models s11, s20, s15, and s40. Therefore, for
pressure-dominated bounce models the amplitude of the gravitational
wave signal directly scales with $ M_\mathrm{ic,b} $ in the obvious
sense that more massive inner cores produce stronger gravitational
wave emission.

What cannot be extracted from
Fig.~\ref{figure:h_max_versus_omega_cap_c_ini} is a clear effect of
the choice of the EoS on $ |h|_\mathrm{max} $, despite the strong
difference in $ \rho_\mathrm{max,b} $ we observe between models using
the Shen EoS and the LS EoS. When plotting the relative change
$ \Delta |h|_\mathrm{max,rel} = |h|_\mathrm{max,LS} /
|h|_\mathrm{max,Shen} - 1 $ obtained by changing from the Shen EoS to
the LS EoS for the same initial model (as presented in
Fig.~\ref{figure:delta_h_max}), the majority of models shows a
decrease of $ |h|_\mathrm{max,LS} $ compared to
$ |h|_\mathrm{max,Shen} $. Only six out of the 68 models (s11A2O05,
s11A2O07, s11A3O05, s20A1O05, s20A1O07, and s40A3O09) listed in
Table~\ref{table:collapse_models} exhibit a larger $ |h|_\mathrm{max} $
when the LS EoS is used.

\begin{figure}[t]
  \epsfxsize = 8.6 cm
  \centerline{\epsfbox{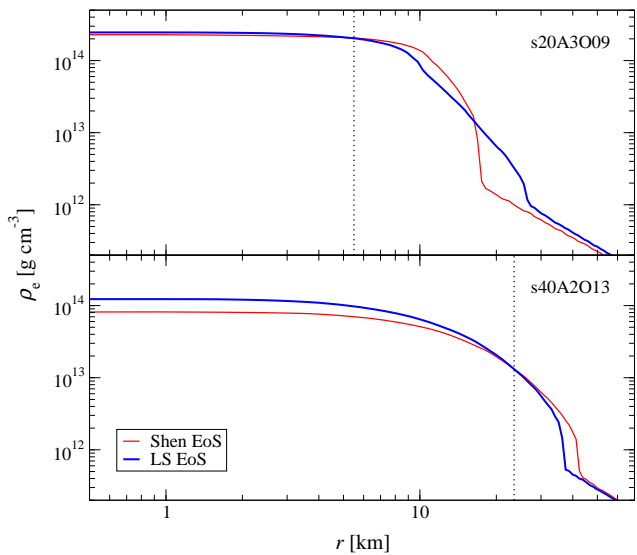}}
  \caption{Radial profiles of the density $ \rho_\mathrm{e} $ in the
    equatorial plane at the time of bounce for model s20A3O09 (top
    panel) and model s40A2O13 (bottom panel) using the Shen EoS (red
    line) and LS EoS (blue line). In the central parts of the
    proto-neutron star (for these models at radii smaller than the
    crossing radius $ r \simeq 5.5 \mathrm{\ km} $ and
    $ r \simeq 23.5 \mathrm{\ km} $, respectively, marked by dotted
    lines) the LS EoS leads to higher densities.}
  \label{figure:density_profile}
\end{figure}

This behavior is similar to the situation discussed by Dimmelmeier et
al.~\cite{dimmelmeier_02_a} who compared collapse dynamics and
gravitational waveforms obtained from Newtonian and general
relativistic collapse simulations with the simple hybrid
EoS. They showed that for $|h|_\mathrm{max}$ the \emph{global}
density distribution in the core at bounce is decisive, not the
\emph{local} maximum density value. In their simulations, the general
relativistic variants consistently produced an increase of
$ \rho_\mathrm{max,b} $ compared to their Newtonian counterparts.
Still, they found that the peak value $ |h|_\mathrm{max} $ of the
gravitational wave amplitude actually decreases for most models when
general relativistic effects are taken into account.

In~\cite{dimmelmeier_02_a}, the negative
$ \Delta |h|_\mathrm{max,rel} $ observed in many models when comparing
Newtonian and relativistic simulations could be attributed to the
``density crossing'' that occurs at some radius inside the inner
core at bounce: The general relativistic simulation of a model yields
a higher density inside that (angle-dependent, due to rotation)
radius, while for larger distances from the center, $ \rho $ is
smaller compared to the Newtonian simulation. Here, we vary the EoS
rather than the description of gravity, but we observe a very similar
density crossing in models that show
$ |h|_\mathrm{max,LS} < |h|_\mathrm{max,Shen} $. In
Fig.~\ref{figure:density_profile} we demonstrate this for models
s20A3O09 (representative for a pressure-dominated bounce) and s40A2O13
(representative for a centrifugal bounce).

\begin{figure}[t]
  \epsfxsize = 8.6 cm
  \centerline{\epsfbox{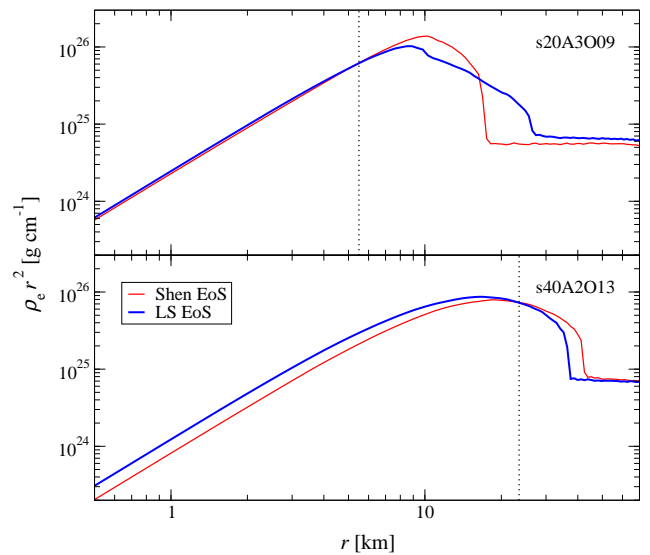}}
  \caption{Radial profiles of the weighted density
    $ \rho_\mathrm{e} r^2 $ in the equatorial plane at the time
    of bounce for model s20A3O09 (top panel) and model s40A2O13
    (bottom panel) using the Shen EoS (red lines) and LS EoS (blue
    lines). The vertical lines mark the crossing radius.}
  \label{figure:weighted_density_profile}
\end{figure}

Following the argument in~\cite{dimmelmeier_02_a}, we plot the
weighted density $ \rho r^2 $ in
Fig.~\ref{figure:weighted_density_profile}, since this is the relevant
quantity in the integrand of the quadrupole gravitational wave
formula. Although the larger $ \rho r^2 $ of the model with the LS EoS
gives a larger quadrupole contribution out to the crossing radius, in
most models the larger $ \rho r^2 $ in the outer parts of the core in
the variant with the Shen EoS more than compensates this and
ultimately leads to a larger integral quadrupole moment and, thus, to
a stronger gravitational wave burst. We note that
in~\cite{dimmelmeier_02_a}, \emph{all} models whose collapse type did
not change exhibited lower peak waveform amplitudes
($ \Delta |h|_\mathrm{max,rel} < 0 $) when going from Newtonian to
general relativistic gravity. In contrast, going from the relatively
stiff Shen EoS to the softer LS EoS results in a less clear trend with
a few models exhibiting $ \Delta |h|_\mathrm{max,rel} > 0 $. This
suggests a less dramatic impact of a change from the Shen EoS to the
LS EoS compared to altering the description of gravity.

For the small set of our models with
$ \Delta |h|_\mathrm{max,rel} > 0 $ we are neither able to identify
any obvious and systematic correlation with model parameters nor do we
find any clear systematics of $ \Delta |h|_\mathrm{max,rel} > 0 $ with
$ \Delta \rho_\mathrm{max,b,rel} $. It appears that the sign and
magnitude of $ \Delta |h|_\mathrm{max,rel}$ depends sensitively and in
a complicated way on the details of the collapse dynamics in each
individual model. Hence, we can explain only why for a \emph{specific}
model differences in the density structure at bounce between the model
variants with the Shen EoS and the LS EoS lead to an observed
$ \Delta |h|_\mathrm{max,rel} $, but cannot predict
$ \Delta |h|_\mathrm{max,rel} $ based on precollapse model parameters.


\subsection{Frequency spectrum of the waveform and variation with the
  equation of state}
\label{subsection:waveform_frequency}

\begin{figure}[t]
  \epsfxsize = 8.6 cm
  \centerline{\epsfbox{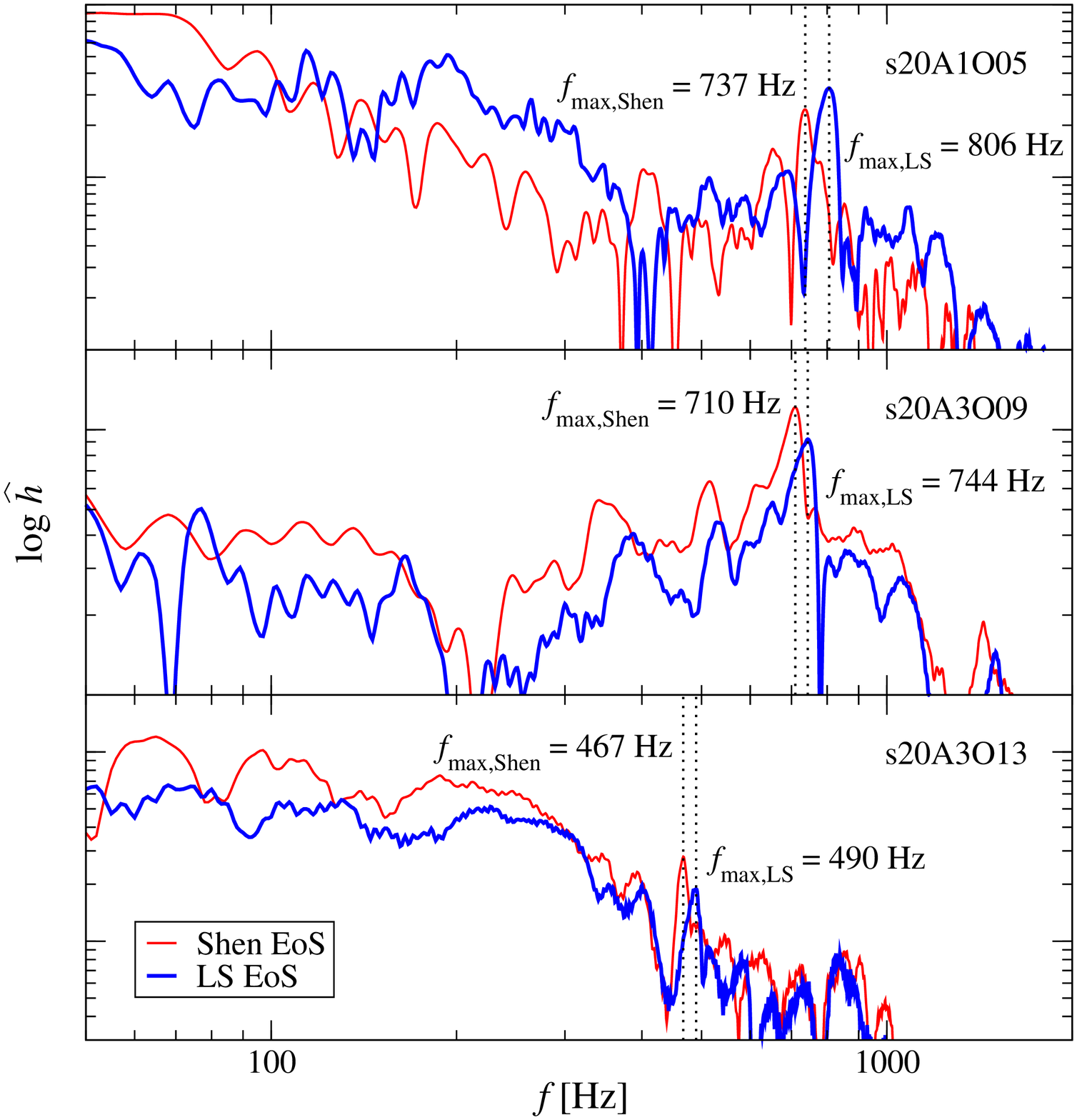}}
  \caption{Spectrum of the gravitational radiation waveform for model
    s20A1O05 (top panel), model s20A2O09 (center panel), and model
    s20A3O13 (bottom panel) using the Shen EoS (red line) and LS EoS
    (blue line). $ \hat{h} $ is the Fourier transform in frequency
    space of the waveform amplitude $ h $. The dotted lines mark the
    frequency $ f_\mathrm{max} $ at the maximum of the waveform
    spectrum, neglecting low-frequency contributions. The scale of the
    vertical axis is one order of magnitude per major tick mark.}
  \label{figure:waveform_spectrum}
\end{figure}

In contrast to the somewhat ambiguous impact of the EoS on the peak
waveform amplitude, the effect of replacing the Shen EoS with the LS
EoS on the waveform peak frequency is unequivocal for models
undergoing a pressure-dominated bounce. The increase in the maximum
density at bounce in the models with the LS EoS always results in a
shift of the main peak in the waveform spectrum to higher frequencies.
In the center panel of Fig.~\ref{figure:waveform_spectrum}, we plot
the waveform spectrum (i.e., the Fourier transform $ \hat{h} $ of
$ h $) for model s20A3O09 as a representative pressure-dominated
bounce model. The  spectrum of this model exhibits a distinct and
narrow high-frequency peak at
$ f_\mathrm{max,Shen} = 710 \mathrm{\ Hz} $ when using the Shen EoS,
while the calculation of the same model with the LS EoS results in a
peak at $ f_\mathrm{max,LS} = 744 \mathrm{\ Hz} $. Thus, for this
particular model, the change in EoS shifts the frequency associated
with the bounce peak by
$ \Delta f_\mathrm{max} = + 34 \mathrm{\ Hz} $. We observe similar
values for $ \Delta f_\mathrm{max} $ in all models undergoing
pressure-dominated bounce.

At frequencies below about $ 200 \mathrm{\ Hz} $, the waveform
spectrum of s20A3O09 exhibits a plateau which is due to the
low-frequency contribution from prompt large-scale postbounce
convection. Such a contribution is present in many models with slow to
moderate rotation, but gradually decreases in magnitude and relevance
with increasing rotation. As pointed out in
Section~\ref{subsection:generic_waveform}, our present numerical
scheme has the tendency to overestimate prompt postbounce convection
compared to full radiation-hydrodynamics calculations.

The waveform of the slowly spinning model s20A1O05, whose spectrum
is shown in the top panel of Fig.~\ref{figure:waveform_spectrum},
is dominated by such prompt postbounce convective motions.
Accordingly, for this model, there is a strong contribution to the
spectrum at low frequencies, even exceeding the still clearly
discernible bounce peak at high frequencies. Nevertheless, also in
this case the shift of the high-frequency bounce peak when replacing
the Shen EoS by the LS EoS is obvious and obeys the systematics
discussed above.

With increasing rotation, centrifugal forces become more relevant and
slow down the late phase of collapse and bounce. As a consequence,
$ f_\mathrm{max} $ always retreats to lower frequencies. This is
apparent in the spectrum of the centrifugal bounce model s20A3O13
shown in the bottom panel of Fig.~\ref{figure:waveform_spectrum}. For
this model, one can still identify the high-frequency bounce peak, but
now at significantly lower frequencies around
$ 400\mbox{\,--\,}500 \mathrm{\ Hz} $. Note that the low-frequency
quasi-continuous part of the spectrum in centrifugally bouncing models
such as s20A3O13 is due to rotationally slowed dynamics and stronger
postbounce oscillations, and should not be confused with the
low-frequency contribution from prompt convection in slowly rotating
models.

\begin{figure}[t]
  \epsfxsize = 8.6 cm
  \centerline{\epsfbox{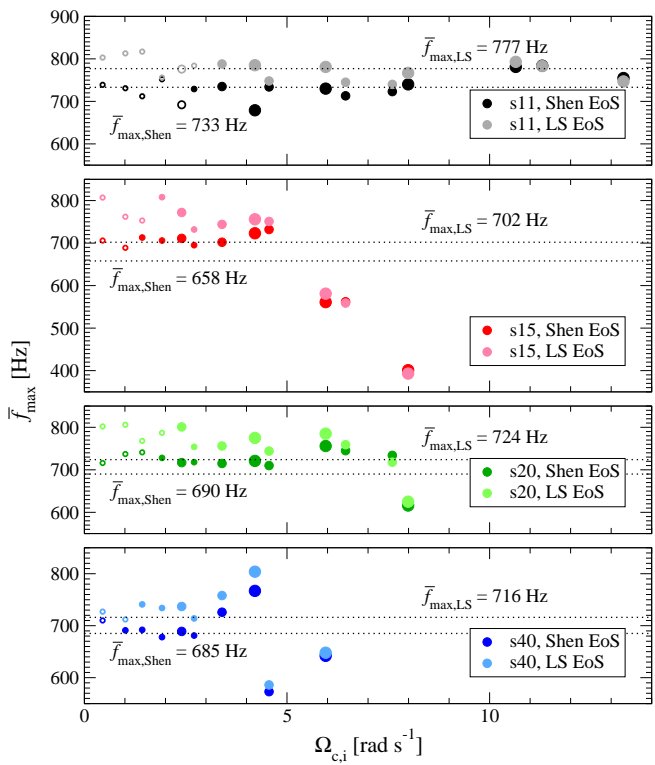}}
  \caption{Frequency $ f_\mathrm{max} $ at the maximum of the waveform
    spectrum for all models with a given progenitor mass versus the
    precollapse central angular velocity $ \Omega_\mathrm{c,i}
    $. Only models which undergo pressure-dominated bounce are
    shown. The dotted lines mark the average $ \bar{f}_\mathrm{max} $
    when using the Shen EoS or LS EoS. The progenitor mass, the EoS,
    the precollapse differential rotation parameter $ A $, and the
    collapse dynamics are encoded as in
    Fig.~\ref{figure:collapse_dynamics}.}
  \label{figure:f_max}
\end{figure}

In Fig.~\ref{figure:f_max}, we plot $ f_\mathrm{max} $ for all models
that undergo pressure-dominated bounce and thus exhibit a clearly
visible high-frequency peak in their spectra that can be associated
with the gravitational wave burst from core bounce. For all models the
systematic increase of $ f_\mathrm{max} $ when changing from the Shen
EoS to the LS EoS is apparent, and only for very few rapidly rotating
models close to the threshold to centrifugal bounce the change of
$ f_\mathrm{max} $ becomes small. In
Table~\ref{table:maximum_frequencies} we summarize the arithmetic mean
$ \bar{f}_\mathrm{max} $ along with the respective absolute and
relative differences between models using the Shen EoS and the LS
EoS. Note that when computing $ f_\mathrm{max} $ we neglect the
contribution below a cut-off frequency
$ f_\mathrm{cut} = 250 \mathrm{\ Hz} $ in order to exclude any
influence from the possibly unphysically strong and prolonged early
postbounce convection. 

In previous work~\cite{dimmelmeier_07_a}, Dimmelmeier et al.\
discussed the detection prospects for the gravitational wave burst
emitted in rotating core collapse models based on the s20 progenitor
and using the Shen EoS. To this end, they simulated a large set of
models with varying precollapse rotation rates $ \beta_\mathrm{i} $ in
the range from $ 0.05\% $ to $ 4\% $, approximately logarithmically
spaced in 18 steps for each of the three rotation profiles A1, A2, and
A3. For the current work, we have repeated the calculations of this
model set (which is extended in terms of precollapse rotation compared to
our standard models stated in Table~\ref{table:initial_models}, but
limited to one progenitor), this time with the LS EoS. While the
models with the Shen EoS that undergo pressure-dominated bounce have
an arithmetic mean peak frequency $ \bar{f}_\mathrm{max,Shen} \sim
718 \mathrm{\ Hz}$~\cite{dimmelmeier_07_a}, we find
$ \bar{f}_\mathrm{max,LS} \sim 758 \mathrm{\ Hz} $ when using the
the LS EoS. Thus for this particular model set the average relative
frequency shift amounts to
$ \Delta \bar{f}_\mathrm{max,rel} \sim 5.6 \% $. Both the average peak
frequencies and their change with EoS are consistent with what we find
for our standard model set using the four different progenitors and a
more restricted variety of precollapse rotation rates.

\begin{table}[t]
  \caption{Average $ \bar{f}_\mathrm{max} $ of the frequency at the
    maximum of the waveform spectrum for all models with a given
    progenitor mass. $ \Delta \bar{f}_\mathrm{max} $ and
    $ \Delta \bar{f}_\mathrm{max,rel} $ are the absolute and relative
    change of the frequency average, respectively, when changing from
    the Shen EoS to the LS EoS.}
  \label{table:maximum_frequencies}
  \begin{ruledtabular}
    \begin{tabular}{lcccc}
      Collapse &
      $ \bar{f}_\mathrm{max,Shen} $ &
      $ \bar{f}_\mathrm{max,LS} $ &
      $ \Delta \bar{f}_\mathrm{max} $ &
      $ \Delta \bar{f}_\mathrm{max,rel} $ \\
      model set &
      [Hz] &
      [Hz] &
      [Hz] &
      [\%] \\ [0.2 em]
      \hline \rule{0 em}{1.2 em}%
      s11 & $ 733 $ & $ 777 $ & $ 44 $ & $ 6.0 $ \\
      s15 & $ 658 $ & $ 702 $ & $ 44 $ & $ 6.7 $ \\
      s20 & $ 690 $ & $ 724 $ & $ 34 $ & $ 4.9 $ \\
      s40 & $ 685 $ & $ 716 $ & $ 31 $ & $ 4.5 $ \\
    \end{tabular}
  \end{ruledtabular}
\end{table}


\section{Detection prospects for the gravitational wave burst signal}
\label{section:detection_prospects}

\begin{figure}[t]
  \epsfxsize = 8.6 cm
  \centerline{\epsfbox{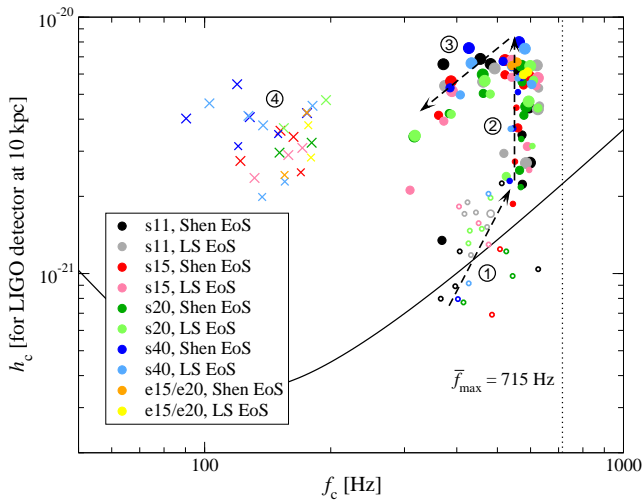}}
  \caption{Location of the gravitational wave burst signals from core
    bounce for all models (including the e15/e20 models) in the
    $ h_\mathrm{c} $--$ f_\mathrm{c} $ plane relative to the
    sensitivity curves of the LIGO, assuming at a distance of
    $ 10 \mathrm{\ kpc} $. The meaning of the arrows 1, 2, and 3 as
    well as area 4 are explained in the main text. The dotted line
    marks the average $ \bar{f}_\mathrm{max} $ of the frequency at the
    maximum of the waveform spectrum. The progenitor model, the EoS,
    the initial rotation parameter $ A $, and the collapse dynamics
    are encoded as in Fig.~\ref{figure:collapse_dynamics}.}
  \label{figure:sensitivity_ligo_all}
\end{figure}

In order to assess the detectability of the burst signal from core
bounce, we compute the (detector-dependent) frequency-integrated
characteristic signal frequency $ f_\mathrm{c} $ and dimensionless
characteristic gravitational wave amplitude $ h_\mathrm{c} $ using
Eq.~(\ref{eq:characteristic_frequency}) and
Eq.~(\ref{eq:characteristic_amplitude}), respectively. We again exclude
frequencies below $ 250 \mathrm{\ Hz} $ from the integrals in an
attempt to filter out dominant contributions from prompt postbounce
convection in slowly rotating models. In
Fig.~\ref{figure:sensitivity_ligo_all}, we plot $ h_\mathrm{c} $
against $ f_\mathrm{c} $ for the current LIGO
detector~\cite{shoemaker_06_a} at a distance of $ 10 \mathrm{\ kpc} $.
For comparison with the detector sensitivity, we include its rms
strain sensitivity curve. Note that the total energy emitted in
gravitational waves ranges from
$ E_\mathrm{gw} \sim 3.5 \times 10^{-10} $ to $ 5.3 \times 10^{-8} $
in units of $ M_\odot c^2 $ (including the contribution from
convection) for our standard models.

The distribution of our standard set of models (as listed in
Table~\ref{table:initial_models}) in the
$ h_\mathrm{c} $--$ f_\mathrm{c} $ plane of
Fig.~\ref{figure:sensitivity_ligo_all} obeys straightforward
systematics. The clustering in frequency of the large number of models
undergoing a pressure-dominated bounce (marked by circles in
Fig.~\ref{figure:sensitivity_ligo_all}) is obvious. Very slowly
rotating models, whose waveforms are dominated by the imprint of
prompt postbounce convection (unfilled circles), exhibit the lowest
values for $ h_\mathrm{c} $, which increases with faster rotation
(along arrow 1), reflecting that the inner core at bounce becomes more
massive (cf.\ Sections~\ref{subsection:waveform_peak_amplitude}
and~\ref{subsection:eos_and_progenitor_influence}). Despite the
frequency cut at $ 250 \mathrm{\ Hz} $ in the integral for
$ f_\mathrm{c} $, the low-frequency contribution from convection in
the spectrum leads to an $ f_\mathrm{c} $ that is lower than the value
obtained for more rapidly rotating models without significant
postbounce convection (filled circles). For the latter model class,
$ h_\mathrm{c} $ simply grows with increasing precollapse rotation
(along arrow 2), now at practically constant $ f_\mathrm{c} $. Even
for these models, $ f_\mathrm{c} $ is always lower than the average
peak frequency $ \bar{f}_\mathrm{max} $ of their waveform spectra,
which amounts to $ 715 \mathrm{\ Hz} $ for the 108 models of our
standard model set (including the e15/e20 models) which exhibit a
pressure-dominated bounce. This is a consequence of the detector
characteristics, whose maximum sensitivity is at much lower
frequencies between $ 100 $ and $ 200 \mathrm{\ Hz} $ and thus
accordingly lowers $ f_\mathrm{c} $ in comparison with a fiducial flat
sensitivity curve.

For rapid rotation, the influence of centrifugal forces on the
collapse dynamics manifests itself as a centrifugal barrier that
limits the characteristic amplitude $ h_\mathrm{c} $ (see also the
discussion in Section~\ref{subsection:rotational_barrier} and
Fig.~\ref{figure:h_max_versus_omega_cap_c_ini}). Simultaneously,
the characteristic frequency $ f_\mathrm{c} $ moves to increasingly
lower values as faster rotation slows down the collapse (along
arrow 3). Models that rotate so rapidly that they undergo a purely
centrifugal bounce (marked by cross symbols in
Fig.~\ref{figure:sensitivity_ligo_all}) constitute a practically
separate class (area 4) in the $ h_\mathrm{c} $--$ f_\mathrm{c} $
diagram somewhat below the maximum value of the amplitude
$ h_\mathrm{c} $, but at considerably lower frequencies
$ f_\mathrm{c} $.

For very rapidly rotating models the imprint of centrifugal effects on
various waveform characteristics (such as $ f_\mathrm{max} $,
$ f_\mathrm{c} $, $ |h|_\mathrm{max} $, or $ h_\mathrm{c} $) is quite
pronounced and permits one to infer on the precollapse rotational
configuration in the case of a successful detection of gravitational
waves from a core collapse event. As already noted
in~\cite{dimmelmeier_07_a}, in the case of moderate or slow rotation,
which is the astrophysically most probable case~\cite{heger_05_a,
  ott_06_c}, the insensitivity of the waveform's frequency
characteristics to variations in the precollapse configuration
significantly obstructs the ``inversion problem'' of gravitational
wave detection, i.e., the constraining of physical parameters of the
precollapse core or of the nascent proto-neutron star from a detected
waveform, leaving only the (e.g., maximum or integrated
characteristic) amplitude as an indicator of the rotational
configuration. In addition, Fig.~\ref{figure:f_max} also implies that
it will be very hard, if not impossible, to constrain other possibly
unknown model parameters aside from rotation (such as EoS or
progenitor mass) from the gravitational waveform of the burst signal
from core bounce alone, since their effect on the burst waveform is
small and no clear trends or systematics are discernible, which adds
to the degeneracy of the inversion problem.

\begin{figure}[t]
  \epsfxsize = 8.6 cm
  \centerline{\epsfbox{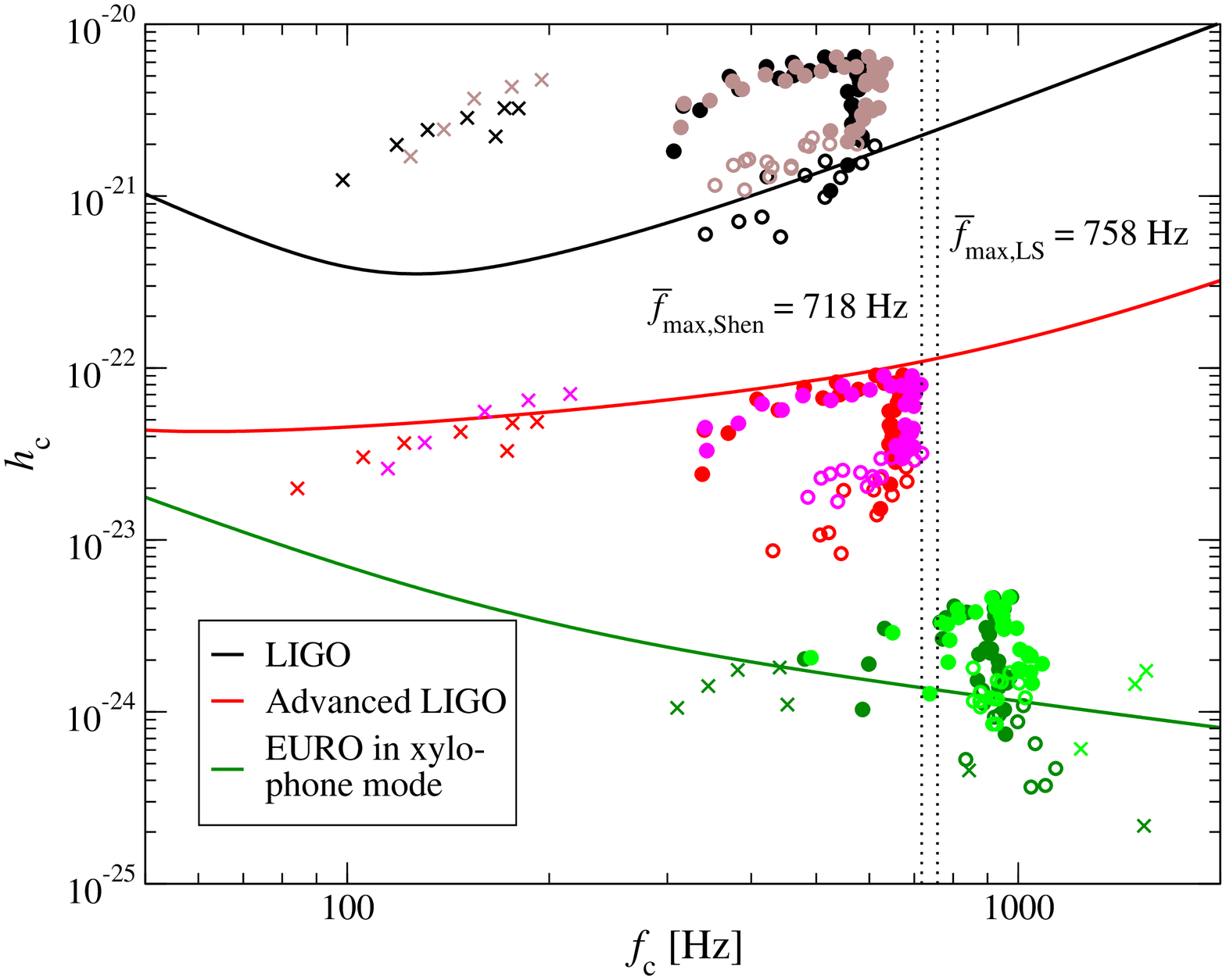}}
  \caption{Location of the gravitational wave burst signals from core
    bounce in the $ h_\mathrm{c} $--$ f_\mathrm{c} $ plane relative to
    the sensitivity curves of various interferometer detectors (as
    color-coded) for an extended set of models with the progenitor s20
    using the Shen EoS (dark hues) or LS EoS (light hues). The sources
    are at a distance of $ 10 \mathrm{\ kpc} $ for LIGO,
    $ 0.8 \mathrm{\ Mpc} $ for Advanced LIGO, and
    $ 15 \mathrm{\ Mpc} $ for EURO. The dotted lines mark the average
    $ \bar{f}_\mathrm{max} $ of the frequency at the maximum of the
    waveform spectrum for the models when using the Shen EoS or LS
    EoS. Only the EoS and the collapse dynamics are encoded as in
    Fig.~\ref{figure:collapse_dynamics}, but not the precollapse
    differential rotation parameter $ A $.}
  \label{figure:sensitivity_eos}
\end{figure}

As an example, we again single out the impact of the EoS on the
waveform frequency while keeping the progenitor model s20 fixed. For
the particular, extended set of models with many precollapse rotation
rates already discussed in Section~\ref{subsection:waveform_frequency},
we show in Fig.~\ref{figure:sensitivity_eos} the location of the
waveform signals in the $ h_\mathrm{c} $--$ f_\mathrm{c} $ plane
for initial LIGO at a distance of $ 10 \mathrm{\ kpc} $, Advanced
LIGO in broadband tuning~\cite{shoemaker_06_a} at a distance of
$ 0.8 \mathrm{\ Mpc} $, and the projected EURO detector in xylophone
mode~\cite{euro_detector} at a distance of $ 15 \mathrm{\ Mpc} $ (cf.\
Fig.~4 in~\cite{dimmelmeier_07_a}). All 54 s20 models
of~\cite{dimmelmeier_07_a} using the Shen EoS along with the newly
computed corresponding models with the LS EoS are shown.

It is obvious that the spread within the group of models with either
the Shen EoS or the LS EoS is larger than the variation due to a
change in the EoS, since the effect of the EoS on the characteristic
signal frequency $ f_\mathrm{c} $ is small (comparable to
$ \Delta \bar{f}_\mathrm{max,rel} $, corresponding to a change of a few
percent). The two EoSs considered here bracket the range from rather
soft (LS EoS) to rather stiff (Shen EoS), and therefore it is unlikely
that employing a larger variety of nonzero-temperature nuclear EoSs
would lead to any more optimistic conclusions.

Based on the relative positions of the models with respect to the
individual detector sensitivities, from
Fig.~\ref{figure:sensitivity_eos} we conclude (in agreement with
previous work~\cite{dimmelmeier_07_a, ott_07_a}) that
initial-LIGO-class detectors are sensitive only to signals coming from
an event in the Milky Way, while Advanced-LIGO-class observatories
could marginally detect events from other galaxies in the Local Group
(e.g., M31 Andromeda at $ \sim 0.8 \mathrm{\ Mpc} $ distance). For the
proposed EURO detector in xylophone mode, we expect a very high
signal-to-noise ratio ($ h_\mathrm{c} $ divided by the detector
sensitivity at $ f_\mathrm{c} $). This detector could also observe
many of the computed signals at a distance of $ 15 \mathrm{\ Mpc} $,
i.e., in the Virgo cluster, for which one expects a favorably high
event rate.


\section{Rotation of the proto-neutron star}
\label{section:rotation_rate}

The calculations presented in this study impose axisymmetry, hence are
unable to track the development of rotationally-induced nonaxisymmetric
structures and dynamics. Nevertheless, we can utilize the results from
our simulations to assess the possibility of rotational
triaxial instabilities during the collapse and early postbounce
phase. In this way we can (i) test the reliability of our present
restriction to axisymmetry and (ii) put constraints on the relevance
of the various types of such instabilities in a core-collapse event.

Nonaxisymmetric rotational instabilities in proto-neutron stars have
long been proposed as strong and potentially long-lasting sources of
gravitational waves. In principle, the gravitational wave emission by
a nonaxisymmetrically deformed proto-neutron star after bounce could
easily exceed (see, e.g., \cite{ott_07_a, ott_06_b}) in total emitted
energy (and, hence, in characteristic strain $ h_\mathrm{c} $) the
gravitational wave burst from core bounce on which this paper is
focussed.

In the context of classical Newtonian theory of fluid equilibria (see,
e.g., \cite{chandrasekhar_69_a}), MacLaurin spheroids (i.e.,
axisymmetric, rigidly rotating, equilibrium configurations of uniform
density) become unstable to nonaxisymmetric deformation when a
nonaxisymmetric configuration with lower total energy exists at a
given rotation rate $ \beta $. MacLaurin spheroids become
\emph{dynamically} unstable to deformation into Riemann ellipsoids at
$ \beta \gtrsim \beta_\mathrm{dyn} = 27\% $. At
$ \beta \gtrsim \beta_\mathrm{sec} = 14\% $ they become
\emph{secularly} unstable to triaxial ellipsoidal deformation in the
presence of dissipative processes (Jacobi ellipsoids via gravitational
wave back-reaction known as the Chandrasekhar--Friedman--Schutz (CFS)
instability~\cite{chandrasekhar_70_a, friedman_78_a}, or Dedekind
ellipsoids via viscous processes). In both the dynamical and the
secular case, the lowest-order deformation in terms of azimuthal
nonaxisymmetric modes proportional to $ \exp (i m \varphi) $ is the
$ m = 2 $ Kelvin (bar-) $ f $-mode, where $ \varphi $ is the azimuthal
angle and the mode order $ m $ is an integer.

Although Newtonian MacLaurin spheroids are highly idealized
configurations, numerical studies (see, e.g., \cite{baiotti_07_a} and
references therein) have shown that the above instability threshold
$ \beta_\mathrm{dyn} $ for the dynamical instability holds
approximately even when differentially rotating compressible fluid
configurations in general relativity are considered. The situation may
be different for the gravitational radiation back-reaction driven
secular instability, since perturbative studies (see, e.g.,
\cite{morsink_99_a}) predict an onset at significantly lower $\beta$
in general relativity than in the Newtonian case. However,
fully relativistic nonlinear hydrodynamic studies of the secular
instability remain yet to be carried out.

Recently, a new kind of dynamical rotational nonaxisymmetric
instability at a value of $ \beta $ much lower than the classical
threshold has been discovered both in numerical and perturbative
studies (see, e.g., \cite{centrella_01_a, shibata_03_b, watts_05_a,
ott_05_a, ou_06_a, saijo_06_a, ott_06_b, ott_07_a, zink_07_a,
cerda_07_b} and references therein). This low-$ \beta $ instability
(making the classical MacLaurin instability a ``high''-$ \beta $
instability) appears to amplify nonaxisymmetric modes at points where
their pattern speed $ \sigma_m $ (the eigenfrequency $ \omega_m $
divided by the azimuthal mode order $ m $) coincides with the local
angular velocity of the fluid~\cite{watts_05_a, ott_05_a, ou_06_a,
  saijo_06_a}.


\subsection{The rotational barrier in core collapse}
\label{subsection:rotational_barrier}

\begin{figure}[t]
  \epsfxsize = 8.6 cm
  \centerline{\epsfbox{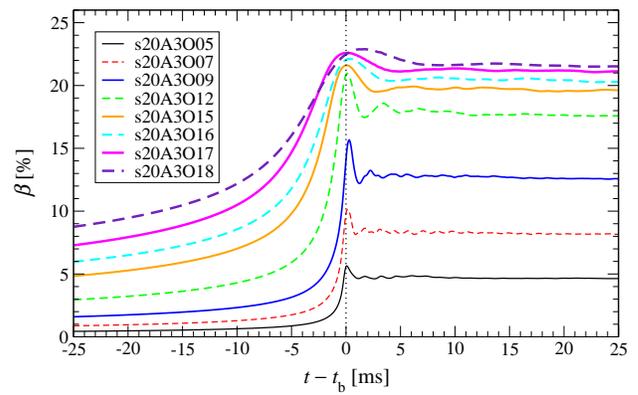}}
  \caption{Time evolution of the rotation rate $ \beta $ around the
    time of core bounce for various models of the s20 progenitor
    series computed with the Shen EoS at fixed precollapse degree $ A
    $ of differential rotation and varying the precollapse
    central angular velocity $ \Omega_\mathrm{c,i} $. Note that we
    have augmented this sequence by three extra models s20A3O16 to
    s20A3O18 (with $ \beta_\mathrm{i} = 3.00 $, $ 3.50 $, and $ 4.00
    $, respectively) not listed in Table~\ref{table:collapse_models}.}
  \label{figure:rotation_rate_evolution}
\end{figure}

From first principles one can derive that the conservation of angular
momentum during the collapse phase results in an increase of the
angular velocity $ \Omega $ of a representative Lagrangian mass
element proportional to $ \varpi^{-2} $, where
$ \varpi = r \, \sin \theta $ is the distance from the rotation
axis. Setting for simplicity $ \varpi $ equal to the spherical radial
coordinate $ r $ (which, of course, only holds in the equatorial
plane), this translates into a scaling of the centrifugal force
proportional to $ r^{-3} $. The gravitational force, on the other
hand, increases only like $ r^{-2} $. Hence, even in this simple
Keplerian picture, one may expect a dominance of the centrifugal force
over gravity at sufficiently small $ r $. In a more elaborate approach,
employing sequences of Newtonian self-gravitating equilibrium
spheroids, Tohline~\cite{tohline_84_a} demonstrated that such a
\emph{rotational barrier} at which the collapsing core becomes
centrifugally stabilized indeed exists in the context of stellar core
collapse. This rotational barrier marks the hard upper limit for the
contraction of the inner core, hence also puts an upper limit
$ \beta_\mathrm{rb} $ on the rotation rate that can be reached when
varying $ \Omega_\mathrm{c,i} $ for a given combination of precollapse
degree of differential rotation and progenitor structure.

Tohline's qualitative conclusions have been confirmed by multiple
numerical studies of rotating collapse (see, e.g.,
\cite{moenchmeyer_91_a, zwerger_97_a, dimmelmeier_02_a, ott_04_a,
ott_06_b} and our present work) while the quantitative results, in
particular the analytic critical rotation rate for centrifugal
stabilization of collapse, do not hold for a dynamical collapse
situation and must be determined via nonlinear hydrodynamic
simulations~\cite{dimmelmeier_07_a}. 

\begin{figure}[t]
  \epsfxsize = 8.6 cm
  \centerline{\epsfbox{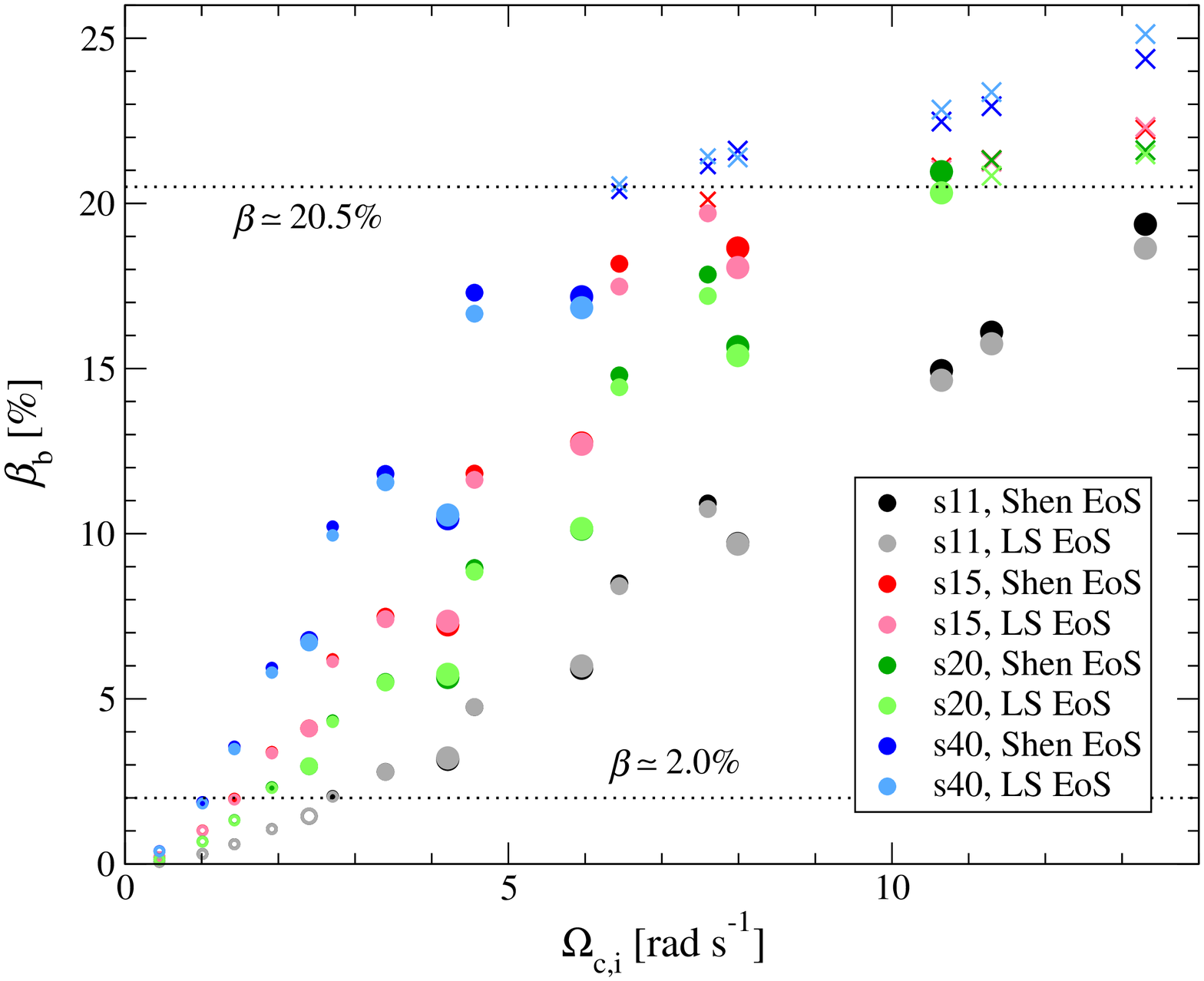}}
  \caption{Rotation rate $ \beta_\mathrm{b} $ at the time of bounce
    for all models versus the precollapse central angular
    velocity $ \Omega_\mathrm{c,i} $. The progenitor mass, the EoS,
    the precollapse differential rotation parameter $ A $, and the
    collapse dynamics are encoded as in
    Fig.~\ref{figure:collapse_dynamics}. The lower horizontal line
    approximately separates pressure-dominated bounce models with and
    without strong prompt postbounce convection, while upper
    horizontal line marks the approximate transition between
    pressure-dominated bounce and centrifugal bounce.}
  \label{figure:beta_b_versus_omega_cap_c_ini}
\end{figure}

In Fig.~\ref{figure:rotation_rate_evolution} we plot the time
evolution of the rotation rate $ \beta $ for a sequence of rotating
collapse models with increasing precollapse central angular velocity
$ \Omega_\mathrm{c,i} $ while all other model parameters are kept
fixed. All models reach their maximum rotation rate
$ \beta_\mathrm{max} $ close to the time of core bounce, hence
$ \beta_\mathrm{max} \simeq \beta_\mathrm{b} $. After bounce, the
inner core re-expands and settles into a new quasi-equilibrium
configuration with $ \beta_\mathrm{pb} < \beta_\mathrm{b} $. Slowly to
moderately rapidly rotating models experience little rotational support,
and in those cases $ \beta_\mathrm{b} $ increases roughly linearly
with $ \Omega_\mathrm{c,i} $ (see also
Table~\ref{table:collapse_models}). For higher values of
$ \Omega_\mathrm{c,i} $, centrifugal forces become relevant and
$ \beta_\mathrm{b} $ saturates at $ \beta_\mathrm{rb} $ as the models
start to bounce centrifugally. For the s20A3 sequence with the Shen
EoS considered here we determine $ \beta_\mathrm{rb} $ to be
$ \sim 23\% $.

\begin{figure}[t]
  \epsfxsize = 8.6 cm
  \centerline{\epsfbox{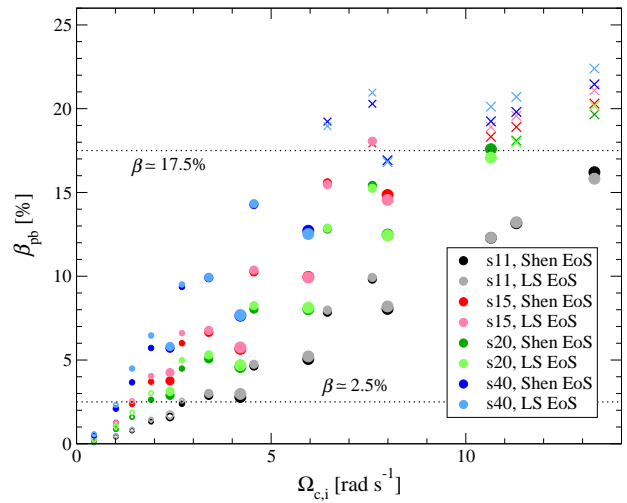}}
  \caption{Rotation rate $ \beta_\mathrm{pb} $ in the late postbounce
    phase for all models versus the precollapse central
    angular velocity $ \Omega_\mathrm{c,i} $. The progenitor mass, the
    EoS, the precollapse differential rotation parameter $ A $, and
    the collapse dynamics are encoded as in
    Fig.~\ref{figure:collapse_dynamics}. As in
    Fig.~\ref{figure:beta_b_versus_omega_cap_c_ini}, the horizontal
    lines again approximately mark the boundaries between different
    bounce dynamics.}
  \label{figure:beta_f_versus_omega_cap_c_ini}
\end{figure}

Figs.~\ref{figure:beta_b_versus_omega_cap_c_ini}
and~\ref{figure:beta_f_versus_omega_cap_c_ini} provide an overview of
the dependences of $ \beta_\mathrm{b} $ and $ \beta_\mathrm{pb} $,
respectively, on $ \Omega_\mathrm{c,i} $ for our entire model set as
listed in Table~\ref{table:collapse_models}. Models that start out in
essentially solid-body rotation (A1) never reach a $ \beta_\mathrm{b}
$ in excess of $ \sim 10\% $ (with the maximum obtained in model
s40A1O13). With increasing $ \Omega_\mathrm{c,i} $ such
rigidly-rotating cores become eventually fully centrifugally supported
already at the onset of collapse and do not collapse at all.
Differentially rotating models may have higher values of
$ \Omega_\mathrm{c,i} $ and thus a more rapidly rotating center, while
the core is still allowed to collapse. As the collapse proceeds,
electron capture reduces the pressure support and the size of the
homologously collapsing inner core stays sufficiently small that
centrifugal forces can become dynamically relevant only in the final
phase of collapse (see the discussion in
Section~\ref{subsection:gr_and_deleptonization_influence}). Thus, for
our model set, the most differentially rotating configuration A3 leads
to the highest values for $ \beta_\mathrm{b} $ and
$ \beta_\mathrm{pb} $. A centrifugal bounce near the rotational
barrier occurs only in a small subset of very rapidly
($ \Omega_\mathrm{c,i} \gtrsim 6.5 \mathrm{\ rad\ s}^{-1} $) and
differentially (A2/A3) rotating models, generally at
$ \beta_\mathrm{b} \gtrsim 20.5\% $.

At a fixed precollapse degree of differential rotation and
$ \Omega_\mathrm{c,i} $, $ \beta_\mathrm{b} $ and
$ \beta_\mathrm{pb} $ increase with a more massive and radially
extended progenitor iron core (cf.\ Table~\ref{table:progenitor_cores}).
This is analogous to the systematics found for the rotational
enhancement of the inner core mass $ M_\mathrm{ic,b} $ at bounce (see
Fig.~\ref{figure:m_rest_inner_core}).

The dependence of both $ \beta_\mathrm{b} $ and $ \beta_\mathrm{pb} $
on the EoS is small and shows little systematic trend. The Shen EoS,
on the one hand, systematically yields a more massive and more extended
inner core that bounces with more dynamically relevant angular
momentum than one obtained with the LS EoS. The LS EoS, on the other
hand, leads to more compact configurations, which provide for stronger
centrifugal spin-up in the final phase of collapse. The competition
between these two effects results in the nonsystematic difference
between the two EoSs seen in
Figs.~\ref{figure:beta_b_versus_omega_cap_c_ini}
and~\ref{figure:beta_f_versus_omega_cap_c_ini}.

\begin{figure}[t]
  \epsfxsize = 8.6 cm
  \centerline{\epsfbox{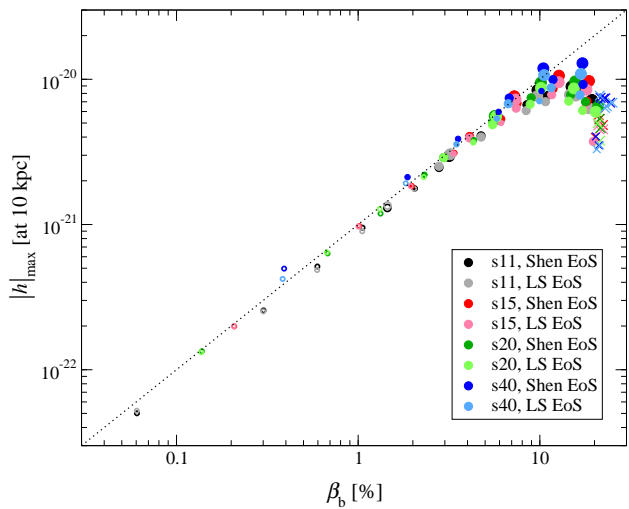}}
  \caption{Peak value $ |h|_\mathrm{max} $ of the gravitational wave
    amplitude at $ 10 \mathrm{\ kpc} $ distance for the burst signal
    (neglecting possibly larger contributions from postbounce
    convection at later times) for all models versus the rotation rate
    $ \beta_\mathrm{b} $ at the time of bounce. At slow to moderately
    rapid rotation, $ |h|_\mathrm{max} $ is proportional to
    $ \beta_\mathrm{b} $ to high accuracy (as marked by the dotted
    line with a slope of $ 1 $ in the log-log plot), while for
    $ \beta_\mathrm{b} \gtrsim 10\% $ centrifugal effects reduce
    $ |h|_\mathrm{max} $.}
  \label{figure:beta_b_and_h_max}
\end{figure}

The centrifugal barrier is also evident in
Fig.~\ref{figure:beta_b_and_h_max}, where we plot the dependence of
the peak value $ |h|_\mathrm{max} $ of the gravitational wave burst
against the rotation rate $ \beta_\mathrm{b} $ at bounce. It is
noteworthy that centrifugal effects are responsible for an upper limit
in $ |h|_\mathrm{max} $ even before the maximum rotation rate
$ \beta_\mathrm{b} \sim 25\% $ is
reached, which reflects the observation that the highest values of
$ |h|_\mathrm{max} \sim 10^{20} $ at $ 10 \mathrm{\ kpc} $ distance
are obtained for models which still undergo a pressure-dominated
bounce, albeit at rapid rotation with $ \beta_\mathrm{b} \sim 10\% $.
Below these rotation rates, $ |h|_\mathrm{max} $ scales linearly with
$ \beta_\mathrm{b} $ with remarkable precision, which is important
information for the inversion problem in the case of a detection.
We find a similar linear dependence of $ |h|_\mathrm{max} $ on the
postbounce rotation rate $ \beta_\mathrm{pb} $. In that case, however,
the linear correlation is not as precise for low rotation rates (as
$ \beta_\mathrm{pb} $ is rather sensitive to angular momentum
redistribution due to convection after core bounce) and, in addition,
the scaling becomes approximately quadratic well before
$ |h|_\mathrm{max} $ reaches its upper limit.

\subsection{\boldmath The prospects for dynamical high-$ \beta $
  instability in iron core collapse}
\label{subsection:high_beta_instability}

We find that none of our models surpass the threshold rotation rate
$ \beta_\mathrm{dyn} $ for the classical dynamical instability (see
Table~\ref{table:collapse_models}). The overall largest $ \beta $ of
$ \sim 25\% $ is reached by model s40A3O15, which has the most massive
and extended progenitor iron core (see
Table~\ref{table:progenitor_cores}) in combination with the strongest
precollapse degree of differential rotation and highest precollapse
central angular velocity considered in this study. This value of
$ \beta_\mathrm{b} \sim 25\% $ comes close to the numerically obtained
instability threshold of $ \beta_\mathrm{dyn} \gtrsim 25.5\% $
reported in~\cite{baiotti_07_a}, but is maintained only for a very
short time, since the core rebounds and settles at a more expanded
quasi-equilibrium state after bounce. Accordingly, its postbounce
rotation rate $ \beta_\mathrm{pb} $ is $ \sim 22\% $, and thus this
model is unlikely to become subject to a dynamical high-$ \beta $
bar-mode instability. As portrayed by
Fig.~\ref{figure:beta_f_versus_omega_cap_c_ini}, the models with less
extreme precollapse conditions in general reach a
$ \beta_\mathrm{pb} $ significantly below $ \sim 20\% $.

Based on the results from our extensive set of simulations, we
consider it unlikely that a proto-neutron star in nature develops a
high-$ \beta $ dynamical instability at or early after core
bounce. On the other hand, during its cooling to the final cold and
condensed neutron star, the proto-neutron star contracts, and, if
angular momentum is conserved and not redistributed or shed by other
means (see, e.g., the discussion in~\cite{heger_05_a, ott_06_c}),
spins up on a timescale of seconds to minutes. While many of the
proto-neutron stars in our model calculations could theoretically
reach $ \beta_\mathrm{dyn} $, it is, however, more likely that the
secular instability driven by dissipation or gravitational radiation
back-reaction, which in proto-neutron stars has a growth timescale on
the order of $ 1 \mathrm{\ s} $~\cite{lai_01_a}, will set in first,
completely diminishing the chances for dynamical high-$ \beta $
instability even in the most rapidly rotating proto-neutron stars.

Finally, we point out that it is in principle possible to construct
precollapse conditions that lead to $ \beta_\mathrm{b} $ and
$ \beta_\mathrm{pb} $ above $ \beta_\mathrm{dyn} $. This may be
achieved by increasing significantly the precollapse degree of
differential rotation and $ \Omega_\mathrm{c,i} $ above the values
used in our most extreme models. However, such configurations
(including already the rotational setup A3 in our models) are very
unlikely to arise in evolution scenarios of single massive stars,
since stellar evolution proceeds sufficiently slowly for redistribution
of angular momentum towards solid-body rotation to occur on
nuclear-burning timescales~\cite{heger_00_a, heger_05_a,
  hirschi_04_a}.


\subsection{\boldmath Differential rotation in the proto-neutron star
  and its relevance for the low-$ \beta $ dynamical instability}
\label{subsection:low_beta_instability}

The low-$ \beta $ dynamical instability appears to develop exclusively
in differentially rotating fluid bodies and has been reported to occur
even for rotation rates as low as $ \sim 1\% $, provided the degree of
differential rotation is sufficiently large~\cite{shibata_03_b}.

The nature of the low-$ \beta $ instability remains to be determined
in detail, yet it has been suggested~\cite{watts_05_a} that it is a
type of dynamical shear instability that operates on the shear energy
stored in differential rotation and radially redistributes angular
momentum via the generation of an azimuthal (nonaxisymmetric, spiral)
structure that propagates outward in radius~\cite{ott_06_b,
saijo_06_a}. In this picture, nonaxisymmetric structure is generated
by transfer of rotational energy from the axisymmetric background
fluid to an azimuthal fluid mode at the location where the background
angular velocity matches the mode pattern speed (i.e., at the corotation
point). This proposed corotation mechanism suggests a close
relationship of the low-$ \beta $ instability observed in simulations
of stellar models with dynamical instabilities in disks such as those
described by Papaloizou and Pringle~\cite{papaloizou_85_a}.

\begin{figure}[t]
  \epsfxsize = 8.6 cm
  \centerline{\epsfbox{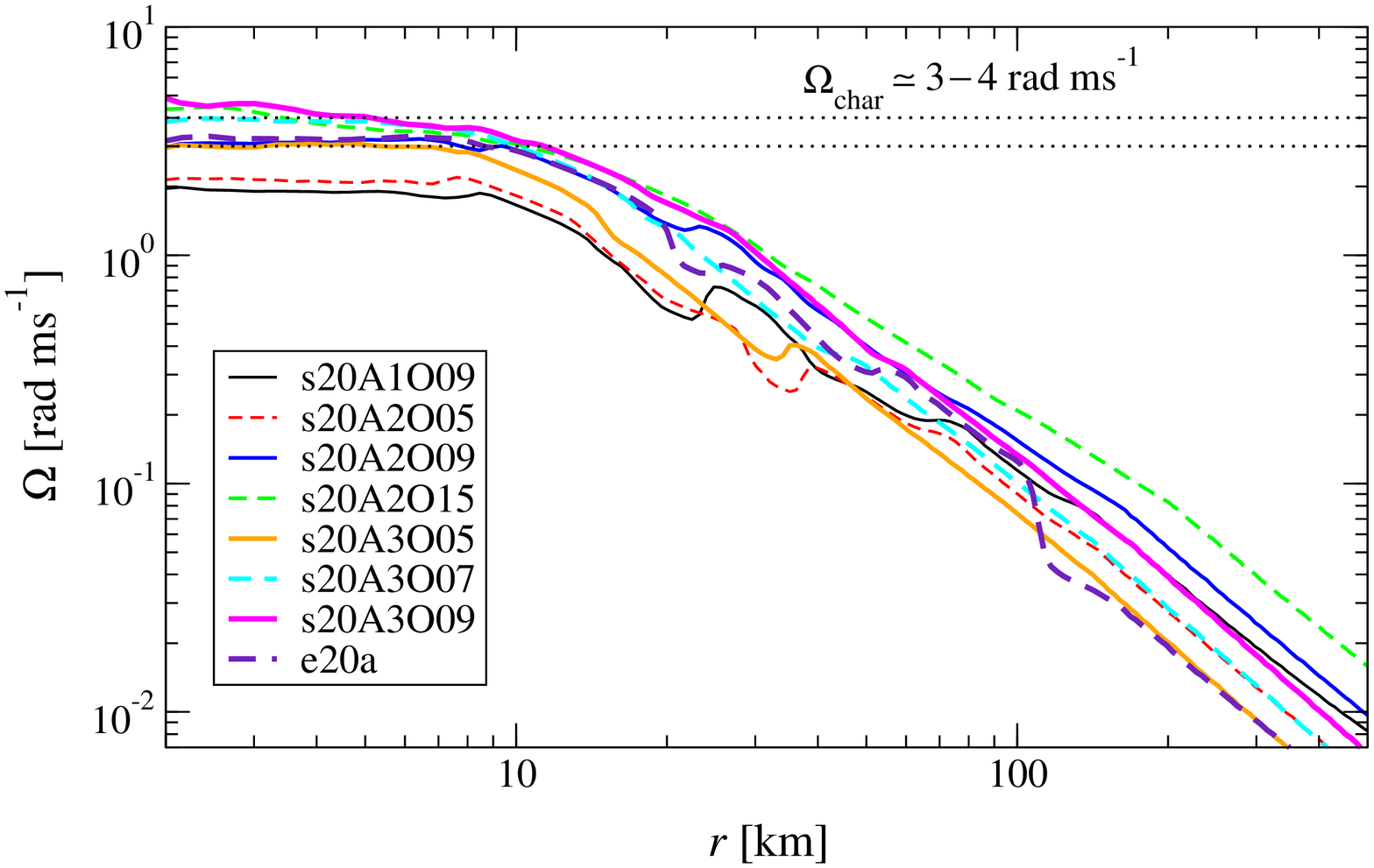}}
  \caption{Radial profile of the angular velocity $ \Omega $ in the
    equatorial plane at $ 20 \mathrm{\ ms} $ after the time of core
    bounce for a representative subset of the models listed in
    Table~\ref{table:collapse_models}. Note that the inner core is in
    approximate solid body rotation out to about $ 10 \mathrm{\ km} $
    while the outer parts of the proto-neutron star and the postshock
    region rotate strongly differentially. The dotted lines mark the
    approximate range for the characteristic angular frequency
    $ \Omega_\mathrm{char} $.}
  \label{figure:omega_r}
\end{figure}

The importance of differential rotation for the low-$ \beta $
instability in stars can now be understood by the combination of two
important factors: First, differential rotation provides the reservoir
of shear energy that can be tapped to generate the nonaxisymmetric
structure. Second, despite a relatively low global rotation rate
$ \beta $, differential rotation allows the \emph{central} regions of
a star to rotate sufficiently rapid to be in corotation with the
lowest-order unstable modes that have pattern speeds of
$ \mathcal{O} (2 \pi / \tau_\mathrm{dyn}) $, where
\begin{equation}
  \tau_\mathrm{dyn} \approx 2 \pi \sqrt{\frac{R^3}{GM}}
  \label{equation:tau_dyn}
\end{equation}
is the dynamical timescale of the rotating star set by the Keplerian
angular velocity~\cite{centrella_01_a, ott_05_a}.

Since solid-body rotation is the state of lowest rotational energy,
neutron stars are very likely to become rigidly rotating within at
most a few dissipative timescales during their post-supernova cooling
evolution. Significant differential rotation may be expected in early
merger remnants of binary neutron stars (e.g., \cite{rosswog_02_a})
and, importantly, is a consequence of rotating iron core collapse to a
proto-neutron star investigated in the present work.

In Fig.~\ref{figure:omega_r} we plot radial profiles of the angular
velocity $ \Omega $ in the equatorial plane at $ 20 \mathrm{\ ms} $
after core bounce for several of our models. As a result of
quasi-homologous contraction, the near uniform precollapse rotational
profile of the inner core is essentially frozen in during
collapse~\cite{ott_06_c}. In the outer core, however, the collapse
proceeds supersonically, resulting in differential rotation at
equatorial radii $ \gtrsim 10 \mathrm{\ km} $. In all models shown in
Fig.~\ref{figure:omega_r}, $ \Omega $ declines by about two orders of
magnitude in the radial interval from $ 10 $ to $ 200 \mathrm{\ km} $,
and roughly obeys a power-law with an exponent in the range of
$ - 1.2 $ to $ -1.4 $. Generally, a stronger degree of precollapse
differential rotation leads to a steeper radial decline of
$ \Omega $ after bounce. When increasing $ \Omega_\mathrm{c,i} $ while
keeping the degree of precollapse differential rotation fixed, the outer
core regions experience more centrifugal support during collapse,
resulting in a shallower postbounce slope for $ \Omega $ (cf.\ model
s20A2O15 in Fig.~\ref{figure:omega_r}).

In general, we find that the central angular velocity
$ \Omega_\mathrm{c} $ after bounce increases monotonically with the
precollapse value $ \Omega_\mathrm{c,i} $. For our models we obtain
values for $ \Omega_\mathrm{c} $ in the nascent proto-neutron star
between about $ 2 $ and $ 6 \mathrm{\ rad\ ms}^{-1} $, which
corresponds to central rotation periods of about $ 1 $ to
$ 3 \mathrm{\ ms} $. Assuming a mass range of the proto-neutron star
of $ \sim 0.6 $ to $ 0.8 \, M_\odot $ for the models considered here
(see Fig.~\ref{figure:m_rest_inner_core}) and a fiducial radius of the
inner core at bounce of $ \sim 20 \mathrm{\ km} $, we obtain dynamical
times of $ 1.7\mbox{\,--\,}2.0 \mathrm{\ ms} $, which yield
characteristic angular frequencies of
$ \Omega_\mathrm{char} \approx 3\mbox{\,--\,}4 \mathrm{\ rad\ ms}^{-1} $.
Since the lowest order unstable mode is likely to have a pattern speed
of the order of $ \Omega_\mathrm{char} $, most models whose angular
velocity we plot in Fig.~\ref{figure:m_rest_inner_core} may indeed
have corotation points with an unstable mode, hence could undergo a
corotation-type low-$ \beta $ instability. Slow rotators (with
$ \Omega_\mathrm{c,i} \lesssim 2 \mathrm{\ rad\ s}^{-1} $) do not
appear to reach a sufficiently high angular velocity in the inner
proto-neutron star core to have corotation points with potentially
unstable modes in the first several tens of milliseconds after bounce.
However, this may change at later times when the proto-neutron star
contracts and spins up.

Finally, we point out that our discussion is based on a very rough
estimate of the pattern speed for the lowest-order unstable azimuthal
mode. More reliable estimates can be made via multi-dimensional
perturbative analysis (see, e.g., \cite{saijo_06_a} in the context of
idealized models) or by performing a large set of numerical
simulations in three dimensions, which we plan to carry out in a
future study.


\section{Summary and conclusions}
\label{section:summary}

In this article we have presented results from a comprehensive set of
collapse simulations of rotating stellar iron cores to proto-neutron
stars, using the axisymmetric general relativistic
hydrodynamics code \coconut. Our simulations treat all the relevant
physics of the collapse phase to good approximation. They include
precollapse iron core profiles from stellar evolutionary
calculations, a highly efficient approximate treatment of
deleptonization, a microphysical finite-temperature EoS, as well as
neutrino pressure contributions. Magnetic fields are not included,
since their relevance in the collapse and early postbounce phases is
very likely negligible in cores with realistic precollapse
fields~\cite{heger_05_a, burrows_07_b, cerda_07_a,
  obergaulinger_06_b}.

The focus of our study is on procuring accurate and reliable waveforms
of the gravitational wave burst signal associated with core bounce and
on understanding the dependence of the signal characteristics on
progenitor star mass, precollapse rotational setup, and nuclear EoS. 
To this end we have performed the to-date most extensive
parameter study of this scenario, covering with more than 100 model
calculations the parameter space spanned by (1)
progenitor mass and model profile (zero-age main sequence masses from
$ 11.2 $ to $ 40\,M_\odot $, presupernova models with and without
rotation), (2) rotational configuration (slow and uniform to rapid and
differential rotation), and (3) nuclear EoS prescription (from
relatively soft to relatively stiff). Importantly, the parameter
space encompasses and even goes beyond all precollapse rotational
configurations that are deemed realistic in the context of collapsing
massive stars.

A central result of this work is the finding that the gravitational
wave burst from core bounce exhibits a generic waveform shape known as
type~I in the literature~\cite{moenchmeyer_91_a, zwerger_97_a},
independent of the model parameters. The multiple centrifugal bounce
dynamics and the corresponding type~II waveform found in previous,
technically less complete studies (see, e.g., \cite{moenchmeyer_91_a,
  zwerger_97_a, dimmelmeier_02_a, ott_04_a}) do not occur in our
models.

We have demonstrated that all models with precollapse core angular
velocities $ \Omega_\mathrm{c,i} $ below $ \sim 5 \mathrm{\ rad\ s}^{-1} $
(corresponding periods longer than about $ 1 \mathrm{\ s} $) reach
nuclear densities and experience a core bounce predominantly due to
nuclear pressure effects. More rapidly rotating cores develop
sufficient rotational support during collapse to undergo either a
mixture of centrifugal and pressure-dominated bounce or a
\emph{single} centrifugal bounce at subnuclear densities. Centrifugal
hang-up much below nuclear density or multiple, damped harmonic
oscillator-like centrifugal bounces do not occur. Therefore, these
models also exhibit a type~I waveform. The detailed analysis of the
collapse dynamics presented in this paper reveals that the
combined effects of general relativity and deleptonization lead to an
increased destabilization of the collapsing core, result in a
relatively small radius and mass $ M_\mathrm{ic,b} $ of the
sonically-connected inner core at bounce (but not small enough to show
the type~III waveform associated with rapid collapse found in some
previous simplistic models), and diminish the dynamical importance of
centrifugal forces during collapse.

The key parameter which determines the peak amplitude
$ |h|_\mathrm{max} $ of the gravitational wave burst has turned out to
be the precollapse central angular velocity $ \Omega_\mathrm{c,i} $.
Slowly rotating cores with
$ \Omega_\mathrm{c,i} \lesssim 1 \mathrm{\ rad\ s}^{-1} $ produce
feeble peak amplitudes on the order of $ 10^{-22} $ at a distance of
$ 10 \mathrm{\ kpc} $. More rapidly rotating cores with
$ 1 \mathrm{\ rad\ s}^{-1} \lesssim \Omega_\mathrm{c,i} \lesssim
6 \mathrm{\ rad\ s}^{-1} $ develop stronger quadrupole deformations and
have a rotationally-increased mass $ M_\mathrm{ic,b} $ at bounce,
resulting in sizeable peak amplitudes in the range of
$ 5 \times 10^{-22} \lesssim |h|_\mathrm{max} \lesssim 10^{-20} $. The
peaks of the waveform spectrum of such cores cluster in frequency
space in the interval of $ 650\mbox{\,--\,}800 \mathrm{\ Hz} $. At
larger $ \Omega_\mathrm{c,i} $, centrifugal effects become strong,
significantly decelerate collapse and bounce, and even lead to a
purely centrifugal bounce in a subset of models. This results in a
general decrease of $|h|_\mathrm{max}$ and a shift of the waveform's
spectral peak to frequencies below $ \sim 400 \mathrm{\ Hz} $ at high
$ \Omega_\mathrm{c,i} $.

We have also shown that, in addition to $ \Omega_\mathrm{c,i} $, the
precollapse core mass in combination with the electron fraction sets
the mass $ M_\mathrm{ic,b} $ of the inner core at bounce, is an important
quantity influencing the strength of the gravitational
burst. Since more massive progenitors generally (though with notable
non-monotonicity in the mass range from about $ 18 $ to
$ 23\,M_\odot $) form larger iron cores, we observe in our model
series a general trend to bigger $ M_\mathrm{ic,b} $ and
larger $ |h|_\mathrm{max} $ with increasing progenitor mass if all
other parameters are kept constant. For instance, the $ 40\,M_\odot $
progenitor yields values of $ |h|_\mathrm{max} $ which are up to 4
times larger than for the lower-entropy $ 11.2\,M_\odot $ counterpart
with the same rotational configuration.

The variations in the degree of differential rotation considered in
this study have only a minor impact on the collapse dynamics and
burst waveform amplitude. Increasing differential rotation at fixed
$ \Omega_\mathrm{c,i} $ generally lowers the centrifugal support of
outer core regions. However, since the dynamically most relevant inner
core at bounce consists of only $ \sim 0.5\mbox{\,--\,}1\,M_\odot $
located within about $ 1000 \mathrm{\ km} $ at the onset of collapse,
the effects of differential rotation on the gravitational wave burst
are small.

Our results further indicate that the nuclear EoS has little influence
on the gravitational wave burst signal. For a given precollapse
configuration, a softer nuclear EoS yields higher densities at bounce
and postbounce times with shorter variation timescales of the
quadrupole moment, but also leads to greater inner core
compactness. In our simulations, the two effects generally cancel,
leading to no systematic trend in the peak waveform amplitude
$ |h|_\mathrm{max} $ with the EoS. The peak of the waveform
spectrum, however, shifts to higher frequencies in the case of a
softer EoS. For the models considered here, this frequency shift
amounts to $ \sim 5.5\% $ on average for models undergoing
pressure-dominated bounce. It is significantly smaller for models
bouncing at subnuclear densities under the influence of centrifugal
effects.

If situated within our Galaxy, a large fraction of our models are
comfortably detectable by current gravitational wave detectors with a
signal-to-noise ratio of up to 6 in the most optimistic case (which
is obtained for the most rapidly rotating models that still undergo
pressure-dominated core bounce). Advanced detectors could observe them
easily out to $ \sim 100 \mathrm{\ kpc} $ and up to several
$ 10 \mathrm{\ Mpc} $ for third-generation detectors.

While such a gravitational wave signal may per se be detectable, the
extraction of detailed physical information from the signal (i.e.,
solving the ``inversion problem'') from the signal will be a
formidable task. The very generic morphology of the burst waveforms
and the clustering in frequency space of most models make it seem
unlikely that a pure waveform-template-based inversion (as, e.g.,
carried out in~\cite{summerscales_08_a} using the waveforms
of~\cite{ott_04_a}) can be successful for determining key physical
parameters to significant precision. Our results, however, suggest
that based on $ |h|_\mathrm{max} $ and the peak frequency
$ f_\mathrm{max} $ of the waveform spectrum alone, it should be
possible to discriminate between purely pressure-dominated bounce
(small to large $ |h|_\mathrm{max} $ at frequencies $ f_\mathrm{max} $
significantly above $ 500 \mathrm{\ Hz} $) and centrifugal bounce
(large $ |h|_\mathrm{max} $ at frequencies $ f_\mathrm{max} $
significantly below $ 500 \mathrm{\ Hz} $). Furthermore, we find that
for not too rapid rotation $ |h|_\mathrm{max} $ can be directly used
to extract the rotation rate $ \beta_\mathrm{b} $ at bounce to good
precision.

Making use of the extensive set of postbounce rotational
configurations obtained with our simulations, we have also studied the
prospects for the development of nonaxisymmetric rotational
instabilities in nascent proto-neutron stars. We find that the
rotational barrier imposed by centrifugal forces prohibits
the spin-up to rotation rates necessary for the classical
dynamical bar-mode instability at high values of $ \beta $. We find,
however, that a large subset of our postbounce models exhibits
sufficiently differential and rapid rotation to become subject to the
recently discovered low-$ \beta $ instability. Still,
three-dimensional simulations as in~\cite{rampp_98_a, ott_05_a,
  ott_07_a, ott_07_b} will be necessary to provide conclusive tests of
our predictions. Furthermore, the interaction and competition of the
low-$ \beta $ instability and other instabilities operating on the
shear energy of differential rotation, for instance the
magneto-rotational instability (see, e.g., \cite{balbus_91_a,
  cerda_07_a}), remain to be studied.

Finally, we point out that this study may be regarded as part --
with the presently highest level of sophistication -- of a
multi-decade effort of our groups~\cite{mueller_82_a,
  moenchmeyer_91_a, zwerger_97_a, dimmelmeier_02_a, ott_04_a,
  dimmelmeier_07_a, ott_07_a} to provide reliable estimates for the
gravitational wave burst emission associated with rotating core
collapse and core bounce. The waveforms presented here are for the
first time not only accurate (i.e., numerically converged), but
reliable and robust, since our calculations take into account all the
necessary physics, including general relativity, deleptonization and a
microphysical EoS. All waveforms are available for download in various
formats in a publicly accessible waveform catalog~\cite{wave_catalog}.

We point out that the gravitational wave emission process considered in
this work operates at measurable strength only if the progenitor core
is rotating a lot more rapidly than expected for ordinary iron cores
(see, e.g., \cite{ott_06_c, heger_05_a}). In slowly rotating
core-collapse supernovae, turbulent convective overturn, instabilities
of the accretion shock, and, possibly, proto-neutron star pulsations are
likely to be the dominant emission processes of gravitational
waves. The characteristics of these emission processes are not as well
understood and will require more extensive and precise modeling
to provide accurate estimates of the complete gravitational wave
signature of core-collapse supernovae.


\acknowledgments

It is a pleasure to thank Shizuka Akiyama, David Arnett, Adam Burrows,
Luc Dessart, Pablo Cerd\'a-Dur\'an, Ian Hawke, Alex Heger, Ewald
M\"uller, Shangli Ou, Jos\'e Pons, Erik Schnetter, Ed Seidel, Bernard
Schutz, Todd Thompson, Joel Tohline, and Burkhard Zink for helpful
comments and inspiring discussions. This work was supported by the
Deutsche Forschungsgemeinschaft through the Transregional
Collaborative Research Centers SFB/TR~27 ``Neutrinos and Beyond'',
SFB/TR~7 ``Gravitational Wave Astronomy'', and the Cluster of
Excellence EXC~153 ``Origin and Structure of the Universe''
(\texttt{www.universe-cluster.de}), by the DAAD and IKY (IKYDA
German--Greek research travel grant), and by the European Network of
Theoretical Astroparticle Physics ENTApP ILIAS/N6 under contract
number RII3-CT-2004-506222. H.D.\ is a Marie Curie Intra-European
Fellow within the 6th European Community Framework Programme (IEF
040464). C.D.O.\ is supported by the Joint Institute for Nuclear
Astrophysics (JINA) under NSF sub-award no.~61-5292UA of NFS award
no.~86-6004791. The authors wish to thank the Max Planck Institute for
Gravitational Physics and the John von Neumann-Institut f\"ur
Computing (NIC) in J\"ulich where the calculations presented in this
paper were performed.


\end{document}